\documentclass[11pt]{article}
\usepackage{verbatim,amsmath,amssymb}
\usepackage{epsfig,color}
\usepackage{geometry}
\usepackage{setspace}
\usepackage{natbib}
\usepackage{amsmath,amssymb,amsthm,mathrsfs,amsfonts,dsfont} 
\usepackage{hyperref}
\usepackage{soul}
\usepackage{float}
\usepackage{graphicx}
\usepackage{scalerel}
\usepackage{todonotes}
\usepackage[font=small,labelfont=bf]{caption}

\definecolor{darkblue}{rgb}{0,0,1}
\hypersetup{pdftex=true, colorlinks=true, breaklinks=true, linkcolor=darkblue, menucolor=darkblue, pagecolor=darkblue, citecolor=darkblue, urlcolor=darkblue}

\input{neco_updated.tex}

\newtheoremstyle{rem}
{6pt}
{6pt}
{\small}
{}
{\bf}
{:}
{.5em}
{}
\theoremstyle{rem}
\newtheorem{remark}{Remark}[section]

\makeatletter
\newcommand\footnoteref[1]{\protected@xdef\@thefnmark{\ref{#1}}\@footnotemark}
\makeatother

\renewcommand{\thefootnote}{\fnsymbol{footnote}}

\begin{document}




\begin{center}
\Large{\bf{A general isogeometric finite element formulation for rotation-free shells with in-plane bending of embedded fibers}}\\

\end{center}

\begin{center}
\large{Thang X. Duong$^{a,b}$, Mikhail Itskov$^{b,}$\footnote[1]{\label{note0} {corresponding authors, email: itskov@km.rwth-aachen.de; sauer@aices.rwth-aachen.de}}, and Roger A. Sauer$^{a,c,d,}$\footnoteref{note0}}\\
\vspace{4mm}

\small{$^a$\textit{Aachen Institute for Advanced Study in Computational Engineering Science (AICES), RWTH Aachen
University, Templergraben 55, 52056 Aachen, Germany}}

\small{$^b$\textit{Department of Continuum Mechanics, RWTH Aachen
University, Templergraben 55, 52056 Aachen, Germany}}

\small{$^c$\textit{Faculty of Civil and Environmental Engineering, Gda{\'n}sk University of Technology, ul. Narutowicza 11/12, 80-233 Gda{\'n}sk, Poland}}

\small{$^d$\textit{Department of Mechanical Engineering, Indian Institute of Technology Kanpur, UP 208016, India}}

\end{center}


%
%
%
%
%

\renewcommand*{\thefootnote}{\arabic{footnote}}
\setcounter{footnote}{0}
\begin{center}
Published\footnote{This pdf is the personal version of an article whose journal version is available at\\ \href{https://onlinelibrary.wiley.com/doi/full/10.1002/nme.6937}{https://onlinelibrary.wiley.com/}} in International Journal for Numerical Methods in Engineering,\\ \href{https://doi.org/10.1002/nme.6937}{DOI: 10.1002/nme.6937}\\
Submitted on 20 October 2021; Revised on 30 January 2022; Accepted on 31 January 2022
\end{center}

\vspace{2mm}
\rule{\linewidth}{.15mm}
{\bf Abstract:}
This paper presents a general, nonlinear {isogeometric} finite element formulation for rotation-free shells with embedded fibers that captures anisotropy in stretching, shearing, twisting and bending -- both in-plane and out-of-plane. These capabilities allow for the simulation of large sheets of heterogeneous and fibrous materials either with or without matrix, such as textiles, composites, and pantographic structures.  The work is a computational extension of our earlier theoretical work \cite{shelltextile} 
{that} extends  existing Kirchhoff-Love shell theory to incorporate the in-plane bending resistance of initially straight or curved fibers.  The formulation requires only displacement degrees-of-freedom to capture all mentioned modes of deformation. To this end, isogeometric shape functions are used in order to satisfy the required $C^1$-continuity for bending across element boundaries.  The proposed formulation can admit a wide range of material models, such as surface hyperelasticity that does not require any explicit thickness integration. To deal with possible material instability due to fiber compression, a stabilization scheme is added. 
Several benchmark examples are used to demonstrate the robustness and accuracy of the proposed computational formulation.
 

\vspace{-5mm}
\rule{\linewidth}{.15mm}
\vspace{1mm}
\section{Introduction}

The computational simulation of fiber reinforced composites has become an essential tool in designing products, for example in the automotive, aerospace, biomedical and sports industry. Besides, computational simulations  play an important role in analyzing the production process of such composite materials itself. For example, for woven and non-crimp fabric composites, this process can include the production of textile fabrics, the stacking of fabric layers, the draping and fixation of the stack to the desired shape in a mould, and the injection of matrix materials into the mould for bonding fibers in the final product.  Likewise, simulations help in designing pantographic structures and reinforcement layouts for reinforced concrete structures.

In the above mentioned applications, finite shell elements based on the classical Cauchy continuum {for the membrane response} are predominantly used to model textile fabrics (e.g.~see \cite{Yu2005,Boisse2008,Boisse97,KhiemNCF2018}). This choice usually provides a good prediction of the overall behavior of fabrics, especially for fibers strongly bonded to the matrix. However, it fails to reproduce localized deformations due to the in-plane bending resistance of embedded fibers. The influence of the in-plane bending stiffness becomes significant when there is a large change in the in-plane curvature. This happens for example in shear bands  occurring when dry fabrics are deformed \cite{Boisse17}. Numerical simulations using Cauchy-based shell formulations will fail to converge to a finite width of the shear bands. Essentially, the classical Cauchy continuum, and numerical methods based on it, are inconsistent with the observed behavior.
 
The in-plane bending stiffness can affect not only  the localized deformation, but also the global deformation.  This is shown in Madeo et al.~\cite{Madeo2016} and Barbagallo et al.~\cite{Barbagallo17} for the bias extension test of  so-called unbalanced woven fabrics, where the in-plane bending stiffness varies between fiber families. As observed in the experiment \cite{Madeo2016,Barbagallo17}, the global deformation is asymmetric.  Numerical simulations with Cauchy-based shell formulations will also fail to produce such shapes.
 
The inability to {properly} respond to in-plane bending deformations is due to the underlying fundamental assumption of the Cauchy continuum  that the corresponding bending moment vanishes at a material point. A more general continuum model is thus required and can be provided by  {Cosserat theories,  e.g.~\cite{Mindlin1962,Koiter63b,toupin_theories_1964},   or  
strain 
gradient theories, see e.g.~\cite{Mindlin65,Germain73}.  Both have been used to explicitly account for fiber bending: Steigmann~\cite{steigmann_theory_2012} presents a Cosserat theory for the bending resistance of fibers embedded  in 3D solids, while other theoretical works  adopt strain gradient theories to describe fiber-reinforced solids \cite{spencer_finite_2007,soldatos2010}, fabric plates \cite{Steigmann2015}, and shells \cite{Steigmann2018}.} 

In the literature, there exist also computational models for 
{gradient theory.} Ferretti et al.~\cite{Ferretti2014} present a computational formulation for a so-called constrained micromorphic theory including a 
{second-gradient}\footnote{i.e.~the second displacement gradient} model, like the one of Germain~\cite{Germain73}, as a special case. In order to reproduce the bias extension test for unbalanced fabrics, Madeo et al.~\cite{Madeo2016} further {extend} the constrained micromorphic continuum model and its corresponding numerical formulation such that it can  capture the change in the relative fiber angles, the variation of the bending stiffness between fiber families, and also the relative slipping of the tows. A finite element formulation for the gradient model of Spencer and Soldatos~\cite{spencer_finite_2007} {is} presented by Asmanoglo and Menzel~\cite{Asmanoglo17}. Here, the $C^1$-continuity requirement for the second-gradient terms is relaxed by additional field variables coupled to the deformation gradient.  

The computational formulations mentioned so far have focused only on plane strain problems. A general second-gradient shell formulation that explicitly accounts for in-plane fiber bending, as considered here, is still missing.
It is worth noting that there are also discrete formulations capable of capturing in-plane bending, either using interacting particles \cite{Antonio2017}, or  grids of  Euler–Bernoulli beams interconnected by pivots at the intersection points \cite{dagostino_continuum_2015}, or interconnected by rotational and translational elastic springs \cite{Madeo2016}.

 


An important development of recent years are high order approximation methods that provide a more accurate and smoother description of computational domains.  In particular, the advent of  so-called \textit{isogeometric analysis} (IGA)  \cite{hughes05} offers  significant advantages over the classical finite element method.  Its ability to describe a surface with high accuracy and smoothness facilitates the recent advancement of so-called \textit{rotation-free} shell formulations. In such formulations, the unknowns per node contain only three displacement degrees-of-freedom, while rotations are obtained from the surface displacement. This is feasible when the discretized geometry is smooth and accurate. Therefore, the combination of IGA with rotation-free shells can increase both accuracy and efficiency of computational formulations. The work of Kiendl et al.~\cite{kiendl09} is the first combining IGA with rotation-free shells. Since then,  rotation-free IGA shells have been steadily advanced, for example to PHT-splines \cite{thanh11}, anisotropic materials \cite{nagy13}, damage \cite{deng15}, biological materials \cite{TEPOLE2015},  fracture \cite{Kiendl16}, liquid shells \cite{liquidshell}, elasto-plasticity \cite{Ambati18}, phase separation \cite{Zimmermann19},  thermo-mechanical coupling \cite{Namvu2019}, 
 {multi-patch constraints} (e.g.~see the recent review in Paul et al.~\cite{Paul2020}), and reduced quadrature \cite{Zou2021}.
%
Balobanov et al.~\cite{Balobanov19} {have presented} a general strain gradient theory  and its corresponding isogeometric finite element formulation for Kirchhoff-Love shells. The formulation requires at least $C^2$-continuity of the geometry, but does not account for in-plane fiber bending explicitly.
%
%
%
%
%

 
A formulation for rotation-free isogeometric shells that can capture in-plane bending of embedded fibers has only recently been presented by Schulte et al.~\cite{Schulte2020}.
While the formulation of Schulte et al.~is formulated for Kirchhoff-Love shell elements, its underlying theory is based on the strain-gradient theory of  Steigmann~\cite{Steigmann2018} for shells with embedded rods. In 
{this} theory, 
{the} strain tensor related to  in-plane curvature is of third order, since it expresses the  relative change in the surface Christoffel symbols. Strickly speaking, the Christoffel symbols are not tensor components since they do not transform as {such}. 
From the material modeling point of view it can thus be inconvenient to formulate invariants of such a strain tensor and interpret their geometrical meaning.
Further, the theory  of  Steigmann~\cite{Steigmann2018} and the implementation of Schulte et al.~\cite{Schulte2020} are restricted to two fiber families that are  initially straight. 
Another IGA-based finite element formulation for the gradient model of Spencer and Soldatos~\cite{spencer_finite_2007} has been presented recently by Witt et al.~\cite{witt_finite_2021}.  However, it is not a shell formulation and it is also restricted to initially straight fibers. 



In Duong et al.~\cite{shelltextile}, we have proposed an advancement  that directly extends Kirchhoff-Love shell theory to incorporate general  in-plane fiber bending. Although this approach follows the straightforward structure of the classical Kirchhoff-Love shell, the resulting theory has no restriction on the initial state of fibers, the number of fiber families, and also the initial angle between them. Another advantage of the approach is that it directly uses second order surface tensors to characterize the deformation, including in-plane bending, which facilitates the induction of invariants.

In this contribution, we present a rotation-free isogeometric finite element formulation based on the theory  by Duong et al.~\cite{shelltextile}. The proposed formulation can capture anisotropy in stretching, shearing, twisting and bending -- both in-plane and out-of-plane.
The 
{formulation is fully presented}
 in the curvilinear coordinate system, which avoids the use of local Cartesian coordinate  transformations at the element level. 
In summary, our contribution contains the following novelties and merits:
  
$\bullet$   It is based on a  generalized Kirchhoff-Love shell theory that captures in-plane bending.

$\bullet$  It uses second order tensors for in-plane bending, which facilitates inducing invariants.

$\bullet$  It is analogous to classical rotation-free isogeometric finite shell element formulations.

$\bullet$ 
{It} admits initially curved fibers, multiple fiber families and general initial fiber angles.

$\bullet$ 
It avoids transforming derivatives into Cartesian coordinates at the element level.

$\bullet$ It includes the full linearization and efficient implementation  for IGA-based finite elements.


  
The remaining presentation of the paper is structured as follows: Sec.~\ref{s:SummaryKL}  summarizes the generalized Kirchhoff-Love shell theory of  Duong et al.~\cite{shelltextile}. 
Sec.~\ref{s:linweakform} presents the linearization of its weak form and the introduction of the new material tangents associated with in-plane bending. Sec.~\ref{s:fe} discusses the {isogeometric} finite element discretization of the formulation. Two material models for simple fabrics and woven fabrics are given in Sec.~\ref{s:two_material_model}. Secs.~\ref{s:num_examples1} and \ref{s:num_examples2} illustrate the performance of the proposed formulation by numerical examples with homogeneous and inhomogeneous deformations, respectively. Sec.~\ref{s:conclusion} concludes the paper.


\section{Summary of generalized Kirchhoff-Love shell theory}{\label{s:SummaryKL}}

This section summarizes the kinematics, stresses, moments, weak form and constitutive equations according to the  generalized thin shell theory of Duong et al.~\cite{shelltextile}.

\subsection{Geometrical description of fiber-embedded surfaces}{\label{s:geodescription}}
The mid-surface $\sS$ of a thin shell at time $t$ is represented in curvilinear coordinates $(\xi^1\,,\xi^2)\in\sP$ by
\eqb{l}
\bx=  \bx(\xi^\alpha,t)~,\quad $with$ \quad \alpha=1,2~.
\eqe
At any point $\bx\in\sS$, a curvilinear basis can be constructed from two (covariant) tangent vectors $\ba_\alpha$ and 
{a} unit normal vector $\bn$ to surface $\sS$. They are defined by
\eqb{l} 
 \ba_\alpha:=\ds\pa{\bx}{\xi^\alpha} = \bx_{,\alpha}~,\quad $and$ \quad \bn := \ds\frac{\ba_1\times\ba_2}{\norm{\ba_1\times\ba_2}}~,
 \label{e:tangentsl}
\eqe
where the comma denotes the parametric derivative. The dual tangent vectors $\ba^\alpha$ are related to the covariant tangent vectors by $\ba_{\alpha}=a_{\alpha\beta}\,\ba^\beta$ and $\ba^{\alpha}=a^{\alpha\beta}\,\ba_\beta$.\footnote{Here and henceforth, the summation convention is applied to repeated Greek indices taking values 1 and 2.} Here, $a_{\alpha\beta}:=\ba_\alpha\cdot\ba_\beta$ and  $a^{\alpha\beta}:=\ba^\alpha\cdot\ba^\beta  $ denote the surface metrics.
 
\begin{figure}[htp]
\begin{center} \unitlength1cm
\begin{picture}(0,7)
\put(-6.5,0.0){\includegraphics[width=0.79\textwidth]{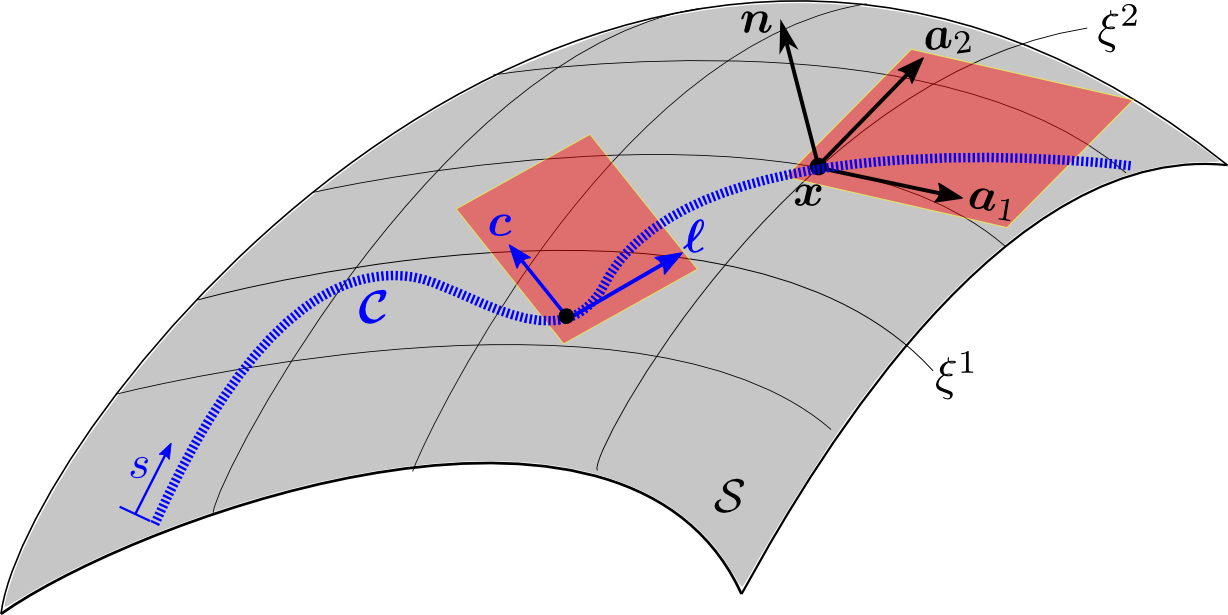}}
\end{picture}
\caption{A fiber bundle represented by curve $\sC$ embedded in shell surface $\sS$. The red planes illustrate tangent planes \cite{shelltextile}}
\label{f:kin}
\end{center}
\end{figure} 
Consider a fiber curve (or a curve of fiber bundles) $\sC$ embedded in surface $\sS$ and given by  $\bx = \bx(s)$ (see Fig.~\ref{f:kin}).  Its normalized tangent vector at location $s$ can be defined by
\eqb{l} 
{\bell}:= \ds \frac{\partial \bx}{\partial s} = \ell_\alpha\,\ba^\alpha = \ell^\alpha\,\ba_\alpha~,
\label{e:belldef}
\eqe
%
%
while the so-called in-plane fiber director $\bc$, perpendicular to $\bell$, can be defined by
\eqb{l} 
\bc := \bn \times{\bell} = c_\alpha\,\ba^\alpha =  c^\alpha\,\ba_\alpha ~.
\label{e:vectorcdef}
\eqe

%


The out-of-plane curvature of surface $\sS$ can be described by the symmetric second order  tensor
\eqb{l} 
\bb :=  b_{\alpha\beta}\,\ba^\alpha\otimes\ba^\beta~,
\label{e:tensorb}
\eqe
with the components expressed by
\eqb{l} 
{b}_{\alpha\beta}:= -\bn_{,\alpha}\cdot\ba_\beta =  \bn\cdot\ba_{\alpha,\beta} = \bn\cdot\ba_{\alpha;\beta}~.
\label{e:bab}
\eqe
Here,
\eqb{lll} 
 \ba_{\alpha,\beta} := \ds\pa{\ba_\alpha}{\xi^\beta} = \bx_{,\alpha\beta} = \Gamma^\gamma_{\alpha\beta}\,\ba_\gamma + b_{\alpha\beta}\,\bn~, \quad$and$\quad
 \ba_{\alpha;\beta} := ( \bn\otimes\bn)\, \ba_{\alpha,\beta}
 \label{e:curvector}
\eqe
are the parametric and covariant derivative of $\ba_\alpha$,  respectively. In Eq.~(\ref{e:curvector}.1), $\Gamma^\gamma_{\alpha\beta}:=\ba_{\alpha,\beta}\cdot\ba^\gamma$ denote the  surface Christoffel symbols.  They can be expressed as  
\eqb{l} 
\Gamma^\gamma_{\alpha\beta}= c^\gamma\, \Gamma^{\mrc}_{\alpha\beta} + \ell^\gamma\, \Gamma^{\uell}_{\alpha\beta} ~,  
\eqe
where 
\eqb{lll} 
\Gamma_{\!\alpha\beta}^{\mrc} \dis \bc\cdot\ba_{\alpha,\beta} = c_\gamma\,\Gamma^\gamma_{\alpha\beta} ~, \\[2mm]
  \Gamma_{\!\alpha\beta}^{\uell} \dis \bell\cdot\ba_{\alpha,\beta}=\ell_\gamma\,\Gamma^\gamma_{\alpha\beta}~.
\label{e:Gama_c_and_ell}
\eqe
Furthermore, in order to characterize in-plane curvatures, the so-called in-plane curvature tensor $ \bbbar$ of fiber $\sC$ is defined as the (negative) symmetric part of the projected surface gradient of  director $\bc$. That is, 
 \eqb{l}
 \bbbar:=  -\frac{1}{2} \left[\bar\nabla_{\!\mrs} \bc  + (\bar\nabla_{\!\mrs} \bc)^\mrT \right] = 
\bar{b}_{\alpha\beta}\,\ba^\alpha\otimes\ba^\beta~,
\label{e:binplane} 
 \eqe
where 
{$\bar\nabla_{\!\mrs} \bullet : =  (\bullet_{,\beta}\cdot \ba^\alpha)\, \ba_\alpha\otimes\ba^\beta$} denotes the projected surface gradient operator.\footnote{ $\bar\nabla_{\!\mrs} \bullet := \bi \,\nabla_{\!\mrs} \bullet$, with
$\bi= \bone - \bn \otimes \bn = \ba_\alpha\otimes\ba^\alpha$ and $\nabla_{\!\mrs} \bullet = \bullet_{,\alpha} \otimes \ba^\alpha$.
}  In Eq.~\eqref{e:binplane}, components $\bar{b}_{\alpha\beta}$ can be computed from
\eqb{l}
\bar{b}_{\alpha\beta} =  -\frac{1}{2} ( c_{\alpha;\beta} + c_{\beta;\alpha}) =  -\frac{1}{2} ( \bc_{,\alpha}\cdot\ba_\beta + \bc_{,\beta}\cdot\ba_\alpha) =   -\frac{1}{2} ( \bar\bc_{,\alpha}\cdot\ba_\beta + \bar\bc_{,\beta}\cdot\ba_\alpha) ~,
\label{e:binplane2} 
\eqe
where 
$
\bc_{,\alpha} =  b_{\alpha\beta}\,c^\beta\,\bn
+  c^\beta_{;\alpha}\,\ba_\beta~$,
%
%
 {and}
\eqb{rlll}
\bar\bc_{,\alpha} := (\ba_\beta\otimes\ba^\beta)\,  \bc_{,\alpha}  =  c^\beta_{;\alpha}\,\ba_\beta~
\label{e:coderivbcbar}
\eqe
is the projection of $\bc_{,\alpha}$ onto the tangent plane.


\subsection{Shell deformation}

Shell deformation is measured with respect to the reference configuration $\sS_0$ at time $t_0$. Analogous to Sec.~\ref{s:geodescription}, we define geometrical objects  on $\sS_0$, such as the tangent vectors $\bA_\alpha$, the normal vector $\bN$, the metric $A_{\alpha\beta}$, the out-of-plane curvature tensor $\bb_0:=B_{\alpha\beta}\bA^\alpha\otimes\bA^\beta$,  the fiber direction $\bL=L^\alpha\bA_\alpha=L_\alpha\bA^\alpha$, the fiber director $\bc_0 = c^0_\alpha\,\bA^\alpha$, and the in-plane curvature tensor $\bbbar_0:= \bar{B}_{\alpha\beta}\,\bA^\alpha\otimes\bA^\beta$. The deformation of the fiber-embedded shell can then be characterized by the following quantities:

1. The surface deformation gradient tensor,
\eqb{l}
\bF:=\ba_\alpha\otimes\bA^\alpha~.
\label{e:def_F}
\eqe
It can be used to construct surface strain measures such as the right Cauchy-Green surface tensor $\bC:=\bF^\mrT\,\bF = a_{\alpha\beta}\bA^\alpha\otimes\bA^\beta$, and the  Green-Lagrange surface strain tensor
\eqb{l}
\bE  :=\ds\frac{1}{2}(\bC - \bI) =\ds \frac{1}{2}\,(a_{\alpha\beta} - A_{\alpha\beta})\, \bA^\alpha\otimes\bA^\beta= E_{\alpha\beta}\,\bA^\alpha\otimes\bA^\beta~.
\label{e:bEtensor}
\eqe

2. The \textit{relative out-of-plane curvature tensor},
\eqb{l}
\bK :=  \bF^T\,\bb\,\bF - \bb_0 = 
  (b_{\alpha\beta} - B_{\alpha\beta})\,\bA^\alpha\otimes\bA^\beta= K_{\alpha\beta}\,\bA^\alpha\otimes\bA^\beta~.
\label{e:Kten}
\eqe
3. The \textit{relative in-plane curvature tensor},
\eqb{l}
\bar\bK :=  \bF^T\,\bbbar\,\bF - \bbbar_0 =  ( \bar{b}_{\alpha\beta} - \bar{B}_{\alpha\beta})\,\bA^\alpha\otimes\bA^\beta = \bar{K}_{\alpha\beta}\,\bA^\alpha\otimes\bA^\beta~.
\label{e:Ktenb}
\eqe


\begin{remark}\label{r:remark2}
Note, that apart from definition~(\ref{e:belldef}.1), the fiber direction vector $\bell$ can also be computed from the (given) reference fiber direction vector $\bL $ via the mapping
\eqb{l}
\lambda\,\bell = \bF\,\bL = L^\alpha\,\ba_\alpha~,
\label{e:bellhatF}
\eqe
where $\lambda$ is the fiber stretch. From this and relation \eqref{e:vectorcdef}, one thus obtains 
\eqb{lll}
c_{\beta;\alpha} \is   - \ell_{\beta}\,   (c^\gamma\,  \hat{L}_{\gamma,\alpha} + \ell^\gamma\,\Gamma^\mrc_{\gamma\alpha})~,\\[2mm]
c^{\beta}_{;\alpha}~ \is   -  \ell^{\beta}\, (c^\gamma\,  \hat{L}_{\gamma,\alpha} + \ell^\gamma\,\Gamma^\mrc_{\gamma\alpha})~,
\label{e:coderivbcDef}
\eqe
on the basis of the definition
\eqb{l}
 \hat{L}_{\alpha,\beta} := a_{\alpha\gamma}\, \hat{L}^\gamma_{,\beta} ~,\quad $with$  \quad  \hat{L}^\alpha_{,\beta}:=\lambda^{-1}\,L^\alpha_{,\beta}~.
\label{e:Lhatdefine}
\eqe
\end{remark}


\begin{remark}
Inserting \eqref{e:coderivbcDef} into \eqref{e:coderivbcbar} gives
\eqb{rlll}
\bar\bc_{,\alpha} = -(c^\gamma\,  \hat{L}_{\gamma,\alpha} + \ell^\gamma\,\Gamma^\mrc_{\gamma\alpha})\,\bell~.
\label{e:coderivbcbar_ell}
\eqe
\end{remark}

%
\begin{remark}
The right Cauchy-Green tensor $\bC$ and the relative curvature tensors $\bK$ and $\bar\bK $ are all symmetric and of second order. They induce various invariants that can be useful for the constitutive modeling. For example,
\eqb{lll} 
\Lambda \dis \bC:\bL\otimes\bL = a_{\alpha\beta}\, L^{\alpha\beta} = \lambda^2 ~, \quad~$with$~\quad L^{\alpha\beta}  :=L^\alpha\,L^\beta~,\\[3mm]
K_\mrn \dis \bK:\bL\otimes\bL = (b_{\alpha\beta} - B_{\alpha\beta})\, L^{\alpha\beta} ~,\\[3mm]
T_\mrg \dis \bK:\bc_0\otimes\bL =\bK:\bL\otimes\bc_0  =  (b_{\alpha\beta} - B_{\alpha\beta})\,L^\alpha\,c_0^\beta~,\\[3mm]
K_\mrg \dis \bar\bK:\bL\otimes\bL = (\bar{b}_{\alpha\beta} -\bar{B}_{\alpha\beta})\,L^{\alpha\beta}~,
\label{e:K_invariant}
\eqe
express the square of the fiber stretch, the so-called \textit{nominal change in normal curvature},  the \textit{nominal change in geodesic torsion}, and  the \textit{nominal change in geodesic curvature} of the curve $\sC$, respectively  (see~\cite{shelltextile}). It should be noted that the measures  $K_\mrn$, $T_\mrg$, and  $K_\mrg$  are not invariants in a strict sense since their sign is not invariant (although their magnitude  still is). Specifically, the sign of  $K_\mrn$ and $T_\mrg$ changes when surface director $\bN$ is flipped, while the sign of $K_\mrg$  depends on the sign of both $\bN$ and $\bL$ due to Eqs.~\eqref{e:vectorcdef} and \eqref{e:binplane2}.
\end{remark}
 %


\subsection{Stress and moment tensors}

Consider cutting the shell $\sS$ virtually apart at $\bx\in\sS$ by the line $\sI(s)$ characterized by the unit tangent vector $\btau:=\partial{\bx}/\partial{s}$ and the unit normal $\bnu:=\btau\times\bn = \nu_\alpha\,\ba^\alpha$. The traction and moment vectors\footnote{\makebox[0.7\textwidth][s]{with the units [force/length] and [moment/length] commonly used in shell theory to avoid thickness integration}} appearing at the cut are general vectors in $\mathbb R^3$ that can be expressed as
\eqb{lll}
\bT \is T^\alpha\,\ba_\alpha + T^3\,\bn~,\\[3mm]
{\bmhat} \is m_\tau\,\btau + m_\nu\,\bnu +  \bar{m}\,\bn ~,
\label{e:tractionTdef}
\eqe
respectively. The last equation implies that the moment vector ${\bmhat}$ includes a moment $\bm := m_\tau\,\btau + m_\nu\,\bnu$ that causes  out-of-plane bending and twisting, and a moment  $\bmbar:=\bar{m}\,\bn$
that causes in-plane bending.
%
%
%
The traction and moment vectors \eqref{e:tractionTdef} induce corresponding internal 
{stresses} and moment tensors of the form
\eqb{lll}
\bsig \is \Nab\,\ba_\alpha\otimes\ba_\beta +  S^\alpha\, \ba_\alpha\otimes\bn~,\\[2mm]
{\bmuhat} \is  \mab\ba_\alpha\otimes\ba_\beta + \bar{m}^\alpha\,\ba_\alpha\otimes\bn~.
\label{e:stresssig_moment}
\eqe
According to Cauchy's theorem, these tensors linearly map the cut normal $\bnu$ to the traction and moment vectors \eqref{e:tractionTdef} as
\eqb{lll}
\bT =  \bsig^\mrT\,\bnu~, \quad $and$ \quad
{\bmhat} = \bmuhat^\mrT\,\bnu~.
\label{e:CauchyForm}
\eqe
Since moment tensor ${\bmuhat}$ (\ref{e:stresssig_moment}.2) is  generally asymmetric, it is  more convenient to work with the corresponding stress couple tensor instead. To this end, Eq.~(\ref{e:CauchyForm}.2)  is rewritten as 
\eqb{lll}
{\bmhat} = \bm + \bmbar =  \bn\!\times\!\bM + \bc\!\times\!\bMbar,
\label{e:MomentCross}
\eqe
where 
\eqb{lll}
\bM = \bmu^\mrT\,\bnu~, \quad $and$\quad \bMbar =  \bmubar^\mrT\,\bnu =  -\bar{m}\,\bell~
\label{e:CoupleStressDef}
\eqe
denote the so-called stress couple vectors for  out-of-plane and in-plane bending, respectively. $\bmu$ and  $ \bmubar$  are the corresponding stress couple tensors. They can be expressed as
\eqb{lll}
\bmu= - \Mab\,\ba_\alpha\otimes\ba_\beta,\quad $and$\quad \bmubar= - \Mbarab\,\ba_\alpha\otimes\ba_\beta~.
\label{e:defbmuandbmubar}
\eqe
Note that, in order to relate the components of traction and moment vectors to the components of the internal stress and stress couple tensors, one can compare \eqref{e:tractionTdef} and \eqref{e:CauchyForm}. This gives
\eqb{llll}
\begin{aligned}[]
T^\alpha~~ \is \nu_\beta\,N^{\beta\alpha}~,\\[2mm]
T^3 ~~\is \nu_\alpha \, S^\alpha ~,~~\\[2mm]
\end{aligned}
\quad \quad \quad \quad
\begin{aligned}[]
m_\nu~~ \is \Mab\,\nu_\alpha\,\tau_\beta~,~~~~~~~~~~~~\\[2mm]
m_\tau ~~\is -\Mab\,\nu_\alpha\,\nu_\beta~,~~~~~~~~~~~~\\[2mm]
\bar{m} ~~ \is   \Mbarab\,\nu_\alpha \,\ell_\beta =  \bar{m}^\alpha\,\nu_\alpha~.~
\end{aligned}
\label{e:tractionT2}
\eqe
\subsection{Weak form and constitutive equations}

Consider the shell $\sS$  subjected to the external body force  $\bff = f^\alpha\,\vaua + p\,\bn$ on $ \sS$ and to the boundary conditions
\eqb{llll}
\bu = \bar\bu & $on$~\partial_u\sS~,  \quad ~ \bT = \bar\bT & $on$~\partial_t\sS~, \quad ~
\bmhat =  \overline{\bmhat} & $on$~\partial_m\sS~.
\label{e:bcweak}
\eqe
Here, $ \bar\bu $ is a prescribed displacement,  $\bar\bT$ is a prescribed boundary traction and $ \overline{\bmhat}$ is a prescribed bending  moment. The equilibrium of the shell 
is then governed by the balance of linear and angular momentum.
Accordingly, the weak form follows as 
\eqb{l}
G_\mathrm{in} + G_\mathrm{int} - G_\mathrm{ext} = 0  \quad\forall\,\delta\bx\in\sV~,
\label{e:wfu}
\eqe
where $\sV$ denotes the set of kinematically admissible variations that satisfies boundary condition~(\ref{e:bcweak}.1), and
\eqb{lll}
G_\mathrm{in} 
\is \ds\int_{\sS_0} \delta\bx\cdot\rho_0\,\dot\bv\,\dif A~, \\[4mm]
G_{\mathrm{int}} \is \ds\frac{1}{2}\int_{\sS_0}\, \tau^{\alpha\beta} \,\delta{a}_{\alpha\beta}\,\dif A + \int_{\sS_0}\Mab_0\,\delta{b}_{\alpha\beta}\,\dif A + \sum_{i=1}^{n_\mrf} \int_{\sS_0}\Mbarab_{0i}\,\delta{\bar{b}}^i_{\alpha\beta}\,\dif A~,\\[4mm]
G_\mathrm{ext} \is \ds\int_{\sS}\delta\bx\cdot\bff\,\dif a 
+ \ds\int_{{\partial\sS}} \delta\bx\cdot\bT\,\dif s + \ds \int_{\partial\sS} \delta\bn\cdot\bM\,\dif s + \ds \sum_{i=1}^{n_\mrf} \int_{\partial\sS} \delta\bc_i\cdot\bMbar_{\!i}\,\dif s~.
\label{e:Giie}
\eqe

Here, $n_\mrf$  denotes the number of fiber families and the quantities indexed by $i$ imply that they are defined for fiber family $i$.
Further, $\tau^{\alpha\beta}$, $M_0^{\alpha\beta} $, and $\bar{M}_{0i}^{\alpha\beta}$ are the components of the nominal effective stress tensor, the nominal stress couple tensor associated with out-of-plane bending, and the nominal stress couple tensor associated with in-plane bending, respectively. They are all symmetric and,  for hyperelastic materials, can be obtained as the derivative of a stored energy function,
\eqb{l}
W = W\big(a _{\alpha\beta}, b_{\alpha\beta},  \bar{b}^i _{\alpha\beta}; ~h_i^{\alpha\beta}\big)~,
\label{e:WEK}
\eqe 
with respect to the corresponding work-conjugate kinematic variables defined in Sec.~\ref{s:geodescription}. That is, the internal virtual work in Eq.~(\ref{e:Giie}.2) can be written as $G_{\mathrm{int}} = \int_{\sS_0}\, \delta W\,\dif A $, since
\eqb{lll}
\delta W= \ds  \frac{1}{2}\, \tau^{\alpha\beta} \,\delta{a}_{\alpha\beta} + \Mab_0\,\delta{b}_{\alpha\beta} + \sum_{i=1}^{n_\mrf} \Mbarab_{0i}\,\delta{\bar{b}}^i_{\alpha\beta}~,
\label{e:WEK_delta}
\eqe
where
\eqb{llrlrlr}
\tau^{\alpha\beta} = \ds2\,\pa{W}{a _{\alpha\beta}}~,\quad\quad
M_0^{\alpha\beta}  =  \ds\pa{W}{b _{\alpha\beta}}~,\quad\quad
\bar{M}_{0i}^{\alpha\beta} = \ds\pa{W}{\bar{b}^i _{\alpha\beta}}~.
\label{e:consticom}
\eqe
In Eq.~\eqref{e:WEK}, 
{$h_i^{\alpha\beta}$} collectively denote the components of any structural tensors characterizing material anisotropy. In the following, 
fiber index $i$ is skipped in $\bar{M}_{0}^{\alpha\beta}$, $\bar{b}_{\alpha\beta}$, $\hat{L}^\beta_{,\alpha}$,  vectors $\bc$, $\bell$, $\bmbar$, and $\bar\bM$ (including their components and derivatives)  to simplify the notation where no ambiguities arise.

For $G_{\mathrm{int}}$ in Eq.~(\ref{e:Giie}.2), one requires the variations (see \cite{shelltextile}) 
\eqb{rrll}
\delta{a}_{\alpha\beta} \is \delta\ba_\alpha\cdot\ba_\beta + \ba_\alpha\cdot\delta\ba_\beta~,\\[3mm]
\delta{b}_{\alpha\beta} \is \bn\cdot\delta\bd_{\alpha\beta}~, \quad$with$ \quad  \delta\bd_{\alpha\beta} := \delta \vauacb - \Gamma_{\alpha\beta}^\gamma\,\delta\ba_\gamma~,\\[3mm]
\bar{M}_0^{\alpha\beta}\,\delta\bar{b}_{\alpha\beta} \is  - \bar{M}_0^{\alpha\beta}\,(\delta\ba_\alpha \cdot \bar\bc_{,\beta}  + \ba_\alpha\cdot\delta\bar\bc_{,\beta})~.
\label{e:variations}
\eqe
 In the last equation, we have used the  symmetry of $\bar{M}_0^{\alpha\beta}.$\footnote{The minus sign in Eq.~(\ref{e:variations}.3) stems from the definition of the in-plane curvature tensor in Eq.~\eqref{e:binplane2}.}
The variation $\delta\bar\bc_{,\alpha}$ follows from Eq.~\eqref{e:coderivbcbar_ell} as 
\eqb{llll}
\delta\bar\bc_{,\alpha} =    \big[\sL^\gamma_\alpha \,(\bn\otimes\bn+\bc\otimes\bc-\bell\otimes\bell) - \sC^\gamma_\alpha\,\bell\otimes\bc - \sN^\gamma_\alpha\,\bell\otimes\bn \big]\,\delta\ba_\gamma - \ell^{\gamma}\,(\bell\otimes\bc) \,\delta\ba_{\gamma,\alpha} ~,
\label{e:varbarbccma}
\eqe
where 
\eqb{lll}
\sL^\gamma_{\alpha}\dis -  \ell^\gamma\, \big(c_{\beta}  \,\hat{L}^\beta_{,\alpha} + \ell^{\beta}\, \Gamma_{\!\beta\alpha}^{\mrc}  \big )  \\[3mm]
\sC^\gamma_{\alpha} \dis  \hat{L}^\gamma_{,\alpha} - \ell^{\gamma}\, \big (\ell_{\beta}  \,\hat{L}^\beta_{,\alpha} + \ell^{\beta}\, \Gamma_{\!\beta\alpha}^{\uell} \big )  \\[3mm]
\sN^\gamma_{\alpha} \dis c^\gamma\, \ell^{\beta}\,b_{\beta\alpha}~.
\label{e:defineAuxScalars}
 \eqe

Further, for the external virtual work (\ref{e:Giie}.3), one requires the variations (see \cite{shelltextile})
\eqb{lll}
\delta\bn \is - \big( c^\alpha\,\bc\otimes\bn + \ell^{\alpha}\,\bell\otimes\bn\big)\, \delta\ba_\alpha~, \\[3mm] 
\delta\bc \is \big(c^\alpha\,\bn\otimes\bn - \ell^{\alpha}\,\bell\otimes\bc\big)\,\delta\ba_\alpha~. 
%
\label{e:lin_bn_bc_bell}
\eqe
With this and Eq.~\eqref{e:MomentCross}, the last term in Eq.~(\ref{e:Giie}.3) can be rewritten into
\eqb{lll}
 \ds \int_{\partial\sS} \delta\bc\cdot\bMbar\,\dif s = \ds \int_{\partial\sS} \delta\bc\cdot (\bmbar\times\bc)\,\dif s = - \ds \int_{\partial\sS} \delta\bc\cdot\bell\,\bar{m}\,\dif s = \ds \int_{\partial\sS} \ell^{\alpha}\,\delta\ba_\alpha\cdot\bc\,\bar{m}\,\dif s~,
 \label{e:barGext}
 \eqe
where $\bar{m}$ is an external bending moment causing in-plane bending. Inserting \eqref{e:barGext} into Eq.~(\ref{e:Giie}.3)  gives (see also \cite{steigmann99,shelltheo})
\eqb{lll}
G_\mathrm{ext} \is 
\ds\int_{\sS_0}\delta\bx\cdot \bff_{\!0} \,\dif A + \ds\int_{\sS}\delta\bx\cdot p\,\bn\,\dif a 
+ \ds\int_{\partial_t\sS} \delta\bx\cdot\bt\,\dif s + [\delta\bx\cdot m_\nu\,\bn\big]~\\[5mm]
\plus  \ds\int_{\partial_{m\tau}\sS}\delta\bn\cdot m_\tau\,\bnu\,\dif s  
+ \ds \int_{\partial_{\bar{m}}\sS} \ell^{\alpha}\,\delta\ba_\alpha\cdot\bc\,\bar{m}\,\dif s~.
\label{e:Giie22}
\eqe
Here, we have assumed an external body force of the form $\bff =  \bff_{\!0}/J + p\,\bn$, where
 $\bff_{\!0}$ denotes a constant body force, and $p$ is an external pressure acting always normal to shell surface $\sS$. Further, $\bt:=\bT - (m_\nu\,\bn)'$ is the effective boundary traction, $m_\tau$ is external bending moment causing out-of-plane, and $m_\nu$ is a point load at corners on Neumann boundaries where $\delta\bx \neq \boldsymbol{0}$.



\section{Linearization of the weak form}{\label{s:linweakform}}

This section presents the linearization of  weak form \eqref{e:wfu}  required for the development of the rotation-free {isogeometric}  finite element shell formulation in Sec.~\ref{s:fe}.  The more important internal virtual work is discussed here, while the external virtual work can be found in Appendix~\ref{s:linGextO}.  We focus on quasi-static conditions, i.e.~the inertial term $\rho_0\,\dot{\bv}$ vanishes.



The linearization of $G_{\mathrm{int}}$ in Eq.~\eqref{e:Giie} requires the increment of $\delta W$, which follows from Eq.~\eqref{e:WEK_delta} as
\eqb{llllllllll}
\Delta\delta W 
\is     \ds\delta\auab\paqq{W}{\auab}{\augd}\Delta\augd 
\plus \ds\delta\auab\paqq{W}{\auab}{\bugd}\Delta\bugd 
\plus \ds\pa{W}{\auab}\Delta\delta\auab ~\\[4mm]
\plus \ds\delta\buab\paqq{W}{\buab}{\augd}\Delta\augd 
\plus \ds\delta\buab\paqq{W}{\buab}{\bugd}\Delta\bugd
\plus \ds\pa{W}{\buab}\Delta\delta\buab~\\[4mm]

\plus \ds\sum_{i=1}^{n_\mrf} \left(\ds \delta\auab\,  \paqq{W}{ \auab  }{\bar{b}^i_{\gamma\delta}}  \Delta\bar{b}^i_{\gamma\delta} \right.

\plus \left. \ds\delta\bar{b}^i_{\alpha\beta} \paqq{W}{\bar{b}^i_{\alpha\beta} }{\augd}\Delta\augd \right)   

\plus  \ds\sum_{i,j=1}^{n_\mrf} \left(  \ds\delta\bar{b}^i_{\alpha\beta} \paqq{W}{\bar{b}^i_{\alpha\beta} }{\bar{b}^j_{\gamma\delta}}\Delta\bar{b}^j_{\gamma\delta} \right) ~\\[5mm]
\plus \ds\sum_{i=1}^{n_\mrf} \left( \ds\delta{b}_{\alpha\beta}\, \paqq{W}{{b}_{\alpha\beta} }{\bar{b}^i_{\gamma\delta} }\Delta\bar{b}^i_{\gamma\delta} \right.
\plus \ds\delta\bar{b}^{i}_{\alpha\beta}\, \paqq{W}{\bar{b}^i_{\alpha\beta} }{{b}_{\gamma\delta} }\Delta{b}_{\gamma\delta} 
%
%
\plus \left. \ds\pa{W}{\bar{b}^i_{\alpha\beta} }\Delta\delta\bar{b}^i_{\alpha\beta}  \right) ~,

\label{e:DxdxW}\eqe
where the term  containing indices $i$ and $j$ accounts for an explicit coupling between fiber families.  
Introducing the material tangents
\eqb{llrlr}
\begin{aligned}[]
c^{\alpha\beta\gamma\delta}\, \dis 4\ds\paqq{W}{\auab}{\augd} ~ \is 2\ds\frac{\partial \tau^{\alpha\beta}}{\partial a_{\gamma\delta}}~,  \\[3mm]
d^{\alpha\beta\gamma\delta} \,\dis 2\ds\paqq{W}{\auab}{\bugd} ~ \is \ds\frac{\partial \tau^{\alpha\beta}}{\partial b_{\gamma\delta}}~,\\[3mm]
e^{\alpha\beta\gamma\delta} \, \dis 2\ds\paqq{W}{\buab}{\augd} ~\is 2\ds\frac{\partial M_0^{\alpha\beta}}{\partial a_{\gamma\delta}}~,\\[3mm]
f^{\alpha\beta\gamma\delta} \, \dis \ds\paqq{W}{\buab}{\bugd} ~\is \ds\frac{\partial M_0^{\alpha\beta}}{\partial b_{\gamma\delta}}~, 
\end{aligned}
\quad \quad \quad \quad\quad\quad
\begin{aligned}[]
\bar{d}_i^{\alpha\beta\gamma\delta} \, \dis 2\ds\paqq{W}{\auab}{\bar{b}^i_{\gamma\delta}} ~\is \ds\frac{\partial \tau^{\alpha\beta}}{\partial \bar{b}^i_{\gamma\delta}}~,\\[3mm]
\bar{e}_i^{\alpha\beta\gamma\delta} \, \dis 2\ds\paqq{W}{\bar{b}^i_{\alpha\beta} }{\augd} ~ \is 2\ds\frac{\partial \bar{M}_{0i}^{\alpha\beta}}{\partial a_{\gamma\delta}}~,\\[3mm]
%
%
%
\bar{f}_{ij}^{\alpha\beta\gamma\delta} \, \dis \ds\paqq{W}{\bar{b}^i_{\alpha\beta} }{\bar{b}^j_{\gamma\delta}} ~ \is \ds\frac{\partial \bar{M}_{0i}^{\alpha\beta}}{\partial \bar{b}^j_{\gamma\delta}}~,\\[3mm]

\bar{g}_i^{\alpha\beta\gamma\delta} \, \dis \ds\paqq{W}{\buab}{\bar{b}^i_{\gamma\delta}} ~\is \ds\frac{\partial M_0^{\alpha\beta}}{\partial \bar{b}^i_{\gamma\delta}}~,\\[3mm]
\bar{h}_i^{\alpha\beta\gamma\delta} \,\dis \ds\paqq{W}{\bar{b}^i_{\alpha\beta}}{{b}_{\gamma\delta}} ~ \is \ds\frac{\partial \bar{M}_{0i}^{\alpha\beta}}{\partial {b}_{\gamma\delta}}~,
\end{aligned}
\label{e:cdef}
\eqe
%
Eq.~\eqref{e:DxdxW} becomes
\eqb{lllllll}
\Delta\delta W 
\is    \ds c^{\alpha\beta\gamma\delta}\,\frac{1}{2}\,\delta\auab\,\frac{1}{2}\,\Delta\augd 
\plus d^{\alpha\beta\gamma\delta}\,\frac{1}{2}\,\delta\auab\,\Delta\bugd 
\plus \tauab\,\frac{1}{2}\,\Delta\delta\auab \\[4mm]

\plus \ds e^{\alpha\beta\gamma\delta}\,\delta\buab\,\frac{1}{2}\,\Delta\augd 
\plus f^{\alpha\beta\gamma\delta} \,\delta\buab\,\Delta\bugd
\plus \Mab_0\,\Delta\delta\buab~ \\[4mm]
\plus \ds\sum_{i=1}^{n_\mrf} \left(  \bar{d}_i^{\alpha\beta\gamma\delta}\,\frac{1}{2}\,\delta a_{\alpha\beta}\,\,\Delta\bar{b}^i_{\gamma\delta} \right. 
\plus  \left.\bar{e}_i^{\alpha\beta\gamma\delta}\,\delta\bar{b}^i_{\alpha\beta}\,\frac{1}{2}\,\Delta\augd \right)
\plus   \ds\sum_{i,j=1}^{n_\mrf} \left(  \bar{f}_{ij}^{\alpha\beta\gamma\delta}\,\delta\bar{b}^i_{\alpha\beta}\, \Delta\bar{b}^j_{\gamma\delta}\right)  \\[4mm]
\plus \ds\sum_{i=1}^{n_\mrf} \left(\bar{g}_i^{\alpha\beta\gamma\delta}\,\delta{b}_{\alpha\beta}\, \Delta\bar{b}^i_{\gamma\delta} \right. 
\plus  \bar{h}_i^{\alpha\beta\gamma\delta}\,\delta\bar{b}^i_{\alpha\beta}\, \Delta{b}^i_{\gamma\delta}
\plus  \left. \bar{M}_{0i}^{\alpha\beta}\,\Delta\delta\bar{b}^i_{\alpha\beta}\right)~.
%
%
\label{e:dxdxWS} 
\eqe
Here and elsewhere, the increments of  kinematical quantities like $\Delta\auab$, $\Delta\buab$, and $\Delta\bar{b}_{\alpha\beta}$ can be taken from their corresponding variations  simply by replacing $\delta$ with $\Delta$.

Considering  the minor symmetries of the material tangents,\footnote{I.e.~$\alpha$ and $\beta$ as well as $\gamma$ and $\delta$ can be exchanged  in the material tangents.} we find 
\eqb{lll}
c^{\alpha\beta\gamma\delta}\,\frac{1}{2}\delta\auab\,\frac{1}{2}\Delta\augd
\is \delta \vaua \cdot \vaub \,c^{\alpha\beta\gamma\delta}\, \vaug \cdot \Delta\vaud~,
\\[2.5mm]

d^{\alpha\beta\gamma\delta}\,\frac{1}{2}\delta\auab\,\Delta\bugd
\is \delta \vaua \cdot \vaub \,d^{\alpha\beta\gamma\delta}\,\bn\cdot
\Delta\bd_{\gamma\delta}~, 
\\[2.5mm]

e^{\alpha\beta\gamma\delta}\,\delta\buab\,\frac{1}{2}\Delta\augd
\is \delta\bd_{\alpha\beta}\cdot\bn\, \,e^{\alpha\beta\gamma\delta}\, \vaug \cdot \Delta\vaud~,
\\[2mm]

f^{\alpha\beta\gamma\delta}\,\delta\buab\,\Delta\bugd
\is\delta\bd_{\alpha\beta}\cdot\bn  \,f^{\alpha\beta\gamma\delta}\, \bn \cdot \Delta\bd_{\gamma\delta}~,
\\[2mm]

\bar{d}^{\alpha\beta\gamma\delta}\,\frac{1}{2}\delta\auab\,\Delta\bar{b}_{\gamma\delta}
\is -\delta \vaua \cdot \vaub \,\bar{d}^{\alpha\beta\gamma\delta}\, (\ba_\delta \cdot\Delta\bar\bc_{,\gamma} + \bar\bc_{,\gamma}\cdot\Delta\ba_\delta)~, 
\\[2.5mm]

\bar{e}^{\alpha\beta\gamma\delta}\,\delta\bar{b}_{\alpha\beta}\,\frac{1}{2}\,\Delta\augd
\is -(\delta \bar\bc_{,\alpha}\cdot\ba_\beta + \delta\ba_\beta\cdot\bar\bc_{,\alpha})\,\bar{e}^{\alpha\beta\gamma\delta}\,\ba_\gamma\cdot\Delta\ba_\delta~,
\\[2mm]

\bar{f}^{\alpha\beta\gamma\delta}\,\delta\bar{b}_{\alpha\beta}\,\Delta\bar{b}_{\gamma\delta}
\is+ (\delta\bar\bc_{,\alpha}\cdot\ba_\beta + \delta\ba_\beta\cdot\bar\bc_{,\alpha})\,\bar{f}^{\alpha\beta\gamma\delta}\,  (\ba_\delta \cdot\Delta\bar\bc_{,\gamma} + \bar\bc_{,\gamma}\cdot\Delta\ba_\delta)~\\[2.5mm]
\bar{g}^{\alpha\beta\gamma\delta}\, \delta\buab\,\Delta\bar{b}_{\gamma\delta}
\is  -\delta\bd_{\alpha\beta}\cdot\bn  \,\bar{g}^{\alpha\beta\gamma\delta}\, (\ba_\delta \cdot\Delta\bar\bc_{,\gamma} + \bar\bc_{,\gamma}\cdot\Delta\ba_\delta)~, 
\\[2.5mm]
\bar{h}^{\alpha\beta\gamma\delta}\,\delta\bar{b}_{\alpha\beta}\,\Delta\bugd
\is -(\delta\bar\bc_{,\alpha}\cdot\ba_\beta + \delta\ba_\beta\cdot\bar\bc_{,\alpha})\,\bar{h}^{\alpha\beta\gamma\delta}\,\bn \cdot \Delta\bd_{\gamma\delta}~,
\label{e:cdef_contr}
\eqe
where Eq.~(\ref{e:variations}) has been used. 
%
%
The linearization of $\delta\auab$ and $ \delta b_{\alpha\beta}$ follows from Eq.~(\ref{e:variations}.1) and (\ref{e:variations}.2)  as \cite{shelltheo}
\eqb{lll}
 \Delta\delta \auab \is \delta\ba_\alpha\cdot\Delta\ba_\beta + \delta\ba_\beta\cdot\Delta\ba_\alpha~,\\[2mm]
 \Delta\delta b_{\alpha\beta} \is \!  -(\delta\bd_{\alpha\beta}\!\cdot\!\ba^\gamma)\,(\bn\!\cdot\!\Delta\ba_\gamma) - (\delta\ba_{\gamma}\!\cdot\!\bn)\,(\ba^\gamma\!\cdot\!\Delta\bd_{\alpha\beta}) - b_{\alpha\beta}\,a^{\gamma\delta}\,(\delta\ba_\gamma\!\cdot\!\bn)\,(\bn\!\cdot\!\Delta\ba_{\delta})~.
\eqe
From Eq.~(\ref{e:variations}.3), we find
\eqb{lll}
\bar{M}_{0}^{\alpha\beta} \Delta\delta\bar{b}_{\alpha\beta} = - \bar{M}_{0}^{\alpha\beta}\,(\delta\ba_\alpha \cdot \Delta \bar\bc_{,\beta}  
+\Delta \ba_\alpha\cdot\delta\bar\bc_{,\beta} + \ba_\alpha\cdot \Delta\delta\bar\bc_{,\beta}) ~,
\label{e:Deltabbarab}
\eqe
due to the symmetry of $\bar{M}_{0}^{\alpha\beta}$. 
Using Eq.~\eqref{e:varbarbccma}, the last term in \eqref{e:Deltabbarab} can be expressed as 
%
%
\eqb{lllll}
\bar{M}_{0}^{\beta\alpha}\,\ba_\beta\cdot\Delta\delta\bar\bc_{,\alpha} = 
 \delta\ba_\gamma\, \bP^{\gamma\beta} \,\Delta\ba_\beta   + \delta \ba_{\beta}\, \bQ^{\beta\gamma\alpha} \,\Delta\ba_{\gamma,\alpha} + \delta\ba_{\gamma,\alpha}\, \bQ^{\beta\gamma\alpha} \, \Delta \ba_{\beta}~,
\eqe
 where we have defined the tensors
\eqb{lllll}
\mP^{\gamma\beta} \dis P^{\gamma\beta}_{\mrc\mrc}\,\bc\otimes\bc +  P^{\gamma\beta}_{\ell\ell}\,\bell\otimes\bell +  P^{\gamma\beta}_{\mrn\mrn}\,\bn\otimes\bn  +   P^{\gamma\beta}_{\ell\mrc}\,(\bell\otimes\bc 
+ \bc\otimes\bell) \\[3mm]
\plus  P^{\gamma\beta}_{\ell\mrn}\,(\bell\otimes\bn+ \bn\otimes\bell) + P^{\gamma\beta}_{\mrn\mrc}\, (\bn\otimes\bc + \bc\otimes\bn) ~,\\[4mm]
\mQ^{\beta\gamma\alpha} \dis \ell^{\beta\gamma}\,\bar{M}_\mrc^\alpha\,\bc\otimes\bc -   \ell^{\beta\gamma}\,\bar{M}_\ell^\alpha (\bc\otimes\bell + \bell\otimes\bc) + c^\beta\,\ell^\gamma\,\bar{M}_\ell^\alpha\,\bn\otimes\bn~.
\eqe
Here, $\bar{M}_\mrc^\alpha:= - \bar{M}_0^{\alpha\beta}\,c_\beta~$, $\bar{M}_\ell^\alpha:= - \bar{M}_0^{\alpha\beta}\,\ell_{\beta}~$, and 
\eqb{lll}
 P^{\gamma\beta}_{\mrc\mrc} \dis  \ds\frac{3}{2}  \bar{M}_\ell^\alpha\, \big( \sL^\gamma_\alpha\,\ell^{\beta} + \sL^\beta_\alpha\,\ell^{\gamma}\big) + 
  \bar{M}_\mrc^\alpha\, \big( \sC^\gamma_\alpha\,\ell^{\beta} + \sC^\beta_\alpha\,\ell^{\gamma}\big)
  ~, \\[5mm]
P^{\gamma\beta}_{\ell\ell} \dis - \bar{M}_\ell^\alpha\,\big( \ell^{\gamma}\, \sL^\beta_\alpha  +   \ell^{\beta}\, \sL^\gamma_\alpha  \big)~,\\[5mm]
P^{\gamma\beta}_{\mrn\mrn} \dis \ds\,\bar{M}_\ell^\alpha\, \sC^\beta_\alpha\,c^\gamma +  \bar{M}_\ell^\alpha\,\sL^\beta_\alpha\,\ell^\gamma -   \bar{M}_\ell^\alpha\,c^\beta\,\Gamma^\gamma_{\alpha\delta}\,\ell^\delta  - \bar{M}_\mrc^\alpha\,\sL^\beta_\alpha c^\gamma-  \bar{M}_0^{\gamma\alpha}\,\sL^\beta_\alpha ~,\\[5mm]
P^{\gamma\beta}_{\ell\mrc} \dis -  \bar{M}_\ell^\alpha\, \big( \sC^\gamma_\alpha\,\ell^\beta + \ell^\gamma\,\sC^\beta_\alpha  \big) +  \bar{M}_\mrc^\alpha\,\big(  \ell^{\gamma}\, \sL^\beta_\alpha  +   \ell^{\beta}\, \sL^\gamma_\alpha  \big)~,\\[5mm]
P^{\gamma\beta}_{\ell\mrn} \dis -\ds\, \bar{M}_\ell^\alpha\, \big( \sN^\gamma_\alpha\,\ell^\beta + \ell^\gamma\,\sN^\beta_\alpha  \big)~,\\[5mm]
P^{\gamma\beta}_{\mrn\mrc} \dis -\ds\, \bar{M}_\ell^\alpha\,b_{\alpha\delta}\,\ell^\delta \big( \ell^{\gamma\beta} + c^{\gamma\beta}  \big) +  \bar{M}_\mrc^\alpha\, \big( \sN^\gamma_\alpha\,\ell^\beta + \ell^\gamma\,\sN^\beta_\alpha  \big)~,
\label{e:definePandQ}
\eqe
where $\sL^\alpha_\beta$, $\sC^\alpha_\beta$ and $\sN^\alpha_\beta$ are given in Eq.~\eqref{e:defineAuxScalars}.

\section{FE discretization}\label{s:fe}

This section presents the {isogeometric} finite element discretization and corresponding linearization of  weak form~\eqref{e:wfu}. An efficient implementation of the FE formulation can then be found in Appendix~\ref{s:efficient}.

\subsection{Surface discretization}
The geometry within an undeformed element $\Omega^e_0$ and its deformed counterpart $\Omega^e$ is interpolated from the positions of control points $\mX_e$ and $\mx_e$, respectively, as
\eqb{lll}
\bX = \mN\,\mX_e~, \quad $and$ \quad
\bx = \mN\,\mx_e~, 
\label{e:interpx}
\eqe
where $\mN(\bxi):= [N_1\bone,\, N_2\bone,\, ...,\, N_{n_\mre}\bone]$ is defined based on isogeometric shape functions \cite{borden11} and $n_\mre$ denotes the number of control points defining the element. From Eq.~\eqref{e:interpx}  follows 
\eqb{lll}
\begin{aligned}
\delta\bx ~~\is \mN\,\delta\mx_e \,~~~, \\[1mm]
\ba_\alpha ~~\is \mN_{,\alpha}\,\mx_e~\, ~,\\[1mm]
\delta\ba_\alpha ~~\is  \mN_{,\alpha}\,\delta\mx_e ~,
\end{aligned}
\quad\quad\quad \quad
\begin{aligned}
\ba_{\alpha,\beta} ~~ \is \mN_{,\alpha\beta}\,\mx_e ~,\\[1mm]
\ba_{\alpha;\beta}~~  \is \mN_{;\alpha\beta}\,\mx_e ~, \\[1mm]
\delta\bar\bc_{,\alpha} ~~\is \mC_{,\alpha}\,\delta\mx_e~,
%
%
\end{aligned}
\label{e:dxe}
\eqe

%
with
\eqb{lllllll}
\mN_{,\alpha} \dis  [N_{1,\alpha}\bone,\, N_{2,\alpha}\bone,\, ...,\, N_{{n_\mre},\alpha}\bone]~,\\[3mm]
\mN_{,\alpha\beta} \dis [N_{1,\alpha\beta}\bone,\, N_{2,\alpha\beta}\bone,\, ...,\, N_{n_\mre,\alpha\beta}\bone]~,\\[3mm]
\mN_{;\alpha\beta} \dis \mN_{,\alpha\beta} - \Gamma^\gamma_{\alpha\beta}\,\mN_{,\gamma}~,\\[3mm]
%
%

 \mC_{,\alpha} \dis 
   \big[ \sL^\gamma_\alpha \,(\bn\otimes\bn+\bc\otimes\bc-\bell\otimes\bell) - \sC^\gamma_\alpha\,\bell\otimes\bc - \sN^\gamma_\alpha\,\bell\otimes\bn \big]\,\mN_{,\gamma}  - \ell^{\gamma}\,(\bell\otimes\bc) \,\mN_{,\gamma\alpha} ~.
\label{e:shapeC}
\eqe
Here, $N_{A,\alpha}=\partial N_A/\partial \xi^\alpha$, and $N_{A,\alpha\beta}=\partial^2 N_A/(\partial \xi^\alpha\partial \xi^\beta)\;(A=1, ..., n_\mre)$. Further  $\sL^\gamma_\alpha$, $ \sC^\gamma_\alpha$, and $\sN^\gamma_\alpha$ are defined by Eq.~\eqref{e:defineAuxScalars}. 
Inserting \eqref{e:dxe} into \eqref{e:variations} gives
\eqb{rrl}
\delta \auab 
\is \delta\mx_e^\mrT\, \big(\mN_{,\alpha}^\mrT\,\mN_{,\beta} + \mN_{,\beta}^\mrT\,\mN_{,\alpha}\big)\,\mx_e~, \\[3mm]
\delta \buab \is \delta\mx_e^\mrT\,\mN_{;\alpha\beta}^\mrT\,\bn~,\\[3mm]
\bar{M}_0^{\alpha\beta}\,\delta\bar{b}_{\alpha\beta} \is - \bar{M}_0^{\alpha\beta}\,(\bar\bc_{,\beta}\cdot \mN_{,\alpha}\,\delta\mx_e  + \ba_\alpha\cdot {\mC}_{,\beta}\,\delta\mx_e)~.
\label{e:variations_app}
\eqe

\subsection{FE force vectors}\label{s:fvectors}

Substituting Eqs.~\eqref{e:dxe} and \eqref{e:variations_app} into Eq.~\eqref{e:wfu} gives the discretized weak form as
\eqb{l}
\ds\sum_{e=1}^{n_\mathrm{el}} \left( G^e_\mathrm{in} + G^e_\mathrm{int} - G^e_\mathrm{ext}\right)  = \delta\mx\cdot\mf = 
 0 \quad\forall\,\delta\mx\in\sV^h~,
\label{e:DiscretizedFEforce}
\eqe
where $n_\mathrm{el}$ is the number of elements, $\mf$ denotes the global FE force vector, and $\sV^h$ denotes the set of kinematically admissible variations for the control points.

In order to obtain the virtual work of the internal FE forces, we insert  interpolation \eqref{e:variations_app} into   Eq.~(\ref{e:Giie}.2). This gives
\eqb{l}
G^e_\mathrm{int} = \delta\mx_e^\mrT\,\big(\mf^e_\mathrm{int\tau} + \mf^e_{\mathrm{int}M} + \mf^e_{\mathrm{int}\bar{M}}\big)~,
\label{e:Pinth}
\eqe
where
\eqb{llll}
\mf^e_\mathrm{int\tau} 
\dis  \ds\int_{\Omega^e_0} \tau^{\alpha\beta} \,\mN_{,\alpha}^\mrT\,\vaub\,\dif A~,\\[5mm]
\mf^e_{\mathrm{int}M} 
\dis  \ds\int_{\Omega^e_0} \, M_0^{\alpha\beta}\,\mN_{;\alpha\beta}^\mrT\,\bn\, \dif A~,\\[5mm]
\mf^e_{\mathrm{int}\bar{M}} 
\dis  -\ds\int_{\Omega^e_0} \, \bar{M}_0^{\alpha\beta}\,( \mN^\mrT_{,\alpha}\,\bar\bc_{,\beta} + {\mC}_{,\beta}^\mrT\,\ba_\alpha)  \, \dif A~.
\label{e:FEforces}
\eqe

Discretization of the external virtual work in Eq.~\eqref{e:Giie22} gives, see Eq.~\eqref{e:barGext} and also \cite{membrane,shelltheo}
\eqb{l}
G^e_\mathrm{ext} = \delta\mx_e^\mrT\,\big(\mf^e_{\mathrm{ext0}} + \mf^e_{\mathrm{ext}p}+\mf^e_{\mathrm{ext}t}+\mf^e_{\mathrm{ext}m} + \mf^e_{\mathrm{ext}\bar{m}}  \big) +  \delta\mx_A\cdot\mf^A_{\mathrm{ext}m_\nu}~,
\label{e:Pexth}
\eqe
where 
\eqb{lllll}
\mf^e_{\mathrm{ext0}}  \dis \ds\int_{\Omega^e_0}\mN^\mrT\,\bff_{\!0}\,\dif A~, \\[5mm]
\mf^e_{\mathrm{ext}p}   \dis \ds\int_{\Omega^e}\mN^\mrT\,p\,\bn\,\dif a~, \\[5mm]
\mf^e_{\mathrm{ext}t}   \dis \ds\int_{\partial_t\Omega^e}\mN^\mrT\,\bt\,\dif s~, \\[5mm]
\mf^e_{\mathrm{ext}m_\tau}   \dis -\ds\int_{\partial_{m\tau}\Omega^e}\mN_{,\alpha}^\mrT\,\nu^\alpha\,m_\tau\,\bn\,\dif s ~,\\[5mm]
\mf^e_{\mathrm{ext}\bar{m}}   \dis  \ds\int_{\partial_{\bar{m}} \Omega^e} \mN_{,\alpha}^\mrT\,\ell^{\alpha}\, \bar{m}\,\bc\,\,\dif s ~, ~~\\[5mm]
\mf^A_{\mathrm{ext}m_\nu}   \dis m_\nu\,\bn_A~~~~~~~~~~~~~~~~~~~~~~
\label{e:fext}
\eqe
are the external FE force vectors. Here, $\mf^A_{\mathrm{ext}m_\nu}$ is a possible corner force at corner node $\mx_A$ due to a 
twisting moment $m_\nu$ applied on a non-smooth boundary (cf.~\cite{shelltheo}, Sec.~6.3).  

{
\begin{remark}
The out-of-plane bending term (\ref{e:FEforces}.2) requires at least second order derivatives of the shape functions. %
As seen in (\ref{e:FEforces}.3) and (\ref{e:shapeC}.4), similar second order derivatives are  now also required for the newly added in-plane bending term. This indicates that  membrane-bending locking,  which is an issue in out-of-plane bending of thin shells, in principle, could now also appear for in-plane bending. To alleviate such locking phenomena,  (although it has not been
done in the present work) various existing (reduced) integration techniques --  see e.g.~\cite{Adam2015,Johannessen2017,Leonetti2018,Zou2021} and references therein  -- can be adapted to in-plane fiber bending, if necessary. Here, all integrals are evaluated by standard Gaussian quadrature.
\end{remark}
}

\subsection{Tangent matrices}
The tangent matrices associated with the internal and external FE forces in \eqref{e:FEforces} and \eqref{e:fext} are derived as follows.

\subsubsection{Tangent matrices of the internal FE forces}
The internal tangent matrices  can be found by linearizing \eqref{e:Pinth}. This gives
\eqb{l}
\Delta G^e_\mathrm{int} = \delta\mx_e^\mrT\,\big( \mk^e_{\mathrm{mat}} + \mk^e_{\mathrm{geo}}   )\,\Delta\mx_e~,
\eqe
where $\mk_{\mathrm{mat}}$ denotes the material tangent
\eqb{l}
 \mk^e_{\mathrm{mat}} = \mk^e_{\tau\tau} + \mk^e_{\tau M} + \mk^e_{M \tau} + \mk^e_{MM} + \mk^e_{\tau \bar{M}} +  \mk^e_{\bar{M} \tau} + \mk^e_{\bar{M}\bar{M}}  + \mk^e_{{M}\bar{M}} + \mk^e_{\bar{M}{M}} ~,
 \label{e:kmate}
\eqe
with
\eqb{llll}
\mk^e_{\tau\tau} \dis \ds\int_{\Omega_0^e} c^{\alpha\beta\gamma\delta}\,\mN^\mrT_{,\alpha}\,(\vaub\otimes\vaug)\, \mN_{,\delta}\,\dif A~, \\[4mm]
\mk^e_{\tau M} \dis \ds\int_{\Omega_0^e} d^{\alpha\beta\gamma\delta}\,\mN^\mrT_{,\alpha}\,(\vaub\otimes\bn)\,\mN_{;\gamma\delta}\,\dif A~,  \\[4mm]
\mk^e_{M \tau} \dis \ds\int_{\Omega_0^e} e^{\alpha\beta\gamma\delta}\,\mN^\mrT_{;\alpha\beta}\,(\bn\otimes\vaug)\, \mN_{,\delta}\, \dif A~, \\[4mm]
\mk^e_{MM} \dis \ds\int_{\Omega_0^e}
f^{\alpha\beta\gamma\delta}\,\mN^\mrT_{;\alpha\beta}\,(\bn\otimes\bn)\,
\mN_{;\gamma\delta}\,\dif A~,\\[4mm]

\mk^e_{\tau\bar{M}} \dis -\ds\int_{\Omega_0^e} \bar{d}^{\alpha\beta\gamma\delta}\,\mN^\mrT_{,\alpha}\,\ba_\beta\otimes  \big(\ba_\delta \, \mC_{,\gamma} + \bar\bc_{,\gamma} \, \mN_{,\delta}\big) \,\dif A~,\\[4mm]
\mk^e_{\bar{M}\tau} \dis -\ds\int_{\Omega_0^e} \bar{e}^{\alpha\beta\gamma\delta}\,
 \big( \mC^\mrT_{,\alpha}\,\ba_\beta  +\mN^\mrT_{,\beta} \,\bar\bc_{,\alpha}\big )\otimes \ba_\gamma \, \mN_{,\delta}    \,\dif A~,\\[4mm]
 \mk^e_{\bar{M}\bar{M}} \dis +\ds\int_{\Omega_0^e} \bar{f}^{\alpha\beta\gamma\delta}\,
 \big( \mC^\mrT_{,\alpha}\,\ba_\beta  +\mN^\mrT_{,\beta} \,\bar\bc_{,\alpha}\big )\otimes \big (\ba_\delta \, \mC_{,\gamma} + \bar\bc_{,\gamma} \, \mN_{,\delta}\big)\,\dif A~,\\[4mm]
\mk^e_{M\bar{M}} \dis -\ds\int_{\Omega_0^e} \bar{g}^{\alpha\beta\gamma\delta}\,\mN^\mrT_{;\alpha\beta}\,  \bn\otimes  \big(\ba_\delta \, \mC_{,\gamma} + \bar\bc_{,\gamma} \, \mN_{,\delta}\big)\,\dif A~,\\[4mm]
\mk^e_{\bar{M}M} \dis -\ds\int_{\Omega_0^e} \bar{h}^{\alpha\beta\gamma\delta}\,
 \big( \mC^\mrT_{,\alpha}\,\ba_\beta  +\mN^\mrT_{,\beta} \, \bar\bc_{,\alpha}\big )\otimes \bn \, \mN_{;\gamma\delta}    \,\dif A~,
\label{e:matKij}
\eqe
while $\mk^e_{\mathrm{geo}}$ denotes the geometrical tangent
\eqb{l}
 \mk^e_{\mathrm{geo}} = \mk^e_{\tau} + \mk^e_{M} + \mk^e_{\bar{M}} ~,
 \label{e:geoKij}
\eqe
 with
\eqb{lll}
\mk^e_{\tau} = \plus \ds\int_{\Omega_0^e} \tauab\,\mN^\mrT_{,\alpha}\,\mN_{,\beta}\,\dif A~, \\[7mm]
\mk^e_{M} = 
\mi \ds\int_{\Omega_0^e} M_0^{\alpha\beta}\, \left[\mN_{,\gamma}^\mrT\,(\bn\otimes\ba^\gamma)\,\mN_{;\alpha\beta} + \mN_{;\alpha\beta}^\mrT\,(\ba^\gamma\otimes\bn)\,\mN_{,\gamma} \right]\,\dif A~\\[5mm]
 \mi \ds\int_{\Omega_0^e} (b_{\alpha\beta}\,M_0^{\alpha\beta})\,a^{\gamma\delta}\,\mN^\mrT_{,\gamma}\,(\bn\otimes\bn)\,\mN_{,\delta}\,\dif A~,
\label{e:matGeosim}
\eqe
and
\eqb{lll}
\mk^e_{\bar{M}} = \mi \ds\int_{\Omega_0^e}  \bar{M}_0^{\alpha\beta} \left( \mN_{,\alpha}^\mrT\,\mC_{,\beta} +  \mC_{,\beta}^\mrT\, \mN_{,\alpha} \right) \dif A -\ds\int_{\Omega_0^e}\mN^\mrT_{,\gamma}\, \bP^{\gamma\beta} \,  \mN_{,\beta}\, \dif A~  \\[5mm]
\mi \ds\int_{\Omega_0^e} \left(\mN^\mrT_{,\beta}\, \bQ^{\beta\gamma\alpha} \,\mN_{,\gamma\alpha} +     \mN^\mrT_{,\gamma\alpha}\, \bQ^{\beta\gamma\alpha}\,  \mN_{,\beta}  \right) \dif A~,
\label{e:kgeo}
\eqe
where  $\bP^{\gamma\beta}$ and $\bQ^{\gamma\beta\alpha}$ are defined by Eq.~\eqref{e:definePandQ}. As expected, $ \mk^e_{\mathrm{mat}}$ and $ \mk^e_{\mathrm{geo}}$  are symmetric.

\begin{remark}
As seen in Eq.~\eqref{e:matKij} and \eqref{e:geoKij},  in-plane bending in general adds five material tangents (the last five terms in \eqref{e:matKij}) and one geometrical tangent, \eqref{e:kgeo}, to the rotation-free shell formulation of Duong et al.~\cite{solidshell}.
\end{remark}

\subsubsection{Tangent matrices of the external FE forces}
By linearizing  and rearranging \eqref{e:Pexth}, one obtains 
\eqb{l}
\Delta G^e_\mathrm{ext} = \delta\mx_e^\mrT\, \mk^e_{\mathrm{ext}} \,\Delta\mx_e + \delta\mx_A^\mrT\, \mk^A_{\mathrm{ext}m_\nu} \,\Delta\mx_A~,
\eqe
which contains the external tangent matrices
\eqb{l}
\mk^e_{\mathrm{ext}}  :=  \mk^e_{\mathrm{ext}p} + \mk^e_{\mathrm{ext}t}+\mk^e_{\mathrm{ext}m_\tau} + \mk^e_{\mathrm{ext}\bar{m}}~.
 \label{e:kexte}
\eqe
Here, $\mk^e_{\mathrm{ext}p}$, $\mk^e_{\mathrm{ext}t}$, $\mk^e_{\mathrm{ext}m_\tau}$,  $\mk^e_{\mathrm{ext}\bar{m}}$, and $\mk^A_{\mathrm{ext}m_\nu}$ are  the tangent matrices associated with $\mf^e_{\mathrm{ext}p}$, $\mf^e_{\mathrm{ext}t}$, $\mf^e_{\mathrm{ext}m}$,  $\mf^e_{\mathrm{ext}\bar{m}}$, and  $\mf^A_{\mathrm{ext}m_\nu}$
 defined in Eq.~\eqref{e:fext}, respectively. Their  expressions are  given in Appendix~\ref{s:disGextO}.

\section{Material model examples}{\label{s:two_material_model}}{\label{s:matmodels}}
This section presents two hyperelastic phenomenological material models for fabrics. The first is a simple fabric model -- motivated by numerical convenience -- that can be used to test numerical aspects of the proposed {isogeometric} finite shell element formulation. The second is a physically-based model for  (plainly) woven fabrics.

Since inducing invariants for  the new in-plane curvature tensor  $\bar\bK $  is very similar to inducing invariants for the out-of-plane curvature tensor $\bK$, the construction of material models for the proposed shell formulation can follow that of classical shells. Further, as in the FE formulation of Duong et al.~\cite{solidshell}, the proposed shell formulation can admit material models expressed directly in terms of the invariants of the surface tensors such as is considered here. The unit of the strain energy density $W$ is thus energy per reference area. This approach facilitates efficient simulations since through-the-thickness integration is not required.\footnote{This does not imply that the thickness is neglected, instead its influence is embedded in the model.} However, our proposed shell formulation can also incorporate material models that are extracted from 3D continua by (numerical) integration over the thickness, see Duong et al.~\cite{solidshell}. 

\subsection{A simple fabric model}

We consider a general fabric consisting of $n_\mrf$ fiber  families that can be initially curved and possibly bonded to a matrix. We assume that the total strain energy function $W$ can be additively decomposed into the strain energies of the matrix deformation  $W_{\mathrm{matrix}}$, fiber stretching $ W_\mathrm{fib\mbox{-}stretch} $,  out-of-plane and in-plane fiber bending $W_\mathrm{fib\mbox{-}bending}$, fiber torsion $W_\mathrm{fib\mbox{-}torsion} $, and the linkage between fiber families $W_\mathrm{fib\mbox{-}angle}$. 
{Accordingly,}
\eqb{llllll}
W =     W_{\mathrm{matrix}} + W_\mathrm{fib\mbox{-}stretch} + W_\mathrm{fib\mbox{-}bending} +  W_\mathrm{fib\mbox{-}torsion} + W_\mathrm{fib\mbox{-}angle}~.
   \label{e:eg_Wsimple}
\eqe
A simple material model is given by 
\eqb{llllll}
W_{\mathrm{matrix}} \is \ds U(J) +  \frac{1}{2}\mu\,\big(I_1 - 2 - 2\,\ln J\big) ~,\\[5mm]
W_{\mathrm{fib\mbox{-}stretch}} \is \ds  \frac{1}{8}\, \sum\limits_{i=1}^{n_\mrf}  \epsilon^i_{\mrL}\,  \big( {\Lambda}_i - 1 \big)^{2} ~, 
\\[5mm]
%
 %
 W_{\mathrm{fib\mbox{-}bending}} \is \ds  \frac{1}{2} \,  \sum\limits_{i=1}^{n_\mrf} \Big[ \beta^i_\mrn\, (K^i_\mrn)^2  +   \beta^i_\mrg\, (K^i_\mrg)^2  \Big]~,\\[5mm]
 W_{\mathrm{fib\mbox{-}torsion}} \is \ds  \frac{1}{2}\,    \sum\limits_{i=1}^{n_\mrf}  \beta^i_{\tau}\,{(T^i_\mrg)^2}, \\[5mm]
   W_{\mathrm{fib\mbox{-}angle}} \is \ds    \frac{1}{4}\,  \sum\limits_{i=1}^{n_\mrf-1}  \sum\limits_{j=i+1}^{n_\mrf}   \epsilon^{ij}_{\mra}\,  \big( {\gamma}_{ij} - \gamma^0_{ij}\big)^2~, 
   \label{e:eg_W1}
\eqe
where $U(J)$ is the surface dilatation energy, and where $\Lambda_i$,  $T^i_\mrg$, $K^i_\mrn$, and   $K^i_\mrg$ are defined in Eq.~\eqref{e:K_invariant} for fiber family $i$. Further, ${\gamma}_{ij}:=\bC:\bL_i\otimes\bL_j$ and ${\gamma}^0_{ij}:=\bL_i\cdot\bL_j$ describe the angle between fiber families $i$ and $j$ in the current and reference configuration, respectively.  Parameters $\mu$, $\epsilon^i_\mrL$, 
  $\epsilon^{ij}_\mra$, $\beta^i_\mrn$, $\beta^i_\mrg$, and $\beta^i_\tau$ denote material constants. 
%
The effective stress and moment components follows from Eq.~\eqref{e:eg_Wsimple} and  \eqref{e:eg_W1} as
\eqb{lll}
\tau^{\alpha\beta} \is \ds  \tau^{\alpha\beta}_ {\mathrm{matrix}} 
   + \frac{1}{2}\,  \sum\limits_{i=1}^{n_\mrf}\epsilon^i_{\mrL}\,(\Lambda_i-1)\,L_i^{\alpha\beta} +  \sum\limits_{i=1}^{n_\mrf-1}  \sum\limits_{j=i+1}^{n_\mrf}  \epsilon^{ij}_{\mra} \big( {\gamma}_{ij} - \gamma^0_{ij}\big)\, (L_i^\alpha\,L_j^\beta)^{\mathrm{sym}}~,\\[6mm]
M_0^{\alpha\beta} \is  \ds \sum\limits_{i=1}^{n_\mrf}\beta^i_\mrn\,K^i_\mrn\, L_i^{\alpha\beta} +    \sum\limits_{i=1}^{n_\mrf} \beta^i_\tau \, T^i_\mrg\, (c_{0i}^\alpha\,L_i^\beta)^{\mathrm{sym}}~,\\[6mm]
\bar{M}_0^{\alpha\beta} \is  \ds \sum\limits_{i=1}^{n_\mrf} \beta^i_\mrg\,K^i_\mrg\, L_i^{\alpha\beta}~,
\label{e:eg_tauab_W1}
\eqe
where $\tau^{\alpha\beta}_ {\mathrm{matrix}} = J \,U'\,a^{\alpha\beta} +  \mu\,(A^{\alpha\beta} - a^{\alpha\beta}) $ is the stress due to the matrix response, and where $(\bullet^{\alpha\beta})^{\mathrm{sym}} = (\bullet^{\alpha\beta} + \bullet^{\beta\alpha})/2$ denotes symmetrization. Further, the material tangents of \eqref{e:eg_tauab_W1} follow from Eq.~\eqref{e:cdef} as
\eqb{lll}
c^{\alpha\beta\gamma\delta} \is \ds  c^{\alpha\beta\gamma\delta}_{\mathrm{matrix}}
+ \ds \sum\limits_{i=1}^{n_\mrf}\epsilon^i_{\mrL}\,L_i^{\alpha\beta}\,L_i^{\gamma\delta} + 2\, \sum\limits_{i=1}^{n_\mrf-1}  \sum\limits_{j=i+1}^{n_\mrf} \epsilon^{ij}_{\mra}\, \big(L_i^\alpha\,L_j^\beta\big)^{\mathrm{sym}}\,\big(L_i^\gamma\,L_j^\delta\big)^{\mathrm{sym}}~,\\[6mm]
f^{\alpha\beta\gamma\delta} \is  \ds \sum\limits_{i=1}^{n_\mrf} \beta^i_\mrn\, L_i^{\alpha\beta}\, L_i^{\gamma\delta} +   \sum\limits_{i=1}^{n_\mrf}  \beta^i_\tau\,  \big  (c_{0i}^\alpha\,L_i^\beta \big)^{\mathrm{sym}}\, \big(c_{0i}^\gamma\,L_i^\delta\big)^{\mathrm{sym}}~,\\[6mm]
\bar{f}_i^{\alpha\beta\gamma\delta} \is  \ds \sum\limits_{i=1}^{n_\mrf} \beta^i_\mrg \, L_i^{\alpha\beta}\,L_i^{\gamma\delta} ~,\\[6mm]
d^{\alpha\beta\gamma\delta} \is e^{\alpha\beta\gamma\delta} = \bar{d}_i^{\alpha\beta\gamma\delta} = \bar{e}_i^{\alpha\beta\gamma\delta} = \bar{g}_i^{\alpha\beta\gamma\delta} = \bar{h}_i^{\alpha\beta\gamma\delta} = 0~,
\eqe
with 
\eqb{ll}
c^{\alpha\beta\gamma\delta}_{\mathrm{matrix}} =   -\big( J\,U' - \mu \big) \, \big(a^{\alpha\gamma}\,a^{\beta\delta} + a^{\alpha\delta}\,a^{\beta\gamma}\big) +   J\big(U' + J\,U'' \big)\, \aab\agd ~. 
\eqe 



\subsection{A woven fabric model}{\label{s:wovenfabricmodel}}
In this section, a physically-based hyperelastic material model of dry woven fabrics is proposed.\footnote{Dry fabrics are fabrics that are not embedded within a matrix material.}  We consider plain weave fabrics with two fiber families. The model will be fitted to the experiment data provided by Cao et al.~\cite{Cao2008}. 

For simplification, we assume that fibers embedded in the apparent textile surface are nearly inextensible in the averaged fiber direction $\bell_i$.\footnote{Due to initial crimping, fibers are not straight initially and hence may appear extensible.} We further assume that fibers in the fabrics are perfectly bonded to each other (i.e.~without inter-fiber sliding), such that hyperelasticity can be assumed.
%
Accordingly, we propose a strain energy of the form 
\eqb{llllll}
W =      W_\mathrm{fib\mbox{-}stretch} + W_\mathrm{fib\mbox{-}bending}  + W_\mathrm{fib\mbox{-}angle} ~, 
   \label{e:eg_WFs}
\eqe
with 
\eqb{llllll}
W_{\mathrm{fib\mbox{-}stretch}} \is \ds \frac{1}{2}\,  \sum\limits_{i=1}^2  \epsilon^i_{\mrL}\big( {{\lambda}_i} - 1 \big)^{2} ~, \\[5mm] 
 W_{\mathrm{fib\mbox{-}bending}} \is \ds  \frac{1}{2} \,  \sum\limits_{i=1}^2 \beta^i_\mrg\, (K^i_\mrg)^2  ~,\\[5mm]
   W_{\mathrm{fib\mbox{-}angle}} \is \ds  \frac{\mu}{2} \,\left(\hat\gamma\,\mathrm{asinh}(\alpha_1\,\hat\gamma) - \frac{1}{\,\alpha_1}\,\sqrt{\alpha_1^2\,{\hat\gamma}^2 + 1}\right)    +\frac{\eta }{2\,\alpha_2}\,{{\mathrm{cosh}(\alpha_2\,\hat\gamma)}} ~,
   \label{e:eg_WFs2}
\eqe
%
where $K_\mrg^i$ is defined in Eq.~(\ref{e:K_invariant}.4) for fiber family $i$, and  $\hat\gamma:=\bell_1\cdot\bell_2$ describes the fiber angle between fiber family $1$ and $2$. Further, $\epsilon^i_\mrL$,  $\beta^i_\mrg$, $\mu$, $\alpha_1$, $\alpha_2$, and $\eta$ are material parameters (see Tab.~\ref{t:WFconstant}). The choice of $W_\mathrm{fib\mbox{-}angle}$ in Eq.~\eqref{e:eg_WFs2} is motivated both physically (i.e.~to reproduce the shear response observed experimentally) and numerically (i.e.~to get a well-behaved and smooth tangent matrix).    The two shearing energy terms in Eq.~(\ref{e:eg_WFs2}.3) phenomenologically reflect two assumed mechanisms of  bonding between yarns of the two fiber families. The first one is due to friction between yarns dominating at small deformations, and the second is due to geometrical interlocking of yarns  at large deformations (or  \textit{yarn-yarn lock-up} \cite{Cao2008}).

\begin{table}[!htp]
\small
\begin{center}
\def\arraystretch{1.5}\tabcolsep=5.0pt
\begin{tabular}{|c|l|l|l| }
\hline
 \parbox[t][][t]{1.9cm}{~Parameter} & \parbox[t]{1.6cm}{value}& \parbox[t]{1.5cm}{unit}   &\parbox[t]{9.5cm}{physical meaning} \\ \hline\hline
$\epsilon^i_{\mrL}$ & 50  & {{N/mm}}  & tensile stiffness of fiber family $i$ 
 \\ \hline
$\beta^i_{\mrg}$ & 4.8   & {N\,mm} &  in-plane bending stiffness of fiber family $i$ \\ \hline
$\mu$ & 1.6  &  {mN/mm}  &  initial shear modulus due to yarn-yarn  friction\\ \hline
$\alpha_1$ &305  & -  & plateau parameter of yarn-yarn  friction \\ \hline
$\eta$ & 2.0 &  {mN/mm}  & shear modulus due to geometrical yarn-yarn interlocking\\ \hline
$\alpha_2$ & 5.4215  & -  & plateau parameter of geometrical yarn-yarn interlocking\\ \hline
\end{tabular}
\end{center}
\vspace{-0.5cm}
\caption{Material parameters for material model \eqref{e:eg_WFs2}. The values are obtained from fitting  \eqref{e:eg_WFs2} to the experimental data of the bias extension test for sample\,\#1 of Cao et al.~\cite{Cao2008}, see Sec.~\ref{s:example_WF}.}
\label{t:WFconstant}
\end{table}
Following from \eqref{e:eg_WFs} and \eqref{e:eg_WFs2},  the effective stress and moment components become 
\eqb{lll}
\tau^{\alpha\beta} \is \ds   \sum\limits_{i=1}^2  \epsilon^i_{\mrL}\,(\lambda_i-1)\, \frac{1}{\lambda_i}\,L_i^{\alpha\beta}  +S\,  l^{\alpha\beta}_{12}~,\\[5mm]
\bar{M}_0^{\alpha\beta} \is  \ds\sum\limits_{i=1}^2   \beta^i_\mrg\,K^i_\mrg\, L_i^{\alpha\beta}~,
\label{e:eg_tauab_WFs}
\eqe
where  we have defined $S(\hat\gamma):=  \mu \,\mathrm{asinh}(\alpha_1\,\hat\gamma) + \eta\,\mathrm{sinh}(\alpha_2\,\hat\gamma)$, and
\eqb{lll}
l^{\alpha\beta}_{12} := \ds   \big(\ell_{1}^{\alpha}\,\ell_{2}^{\beta}\big)^{\mathrm{sym}} - \frac{\hat\gamma}{2}\,\big( \ellab_1 + \ellab_2\big)~.
\eqe
By applying Eq.~\eqref{e:cdef} to material model \eqref{e:eg_WFs}, we further find the material tangents as
\eqb{lll}
c^{\alpha\beta\gamma\delta} \is  \ds    \ds \sum\limits_{i=1}^2\, \epsilon_\mrL^i \, \lambda_i^{-3}\,L_i^{\alpha\beta}\,L_i^{\gamma\delta} + 2\,S\,l^{\alpha\beta\gamma\delta}_{12} + 2\,S'\,l^{\alpha\beta}_{12}\,l^{\gamma\delta}_{12}~,\\[1mm]
\bar{f}^{\alpha\beta\gamma\delta} \is   \ds \sum\limits_{i=1}^2 \beta^i_{\mrg}\, L_i^{\alpha\beta}\,L_i^{\gamma\delta}~,\\[6mm]
{d}^{\alpha\beta\gamma\delta} \is {e}^{\alpha\beta\gamma\delta} = {f}^{\alpha\beta\gamma\delta}  = {g}^{\alpha\beta\gamma\delta} = {h}^{\alpha\beta\gamma\delta} =\bar{ h}^{\alpha\beta\gamma\delta}  = \bar{e}^{\alpha\beta\gamma\delta} = 0~,
\eqe
%
where $S' = \mu\,\alpha_1\,\big(1/\sqrt{\alpha_1^2\,{\hat\gamma}^2 + 1}\big) +  \eta\,\alpha_2\,\cosh(\alpha_2\,\hat\gamma)$, and
\eqb{lll}
l^{\alpha\beta\gamma\delta}_{12}:= \ds \pa{l^{\alpha\beta}_{12}}{a_{\gamma\delta}} = -\big(\ell_{1}^{\alpha}\,\ell_{2}^{\beta}\big)^{\mathrm{sym}}\,\frac{1}{2}(\ell^{\gamma\delta}_1 + \ell^{\gamma\delta}_2) -  \frac{1}{2}\,(\ellab_1 + \ellab_2)\,l_{12}^{\gamma\delta} +  \frac{\hat{\gamma}}{2}\,(\ellab_1\,\ell^{\gamma\delta}_1 + \ellab_2\,\ell^{\gamma\delta}_2)~.
\eqe

\subsection{An effective (stabilized) fiber compression model}{\label{s:stabil}}
In most textile materials, fibers buckle under axial compression. If the buckling is microscopic,\footnote{I.e.~at the length scale of a single fiber.} the fibers macroscopically appear to have much smaller stiffness in compression than in tension.
For simplification, the (microscopic) buckling is usually not simulated explicitly and the compressive stiffness is usually neglected in the construction of material models. For instance, one can simply set $\epsilon_\mrL^i = 0$ for $\lambda_i<1$ in material models \eqref{e:eg_W1} and \eqref{e:eg_WFs2}.

Although the latter simplification does not  affect the accuracy much, it can still lead to a material instability\footnote{I.e.~a instability due to a lack of stiffness in a particular direction at a material point.}  in quasi-static computations. Therefore, if no other medium, e.g.~matrix, effectively supports the fibers, a stabilization technique may be necessary. For this purpose, 
 %
%
we consider the additional stabilization term 
\eqb{lll}
W_\mathrm{fib\mbox{-}stab} =  \ds  {\frac{1}{2}}\,\epsilon^\mre_{\mathrm{stab}}\, \sum\limits_{i=1}^{n_\mrf}  \big( {\lambda}_i - 1 \big)^{2}  +  \ds   { \frac{1}{2}}\, \epsilon^\mrv_{\mathrm{stab}}\, \sum\limits_{i=1}^{n_\mrf}  \big( \tilde\lambda_i -  1 \big)^2  
 \label{e:Wstabvis1}
\eqe
in the strain energy in case $\lambda_i<1$. Here, $\epsilon^\mre_\mathrm{stab}$ and $\epsilon^\mrv_\mathrm{stab}$ are  stabilization parameters, and
\eqb{lll}
 \tilde\lambda^2_i:=a_{\alpha\beta}\,\ellab_{i\mbox{-}\mathrm{pre}} = \ds\frac{a_{\alpha\beta}\,L_i^{\alpha\beta}}{a^{\mathrm{pre}}_{\gamma\delta}\,L_i^{\gamma\delta}}
 \eqe
  denotes the {square of the} instantaneous fiber stretch measured with respect to the configuration at the preceding computational load or time step. Term \eqref{e:Wstabvis1} leads to the stabilization stress
\eqb{lll}
\tau^{\alpha\beta}_\mathrm{fib\mbox{-}stab} \is \ds   \epsilon^\mre_{\mathrm{stab}} \ds\sum\limits_{i=1}^{n_\mrf} (\lambda_i-1) \frac{1}{\lambda_i}\,L_i^{\alpha\beta} +  \ds   \epsilon^\mrv_{\mathrm{stab}} \sum\limits_{i=1}^{n_\mrf}  (\tilde\lambda_i-1)\frac{1}{\tilde\lambda_i} \,\ellab_{i\mbox{-}\mathrm{pre}}
 \label{e:Wstabvis1_stress}
\eqe
and its tangent
\eqb{lll}
 c^{\alpha\beta\gamma\delta}_\mathrm{fib\mbox{-}stab} \is \ds   \epsilon^\mre_{\mathrm{stab}} \sum\limits_{i=1}^{n_\mrf}\, \lambda_i^{-3}\, L_i^{\alpha\beta}\,L_i^{\gamma\delta} +   \epsilon^\mrv_{\mathrm{stab}} \sum\limits_{i=1}^{n_\mrf}\, \tilde\lambda_i^{-3}\, \ellab_{i\mbox{-}\mathrm{pre}} \,\ell^{\gamma\delta}_{i\mbox{-}\mathrm{pre}}~.
 \label{e:Wstabvis1_tangent}
\eqe


\begin{remark}
Note that the stress  in Eq.~\eqref{e:Wstabvis1_stress} (and its tangent in Eq.~\eqref{e:Wstabvis1_tangent}) are  added into the system only for fiber compression, i.e.~$ \lambda_i < 1$ with $\epsilon_\mrL^i = 0$.
\end{remark}

\begin{remark}
The first term in \eqref{e:Wstabvis1} {describes the elastic response} of fibers (or a bundle of fibers) in compression. In some cases, this  compressive resistance 
can be physically justified. E.g.~for woven fabrics, a small resistance stems from  the small local out-of-plane fiber bending stiffness  due to initial crimping of yarns in the fabrics. On the other hand, the first term in Eq.~\eqref{e:Wstabvis1} can also be used as a penalty regularization for enforcing  near incompressibility of fibers (if required), by simply setting  $\epsilon^\mre_\mathrm{stab}$ to a large value. 
\end{remark}

\begin{remark}
The second term in  \eqref{e:Wstabvis1}  provides numerical damping to fibers in compression. It stems from the potential 
\eqb{lll}
W^i_\mrv :=   \ds \frac{1}{2}\, \eta \,\,    ( \dot{ \tilde{\lambda}   }_i)^2 ~,
 \label{e:WstabvisO}
\eqe
where $\eta$ denotes the so-called (instantaneous) stretching viscosity, 
  and the approximation $\dot{ \tilde{\lambda}   }_i \approx  (\tilde{\lambda}_i - 1)/\Delta t$, with  $\Delta t $ being time step size, has been used. {This approximation is first order accurate since $\tilde{\lambda}_i$  is the stretch w.r.t.~the previous time step}. Therefore, $\epsilon^\mrv_{\mathrm{stab}}$ relates to  the stretching viscosity by  $\eta:=2\, \epsilon^\mrv_{\mathrm{stab}} \, \Delta t^2$. 
\end{remark}

\begin{remark}
The second term in \eqref{e:Wstabvis1} is influenced not only by the parameter  $\epsilon^\mrv_\mathrm{stab}$ but also depends upon step size $\Delta t$. That is, for a fixed $\epsilon^\mrv_\mathrm{stab}$, 
the stretch $\tilde\lambda_i$ approaches $1$ when $\Delta t$ decreases. Hence, the stored energy  $\epsilon^\mrv_{\mathrm{stab}}\, \big( \tilde\lambda_i -  1 \big)^2$ consistently approaches zero as $\Delta t \rightarrow 0$.
\end{remark}
%

%
%
%
\vspace{-0.5cm}
\section{Numerical examples: Homogeneous deformation}{\label{s:num_examples1}}
\vspace{-0.2cm}
This section verifies the proposed {isogeometric} finite element shell formulation via several benchmark examples characterized by homogeneous deformations. 
The FE simulation results  are compared with exact solutions provided by Duong et al.~\cite{shelltextile}. For all these examples, the domain is discretized by a single quadratic NURBS patch. Material model \eqref{e:eg_Wsimple}--\eqref{e:eg_W1} is used, with its parameters  specified separately for each example.

For unit normalization, we use a reference length $L_0$, and a reference surface stress $\epsilon_0$, which has the unit [force$/$length].   Therefore,  the unit of surface strain energy density $W$, reaction forces, and reaction moments is $[\epsilon_0]$, $ [\epsilon_0\,L_0]$, and $[\epsilon_0\,L_0^2]$, respectively.
The units of the material parameters in model~\eqref{e:eg_Wsimple}--\eqref{e:eg_W1} then  follow as: $[\epsilon_0]$ for  meambrane stiffnesses $\mu,~\epsilon_\mrL,~\epsilon_\mra$, and $[\epsilon_0\,L_0^2]$ for bending stiffnesses $\beta_\mrn,~\beta_\tau,~\beta_\mrg$.

\vspace{-0.1cm}
\subsection{Uniaxial tension}
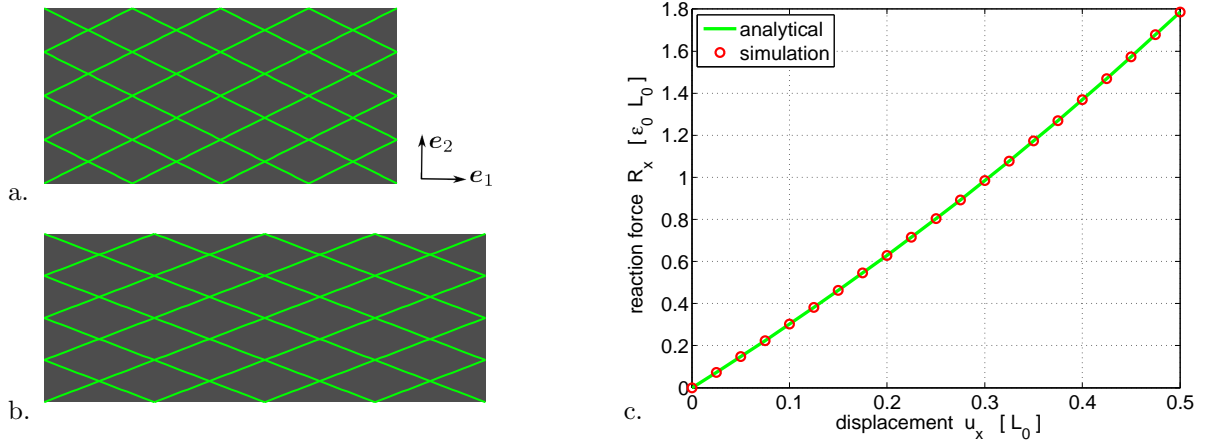
\begin{figure}[H]
\begin{center} \unitlength1cm
\begin{picture}(0,5.5)

\put(-8.3, 1.4){\includegraphics[width=0.47\textwidth]{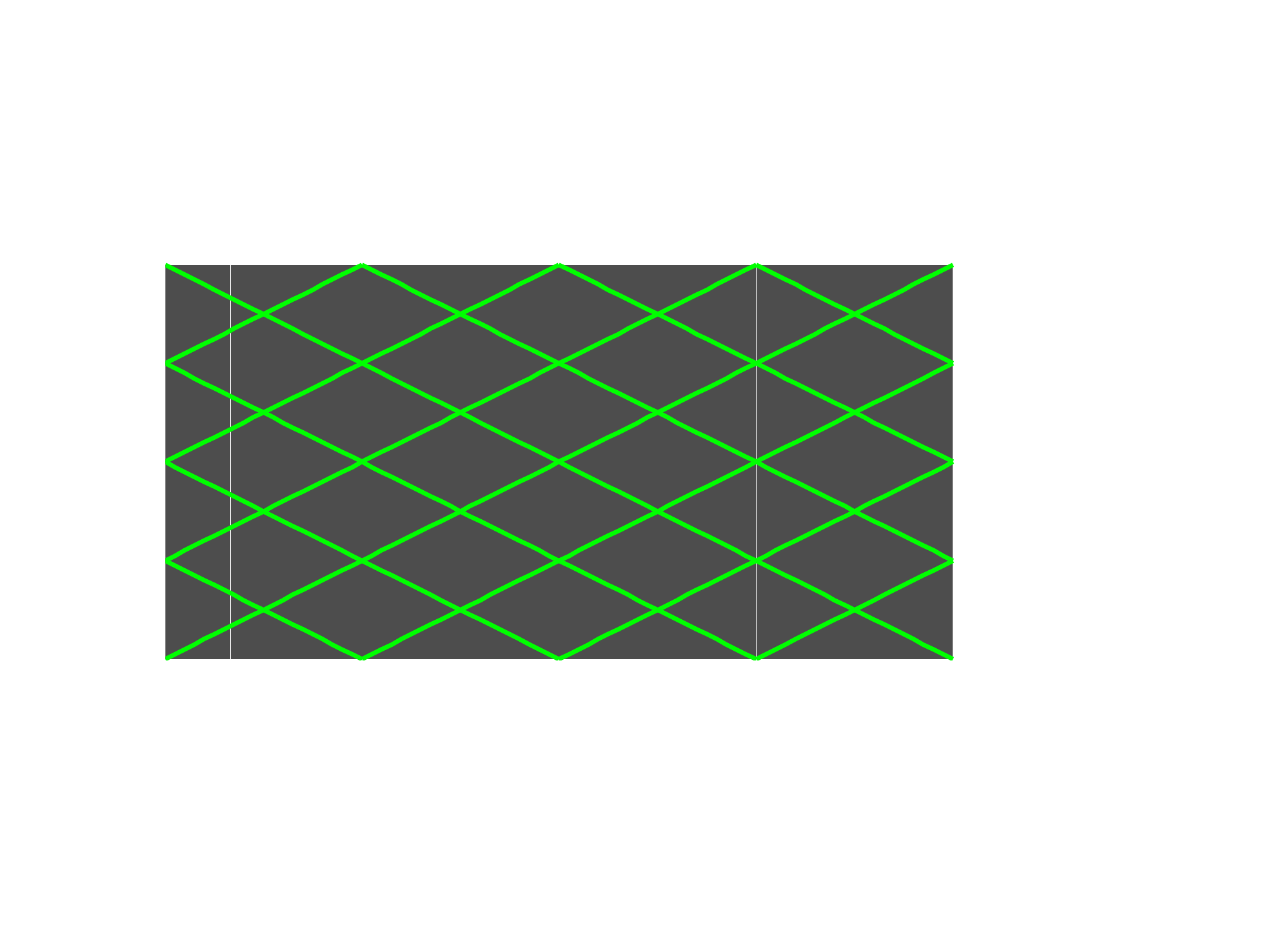}}

\put(-8.3,-1.5){\includegraphics[width=0.47\textwidth]{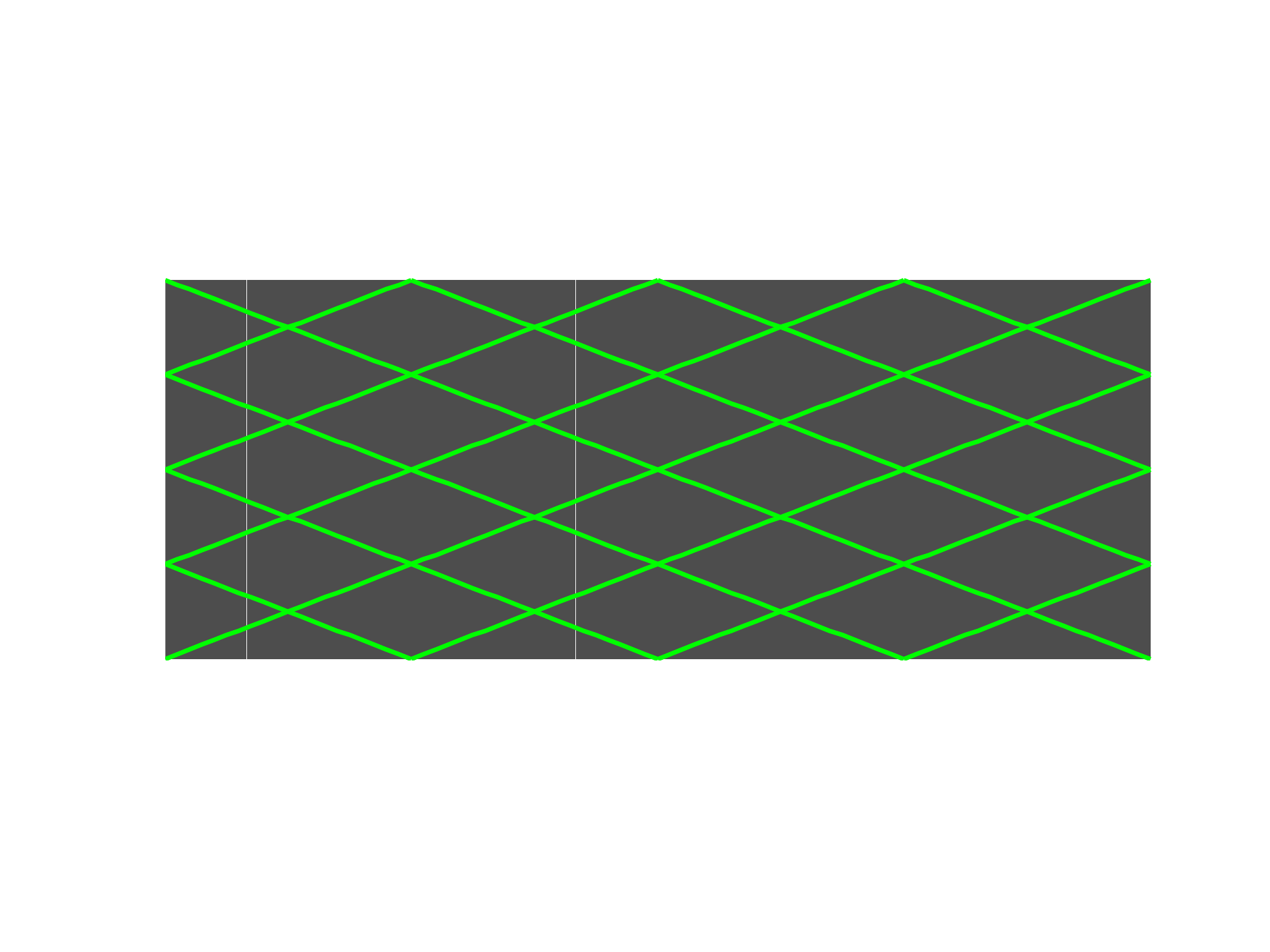}}
\put(-2.4,3.1){\includegraphics[width=0.06\textwidth]{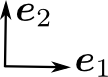}}
\put(0.12,-0.3){\includegraphics[width=0.52\textwidth]{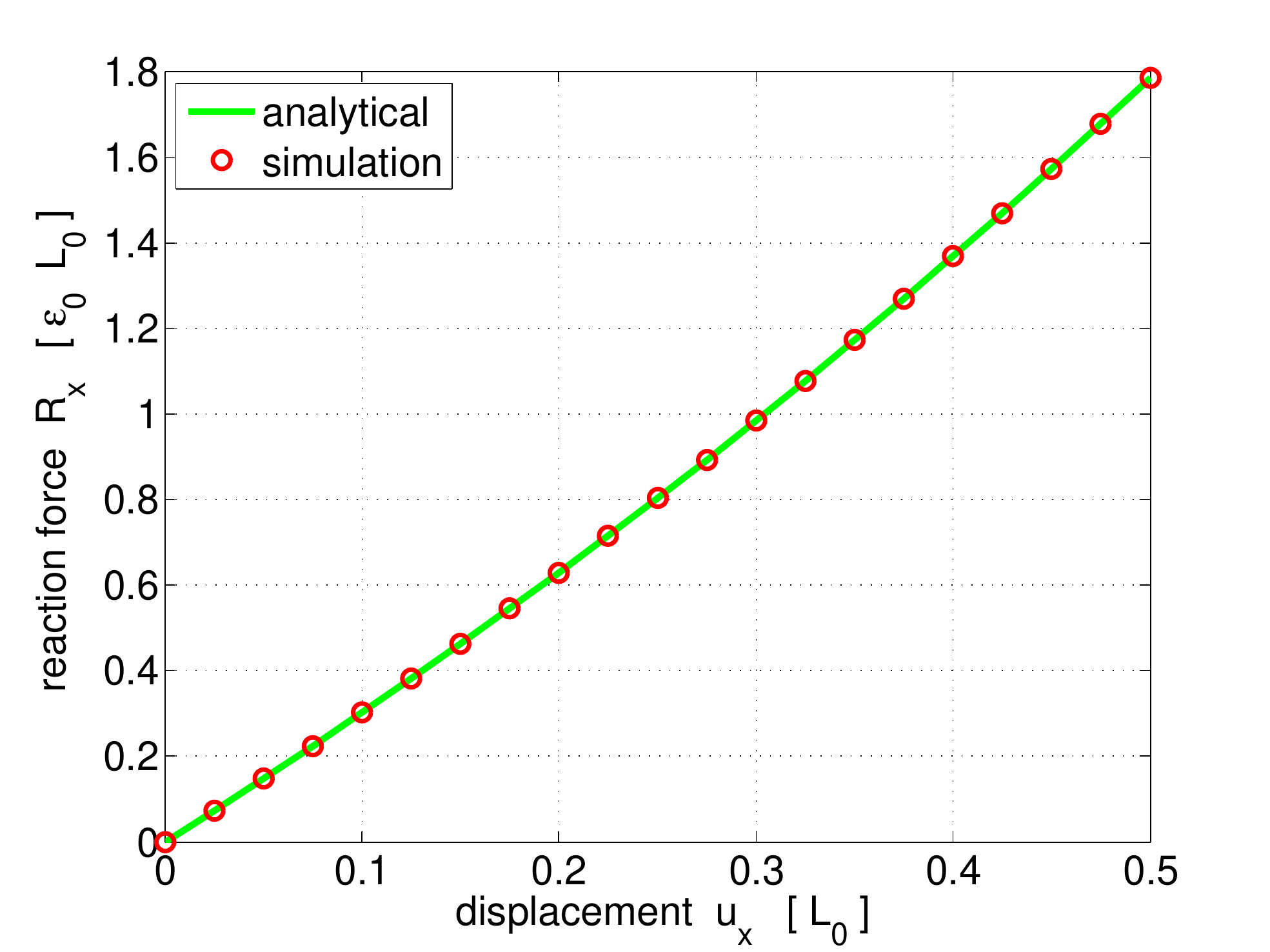}}

\put(-7.8,2.9){{\small{a.}}}
\put(-7.8, 0.0){{\small{b.}}}

\put(0.3,0){{\small{c.}}}

\end{picture}
\caption[caption]{Uniaxial tension: {a.}~Initial and {b.}~deformed configurations with two fiber families (in green). 
{c.}~Comparison with the analytical solution of Duong et al.~\cite{shelltextile} 
for the reaction force $R_x$ vs.~displacement $u_x$ at $X = 2\,L_0$.
Here, $\epsilon_\mrL/2= \mu=\epsilon_\mra = \epsilon_0$.  
}
\label{f:uniax1}
\end{center}
\vspace{-0.5cm}
\end{figure} 

The first example considers uniaxial tension of a rectangular sheet of 
{size} $2L_0\times L_0$ as shown in Fig.~\ref{f:uniax1}a-b. The sheet consists of two fiber families with 
{initial} directions {$\bL_1 = (2\,\be_1 + \be_2)/\sqrt{5}$ and  $\bL_2 = (2\,\be_1 - \be_2)/\sqrt{5}$}. 
{The top edge is free, while the left and bottom edges are fixed along  $\be_1$ and $\be_2$, respectively.} The sheet is pulled by applying the displacement $u_x$ on the right edge in 
{the $\be_1$ direction}. 
In this test we use the material parameters $\mu = \epsilon^{12}_\mra =\epsilon_0$, $\epsilon^i_\mrL = 2\,\epsilon_0$, while $U(J)$ is set to zero, and $\beta^i_\mrn$, $\beta^i_\mrg$, and $\beta_\tau^i$ have no influence. 


Fig.~\ref{f:uniax1}c shows the FE results in comparison to the exact solution. Our implementation is verified by obtaining an error within machine precision for a single finite element.
%
%

\subsection{{Pure shear}}

\begin{figure}[H]
\begin{center} \unitlength1cm
\begin{picture}(0,4.8)
\put(-8.6,-0.2){\includegraphics[height=0.28\textwidth]{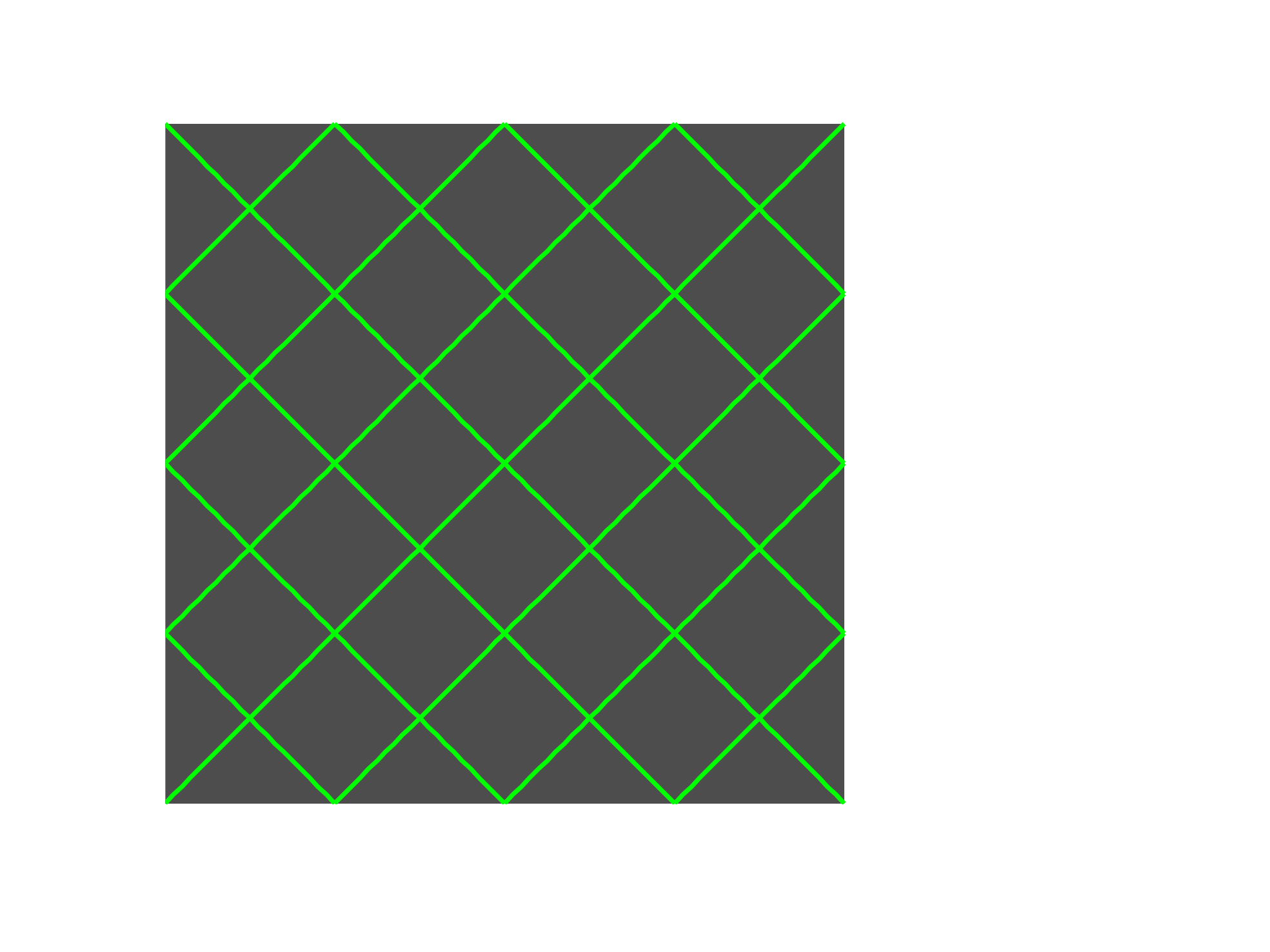}}
\put(-3.4,3){\includegraphics[width=0.07\textwidth]{figs_examples/be_alpha.png}}
\put(-4.4,-0.2){\includegraphics[height=0.28\textwidth]{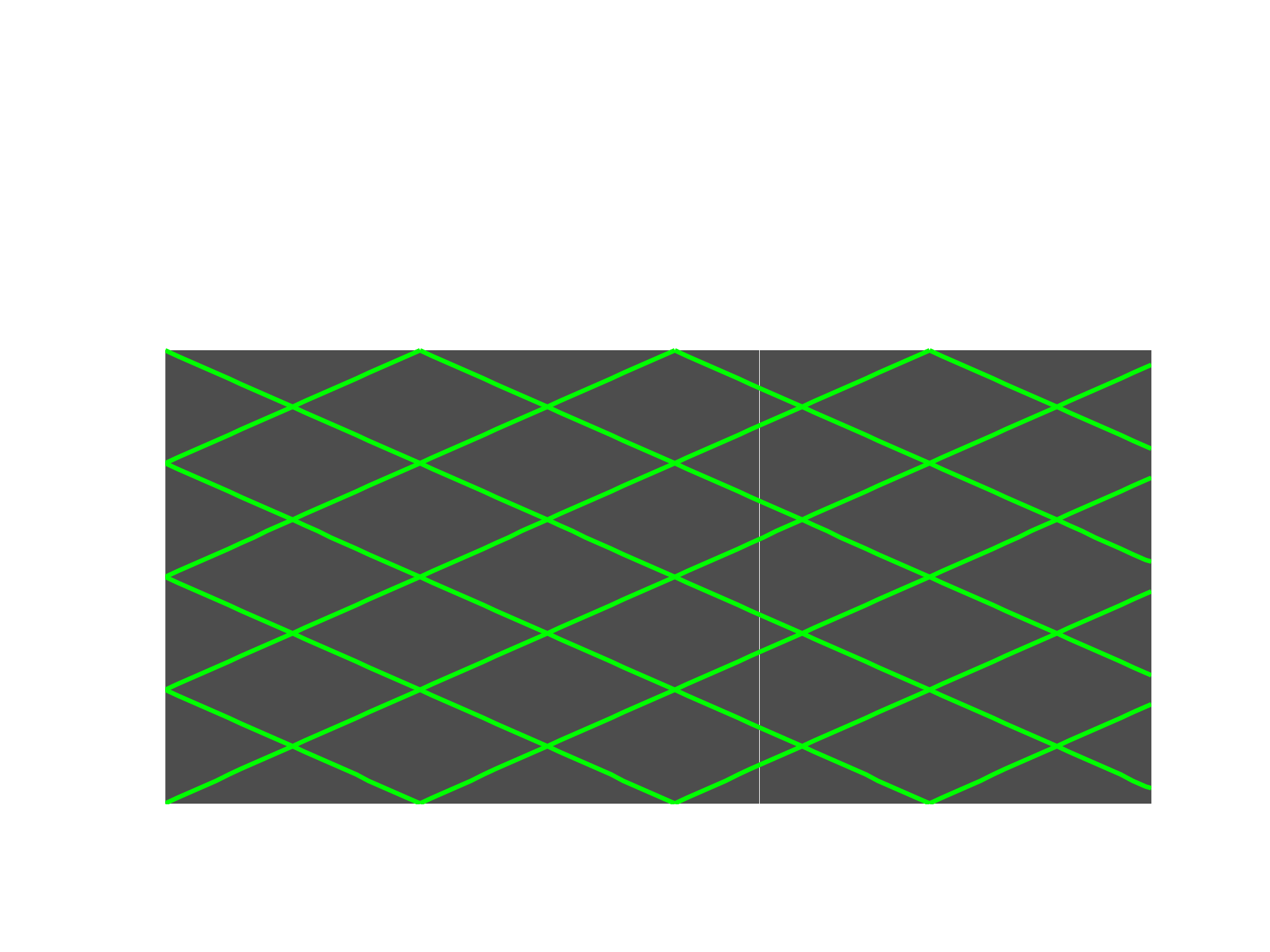}}
\put(1.62,-0.15){\includegraphics[width=0.43\textwidth]{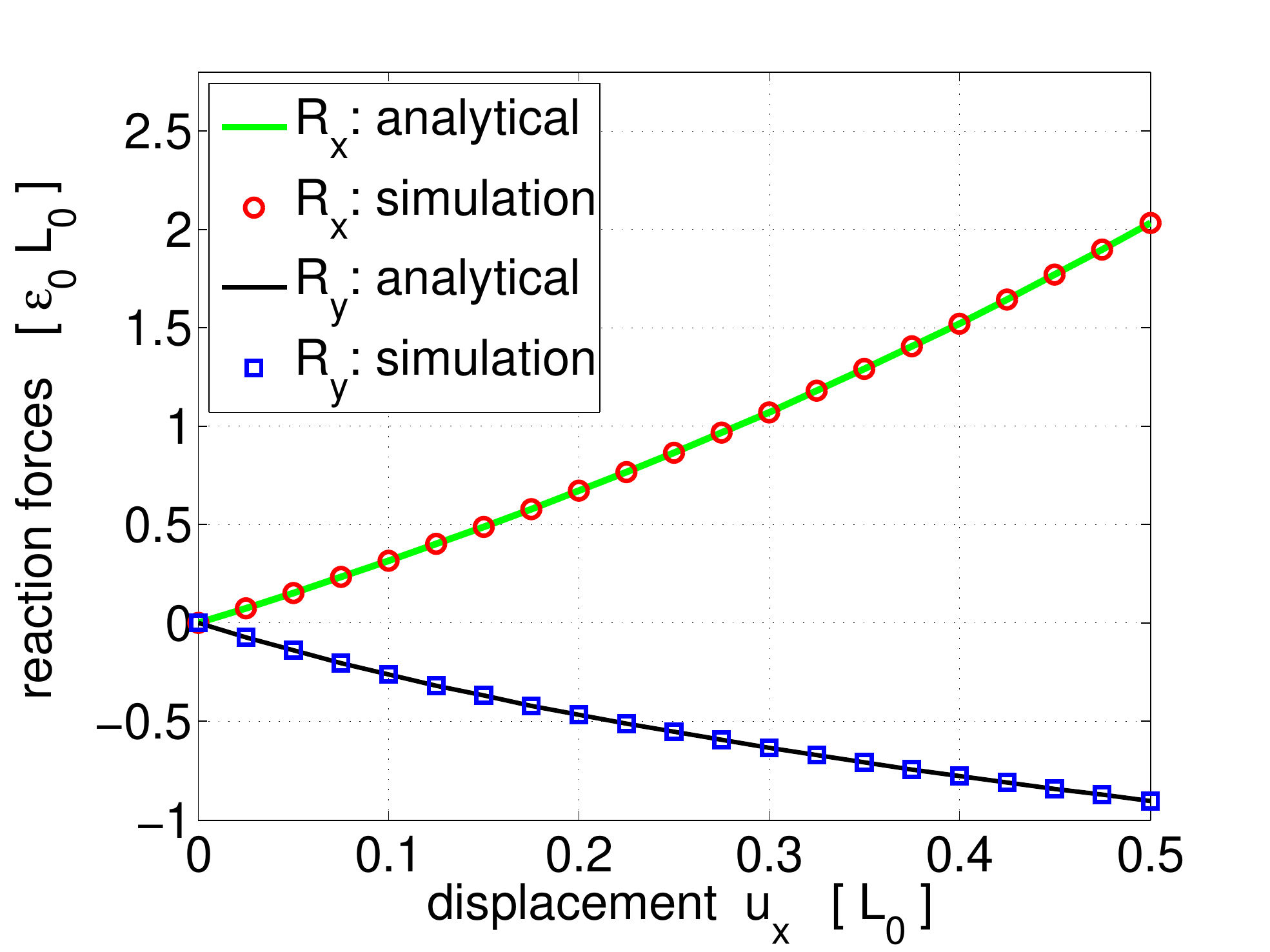}}
\put(-7.9,-0.1){{\small{a.}}}
\put(-3.8,-0.1){{\small{b.}}}
\put(1.7,-0.1){{\small{c.}}}
\end{picture}
\caption[caption]{Pure shear: {a.}~Initial and {b.}~deformed configurations with two fiber families. {c.}~Comparison with the exact solution \cite{shelltextile} for the reaction forces vs.~displacement $u_x$ at $X = L_0$.  Here, $\epsilon_\mrL/2=\epsilon_\mra= \mu=\epsilon_0$. }
\label{f:biiax1}
\end{center}
\end{figure}

Next, our implementation is tested for pure shear. A square sheet with diagonal fibers is considered as shown in Fig.~\ref{f:biiax1}a. A Dirichlet boundary condition is applied on all edges, such that the sheet with initial dimension $L_0\times L_0$ is deformed into the rectangular shape $\ell\times h$, with 
{$\bar\lambda := \ell/L_0 = L_0/h$} (Fig.~\ref{f:biiax1}b). The material parameters are $\mu = \epsilon_0$, $\epsilon^i_\mrL = 2\,\epsilon_0$, and  $\epsilon^{12}_\mra = \epsilon_0$. Further, $U(J)$ is set to zero, while $\beta^i_\mrn$, $\beta^i_\mrg$, and $\beta_\tau^i$ have no influence. 
The exact solution for the reaction forces is \cite{shelltextile}
 \eqb{lll}
  R_x \is \ds h\,\left[\mu\,( \bar\lambda^2 - 1) + \frac{1}{4}\,\epsilon_\mrL\,( {\bar\lambda}^4 - 2\,{\bar\lambda}^2 + 1) + \frac{1}{4}\,\epsilon_\mra\,({\bar\lambda}^4 - 1)\right]~,\\[3mm]
  R_y \is \ds \ell\,\left[\mu\, \left( \frac{1}{\bar\lambda^2} - 1\right) + \frac{1}{4\,\bar\lambda^4}\,\epsilon_\mrL\,( {\bar\lambda}^4 - 2\,{\bar\lambda}^2 + 1) - \frac{1}{4\bar\lambda^4}\,\epsilon_\mra\,({\bar\lambda}^4 - 1)\right]~.
 \eqe
Fig.~\ref{f:biiax1}c shows the comparison between the simulation and the exact solution. Again, we obtain an error within machine precision for a single finite element.

%
%
%

\subsection{Picture frame test}

\begin{figure}[H]
\begin{center} \unitlength1cm
\begin{picture}(0,5.6)
\put(-9.5,0){\includegraphics[width=0.58\textwidth]{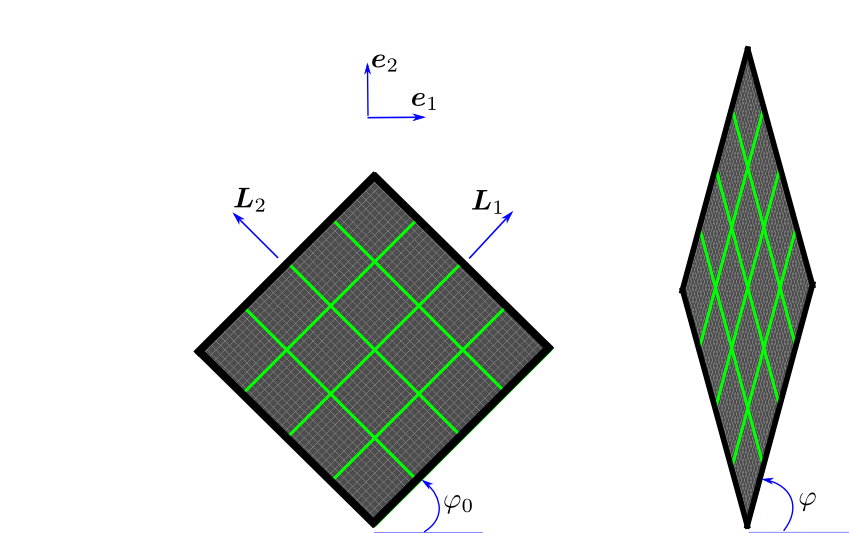}}
\put(0.5,-0.15){\includegraphics[width=0.50\textwidth]{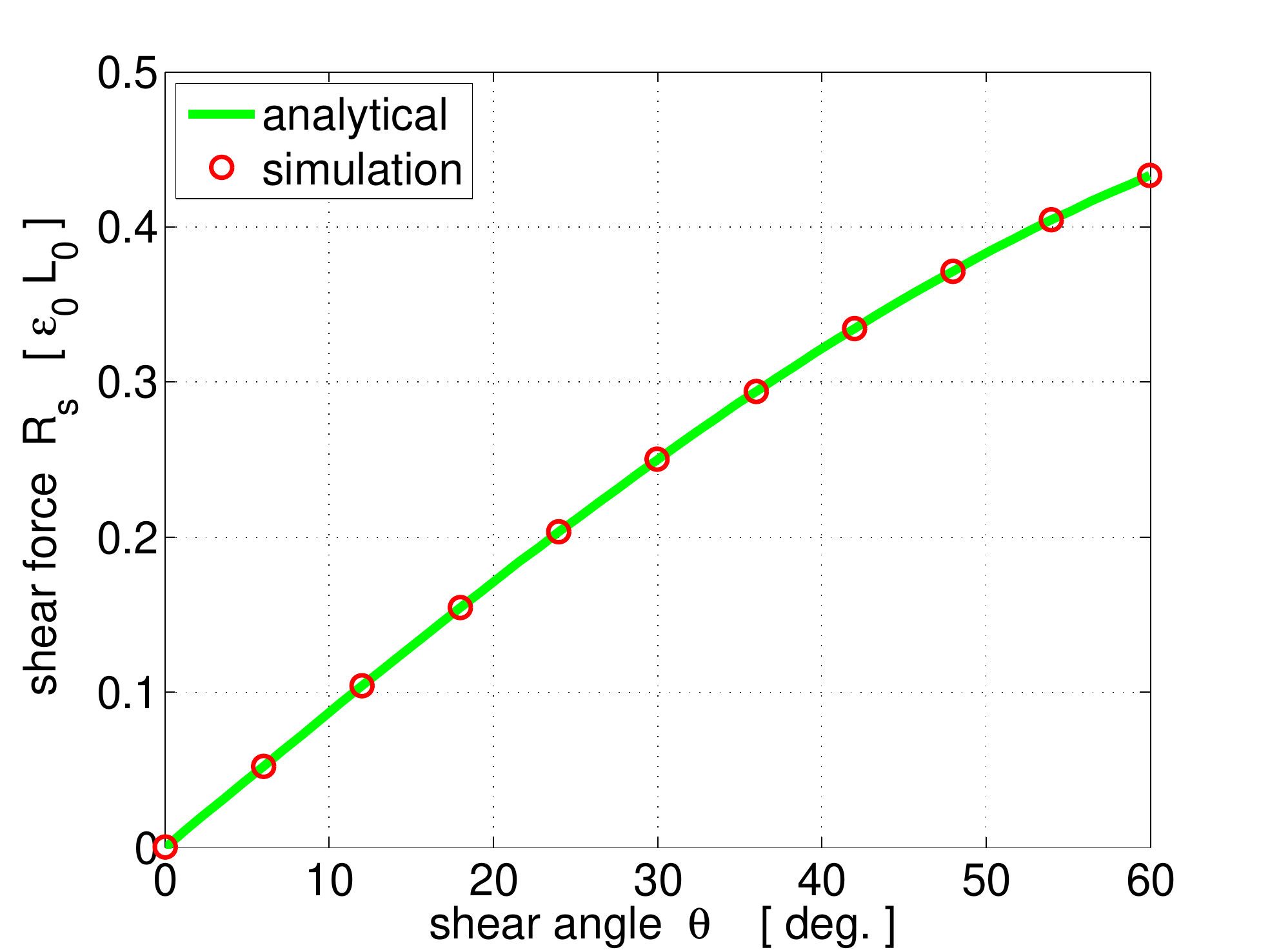}}
\put(-6.0,-0.1){{\small{a.}}}
\put(-1.8,-0.1){{\small{b.}}}
\put(0.6,-0.1){{\small{c.}}}
\end{picture}
\caption[caption]{Picture frame test: {a.}~Initial and {b.}~deformed configurations with two fiber families. {c.}~Comparison with the exact solution \cite{shelltextile} for the shear force vs.~shear angle $\theta:= 2\,\varphi- 90^\circ$.  Here, $\epsilon^{12}_\mra =\epsilon_0$.}
\label{f:pic1}
\end{center}
\end{figure}
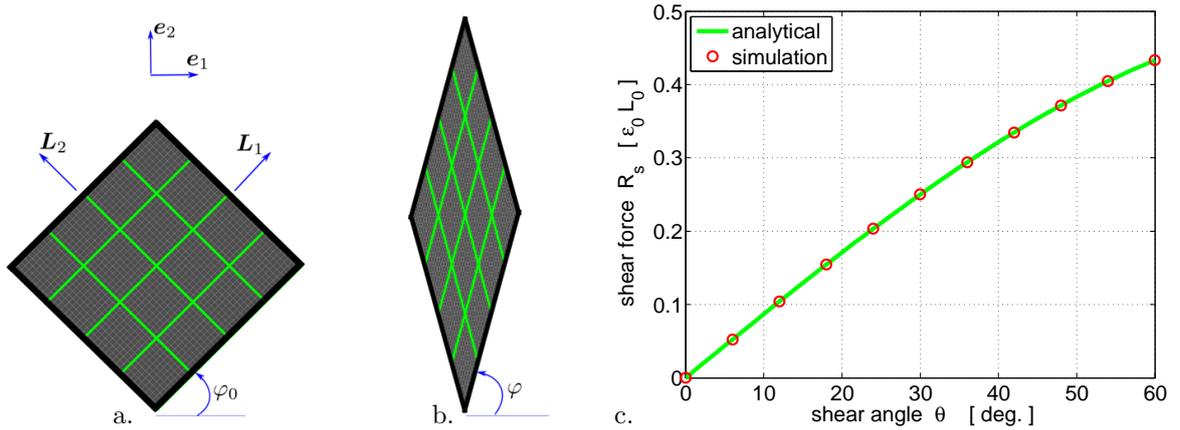

The third example verifies our implementation in the picture frame test. A $L_0\times L_0$ square sheet with  two fiber families is considered as shown in Fig.~\ref{f:pic1}a. The picture frame deformation (see Fig.~\ref{f:pic1}b) is obtained by applying the Dirichlet boundary condition  $\bar\bx (\varphi,\bar\bX) = \sqrt{2}\,\big(\cos\varphi\,\be_1\otimes\be_1 + \sin\varphi\,\be_2\otimes\be_2\big)\,\bar\bX$ for every boundary node $\bar\bX$ of the frame.  The material parameters are taken as $\epsilon^{12}_\mra = \epsilon_0$,  $U(J)=\mu=0$, while  $\epsilon^i_\mrL$, $\beta^i_\mrn$, $\beta^i_\mrg$, and $\beta_\tau^i$ have no influence. The exact solution of 
{the shear force} (i.e.~the tangential reaction) at an edge of the sheet is $R_\mrs = -\epsilon_\mra^{12}\,\cos(2\,\varphi)\,L_0/2$, see e.g.~\cite{shelltextile}. Fig.~\ref{f:pic1}c shows  agreement between the simulation and the exact solution. Again, the error is  within machine precision for a single finite element.

\vspace{-0.8cm}
\subsection{Annulus expansion}
\begin{figure}[H]
\begin{center} \unitlength1cm
\begin{picture}(0,4.4)

\put(-8.0,0.5){\includegraphics[width=0.25\textwidth]{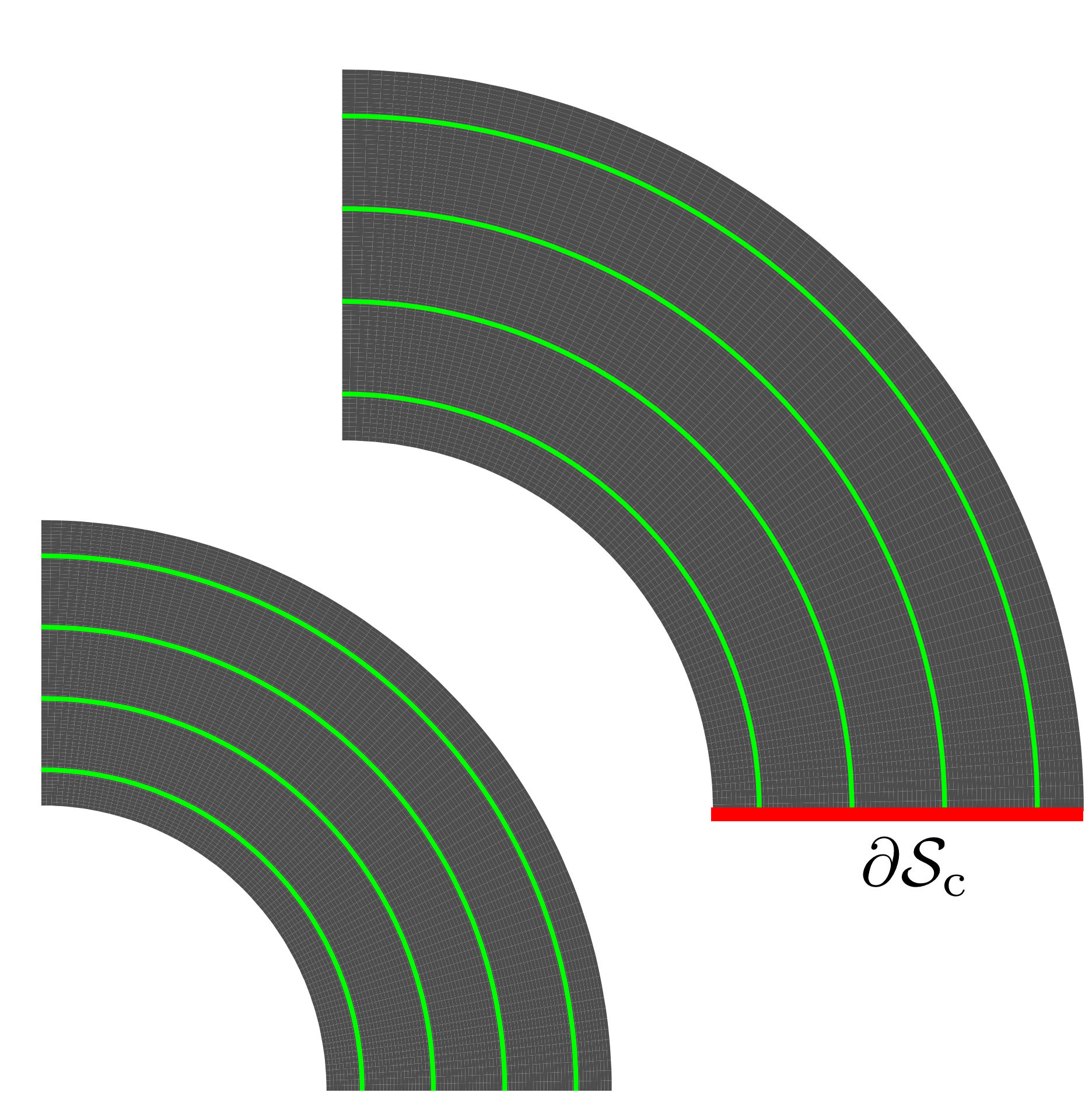}}

\put(-3.7,0){\includegraphics[width=0.38\textwidth]{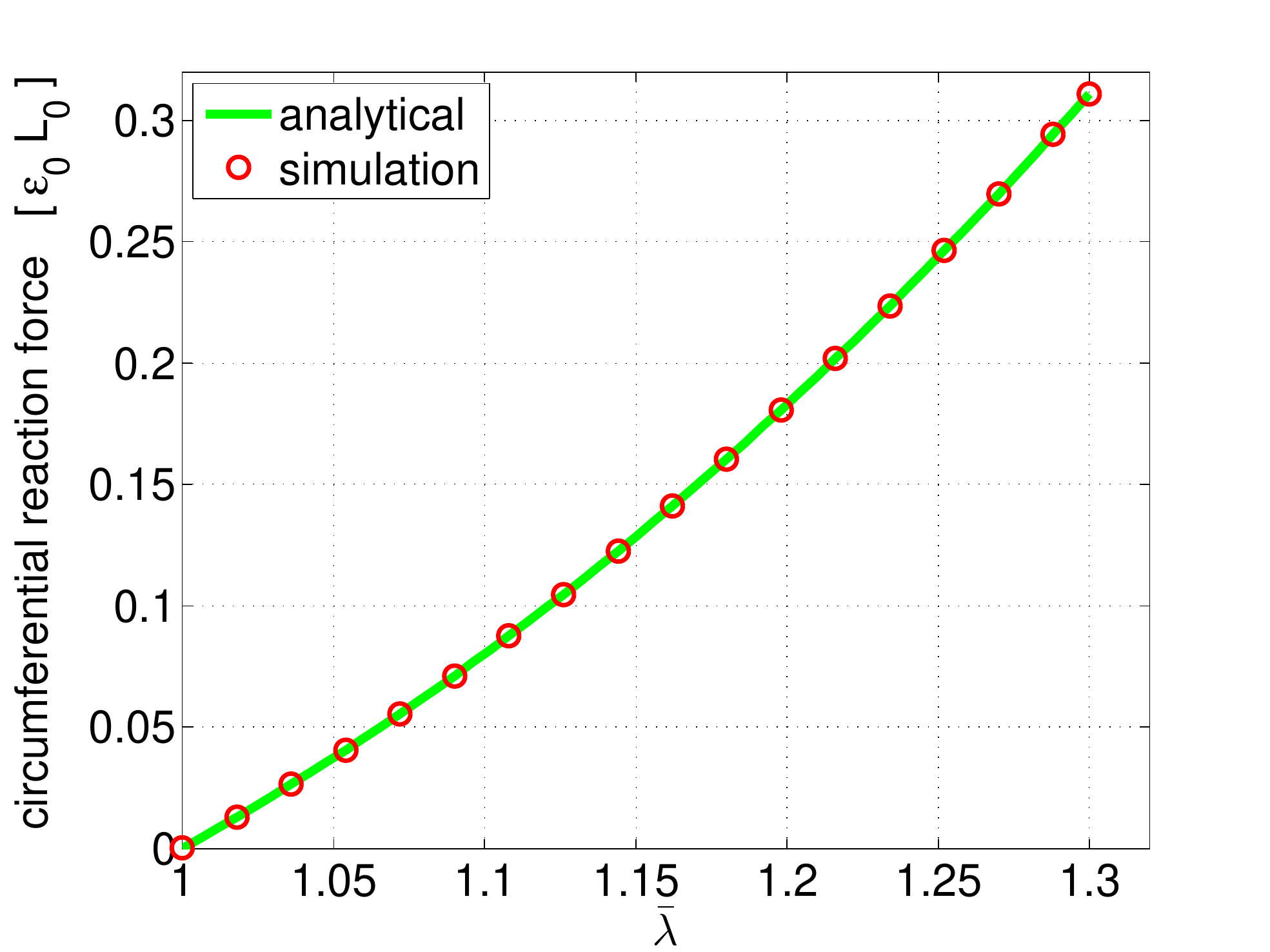}}
\put(2.4,0){\includegraphics[width=0.38\textwidth]{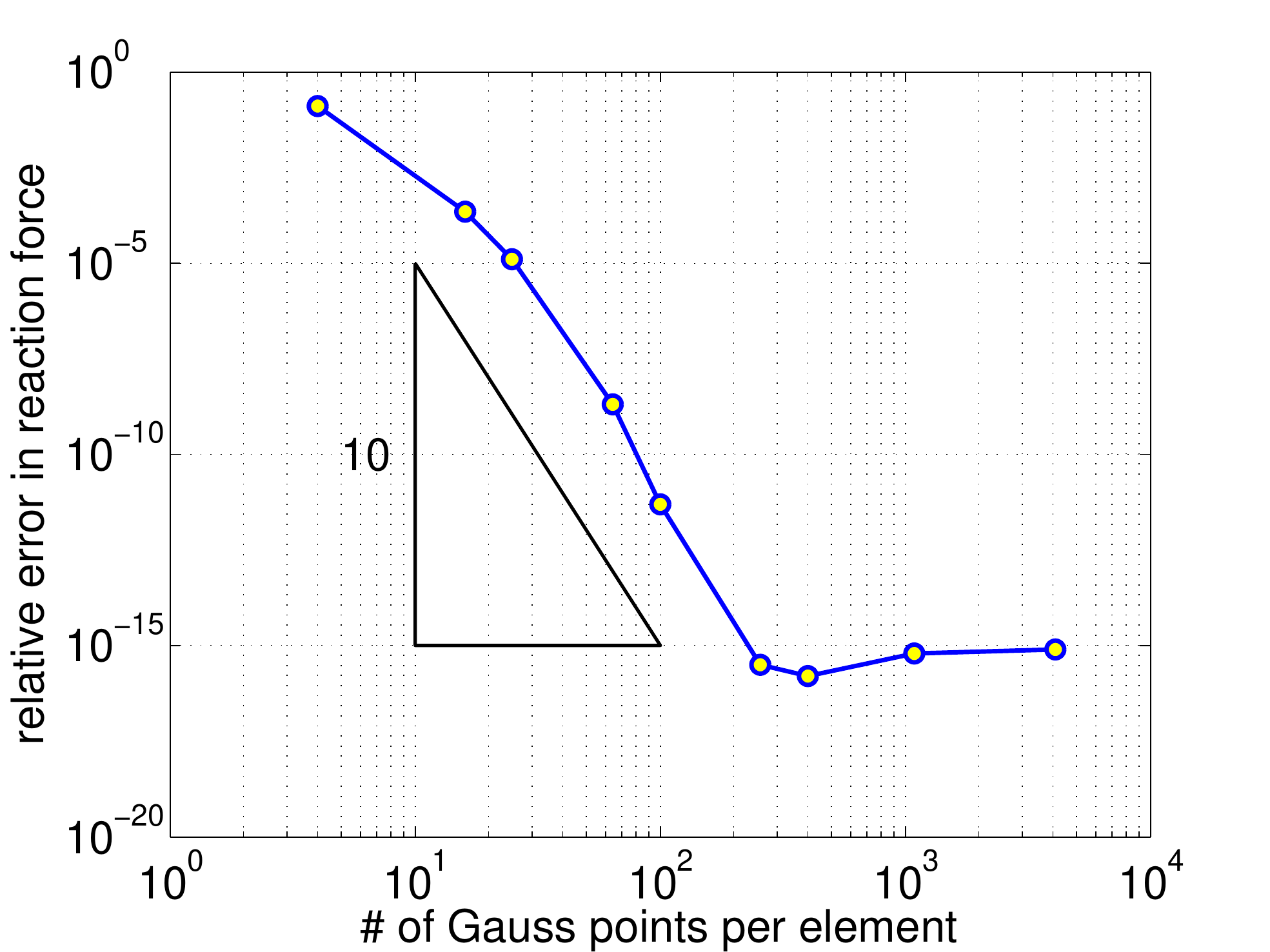}}

\put(-7.9,1.2){{\small{a.}}}
\put(-7.2,2.9){{\small{b.}}}

\put(-3.65,0){{\small{c.}}}
\put(2.3,0){{\small{d.}}}

\end{picture}
\caption[caption]{Annulus expansion: a.~Initial and b.~deformed configurations at $\bar{\lambda} = 1.3$ with distributed fibers. c.~Circumferential reaction force vs. $\bar\lambda$ compared to the analytical solution \cite{shelltextile}. d.~Relative error in the reaction force  vs. the number of Gauss points.
Here, the error is defined by $ |R_\mathrm{num} - R_\mathrm{exact}|/R_\mathrm{exact}$, where $R_\mathrm{num} =  \int_{\partial\sS_\mrc} \bnu\,\bsig\,\bnu\,\dif s$ and $R_\mathrm{exact}$ are the circumferential reaction forces according to the FE solution and the analytical solution, respectively,  while  $\bnu$ denotes the normal vector of the interface $\partial\sS_\mrc$.
}
\label{f:disk1}
\end{center}
\end{figure}
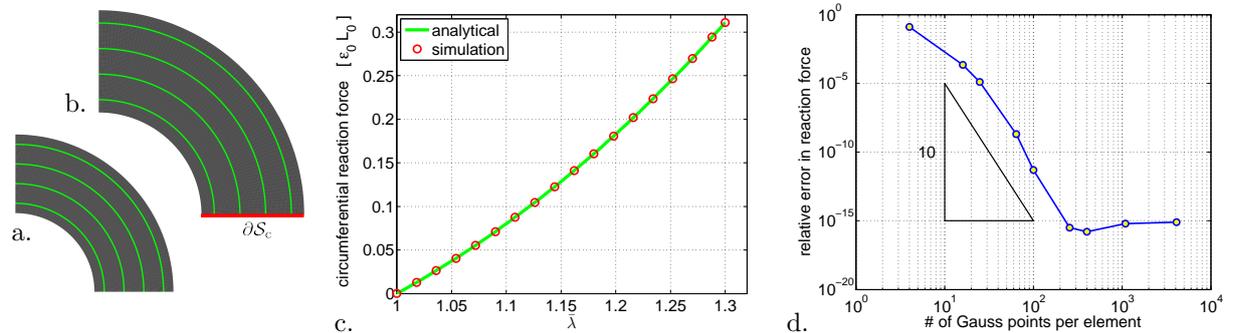

{\label{s:disk_expan}}
The fourth example considers the homogeneous expansion of an annulus containing distributed {circumferential} fibers embedded in a matrix material. Due to the symmetry, only one fourth of the annulus is simulated as shown in Fig.~\ref{f:disk1}a.  {In the reference configuration}, the annulus has inner radius $R_\mri=L_0/2$ and outer radius $R_\mro = L_0$. A Dirichlet boundary condition is applied on the inner and outer boundary, such that they both expand with 
the stretch $\bar\lambda:=r_\mri/R_\mri = r_\mro/R_\mro$, see Fig.~\ref{f:disk1}b.

In order to induce a homogeneous deformation within the annulus, a graded matrix material with the surface dilatation energy $U(J) = \frac{1}{2}\,K\,(J-1)^2$, where $
K(R) = (\epsilon_\mrL/2)\,\ln R$, is required in Eq.~\eqref{e:eg_W1}, see \cite{shelltextile}. The material parameter $\epsilon_\mrL = 2\,\epsilon_0$ is used, while $\beta_\mrn$, $\beta_\tau$, and $\beta_\mrg$
 have no influence since {$K_\mrn=T_\mrg=K_\mrg=0$} during deformation \cite{shelltextile}. Fig.~\ref{f:disk1}c shows that the reaction force vs.~stretch curve is in good agreement with the exact solution. The error is within machine precision for a single finite element as long as numerical integration is sufficiently accurate as is shown in Fig.~\ref{f:disk1}d. 
 


\subsection{Pure bending}

\begin{figure}[H]
\begin{center} \unitlength1cm
\begin{picture}(0,9.0)
\put(1.5,4.5){\includegraphics[width=0.42\textwidth]{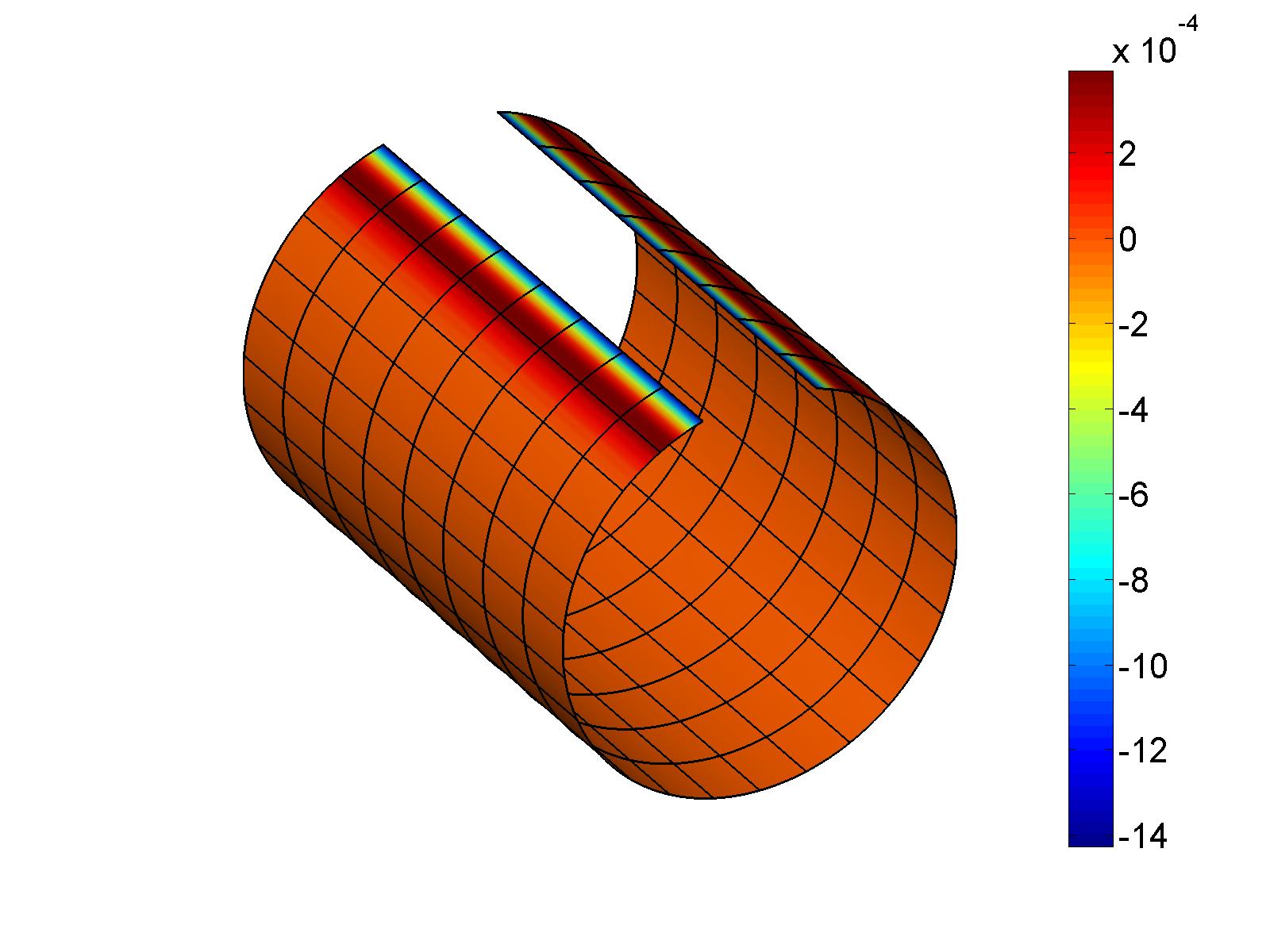}}
\put(-8.1,4.5){\includegraphics[width=0.63\textwidth]{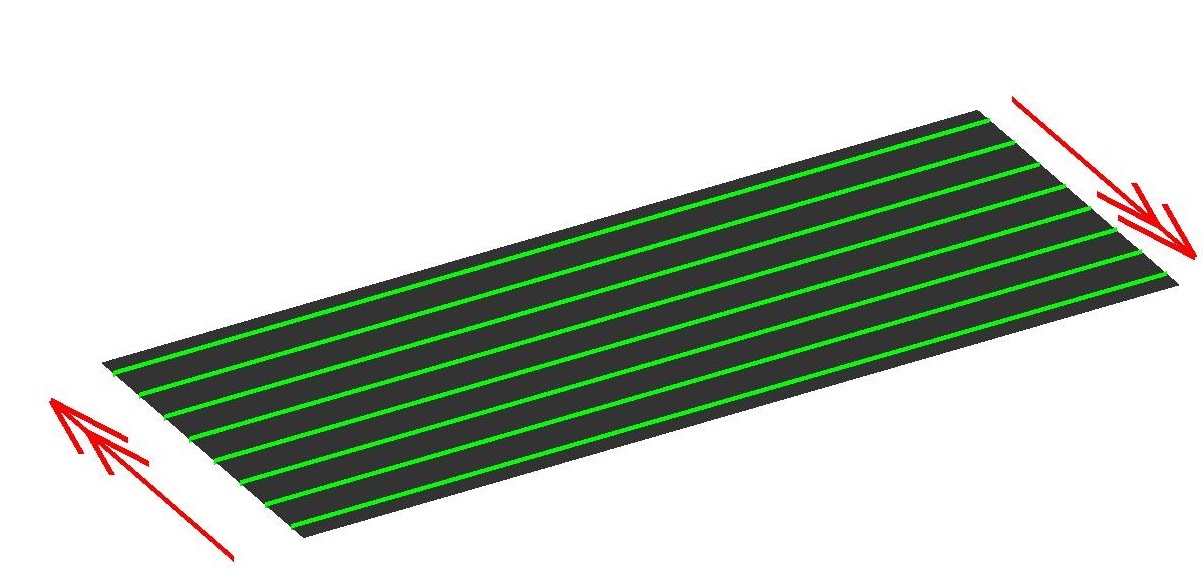}}

\put(-1.1,4.9){\includegraphics[width=0.07\textwidth]{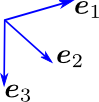}}

\put(-8.0,0){\includegraphics[width=0.35\textwidth]{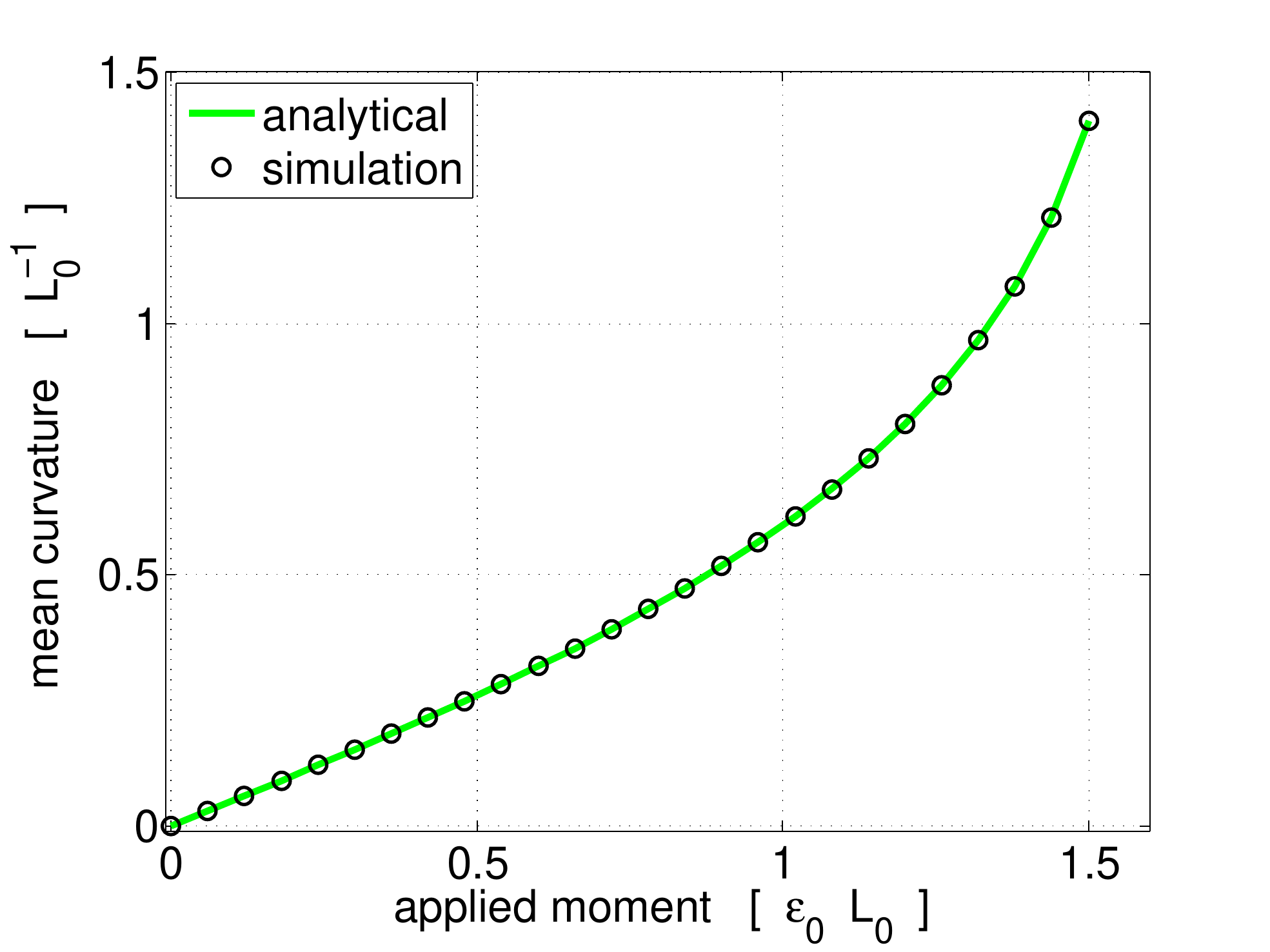}}
\put(-2.7,0){\includegraphics[width=0.35\textwidth]{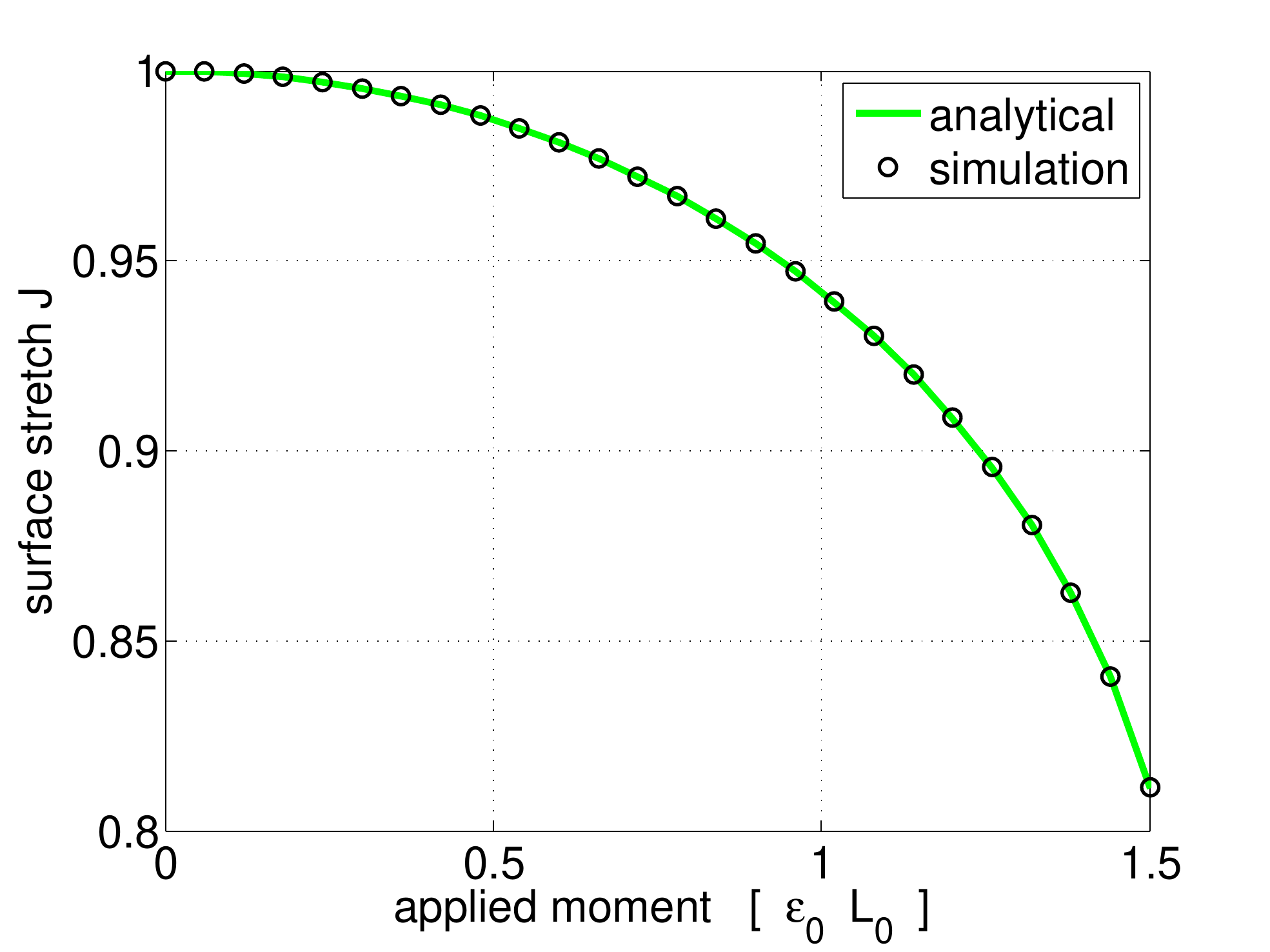}}

\put(2.7,-0.1){\includegraphics[width=0.36\textwidth]{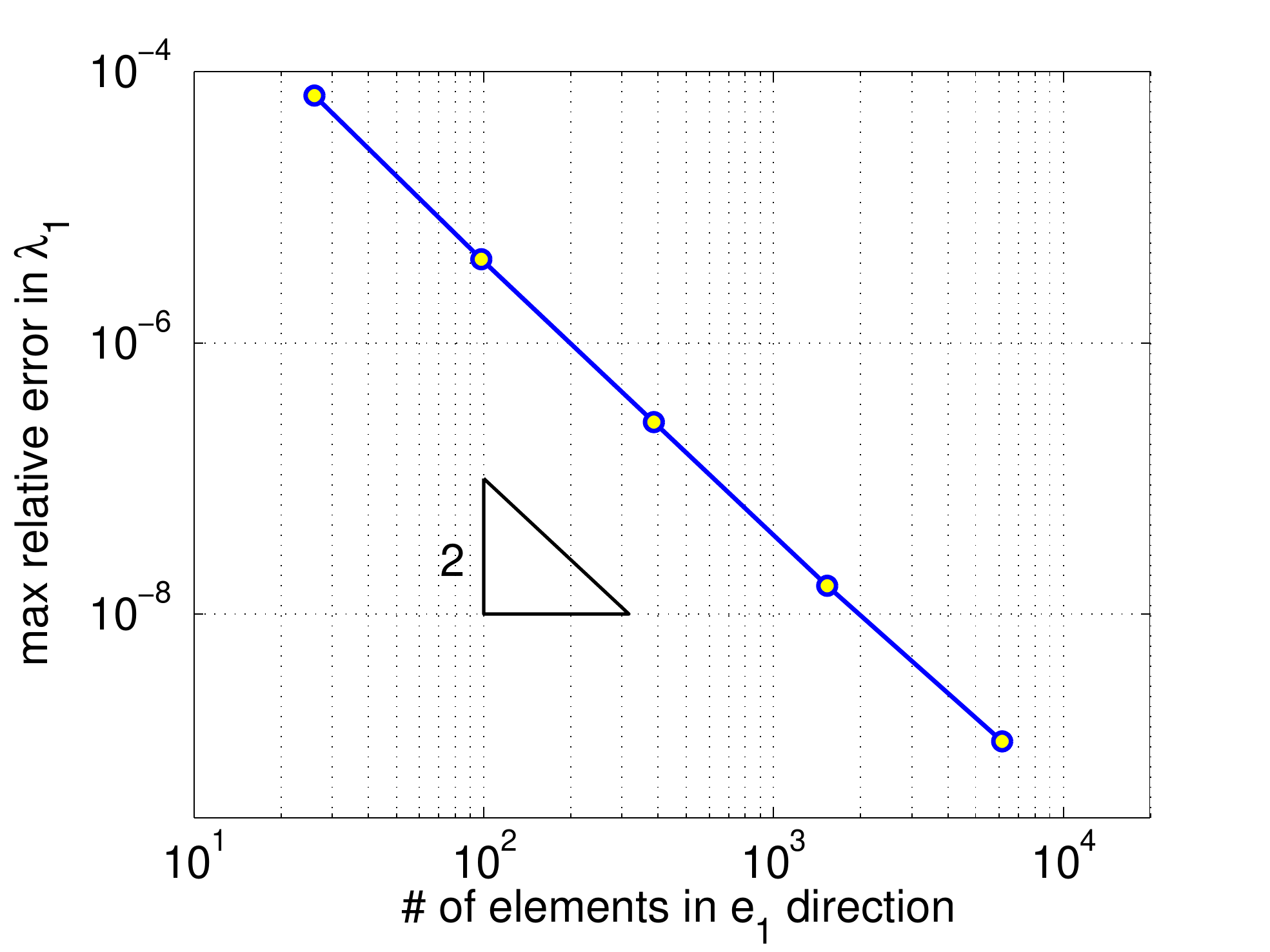}}

\put(-7.3,6.6){{\small{a.}}}
\put(2.7,6.5){{\small{b.}}}

\put(-7.9,-0.2){{\small{c.}}}
\put(-2.6,-0.2){{\small{d.}}}
\put(2.8,-0.2){{\small{e.}}}

\end{picture}
\caption[caption]{Pure bending of a flat sheet: a.~Initial configuration $(2.5 \,L_0 \times L_0)$ with fibers distributed along $\be_1$.  b.~Deformed configuration colored by the relative error in the mean curvature. Comparisons with the exact solution are shown in  {c.}~for the mean curvature and in {d.}~for surface stretch $J =\det_\mrs \bF$.  e.~Mesh convergence of the {maximum} relative error of $\bar\lambda_1$ (over the sample domain). 
Here,   $\mu = 10\,\epsilon_0$  and  $\beta_\mrn = \epsilon_0\,L_0^2$. }
\label{f:purebend1}
\end{center}
\end{figure}
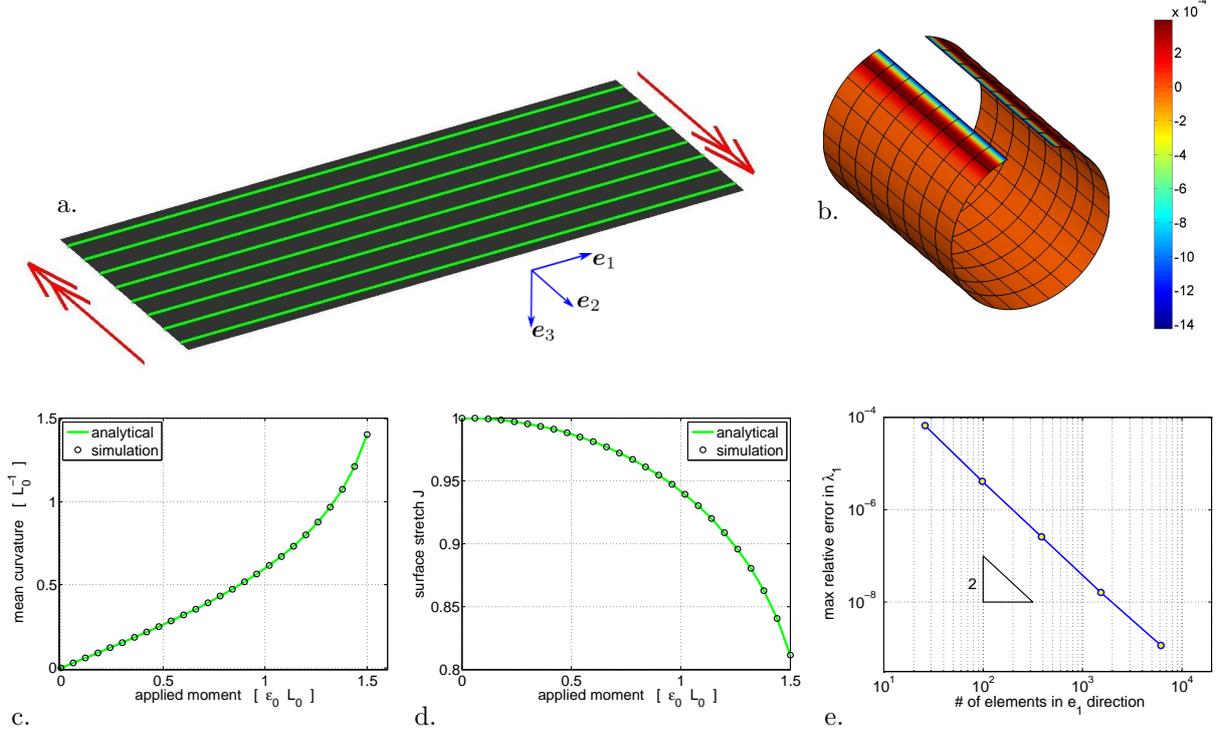

The fifth example considers pure bending of a flat rectangular sheet of size $2.5\,L_0\times L_0$ subjected to the distributed moment $M_{\mathrm{ext}}$  (unit [moment/length]) along the two shorter edges as shown in Fig.~\ref{f:purebend1}a. The sheet contains a single fiber family in the $\be_2$ direction embedded in a matrix material. Here, the material parameters are taken as  $\mu = 10\,\epsilon_0$ ,  $\beta_\mrn = \epsilon_0\,L_0^2$ , and $\epsilon_\mrL=0$, while $\beta_\mrg$ and $\beta_\tau$ have no influence.  The external moment deforms the sheet into a {cylindrical} segment as seen in Fig.~\ref{f:purebend1}b. 

%
According to Duong et al.~\cite{shelltextile},  the relationship 
\eqb{lll}
H = \ds \frac{M_{\mathrm{ext}}}{2\,\beta_\mrn\, \bar{\lambda}_1^4}
\eqe
between the mean curvature $H$ and external moment $M_{\mathrm{ext}}$ is obtained. Additionally, the exact solution for the stretch along the longer direction (due to high order effects) is 
\eqb{lll}
\bar\lambda_1 = \ds \sqrt{\frac{1}{2} + \sqrt{ \frac{1}{4} -  \frac{1}{\mu\,\beta_\mrn} \,M^2_{\mathrm{ext}} }  } ~,\quad\quad $with$\quad  M^2_{\mathrm{ext}}\leq \ds \frac{1}{4}\,\mu\,\beta_\mrn~,
\eqe
while the stretch along the shorter direction is $\bar\lambda_2 = 1$. 

Figs.~\ref{f:purebend1}c-d demonstrate good agreement between the exact and FEM solution for the mean curvature $H$ and surface stretch $J=\bar\lambda_1\,\bar\lambda_2$. The convergence with mesh refinement plotted in Fig.~\ref{f:purebend1}e verifies the consistency of the {isogeometric} finite element implementation.

\section{Numerical examples: Inhomogenous deformation} {\label{s:num_examples2}}
This section demonstrates the performance of the proposed shell formulation by  two tests: the bias extension and torsion tests of dry fabric sheets.
Further, we fit material model \eqref{e:eg_WFs}--\eqref{e:eg_WFs2} to the existing experimental data of Cao et al.~\cite{Cao2008}, and study the influence of in-plane bending.



\subsection{Bias extension of woven fabrics}{\label{s:example_WF_all}}
The first example studies the bias extension test for plain weave fabrics using material model  \eqref{e:eg_WFs2} within the proposed shell formulation.
\subsubsection{Bias extension of balanced weave fabrics: fitting to experimental data}{\label{s:example_WF}}

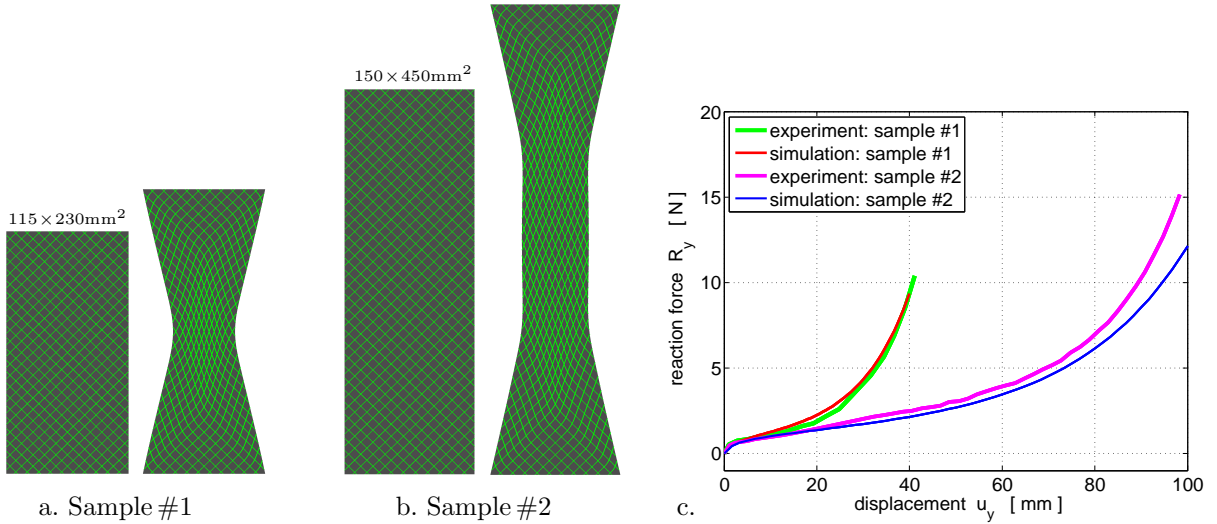
\begin{figure}[H]
\begin{center} \unitlength1cm
\begin{picture}(0,7.4)
\put(-7.9,0.14){\includegraphics[height=0.29\textwidth]{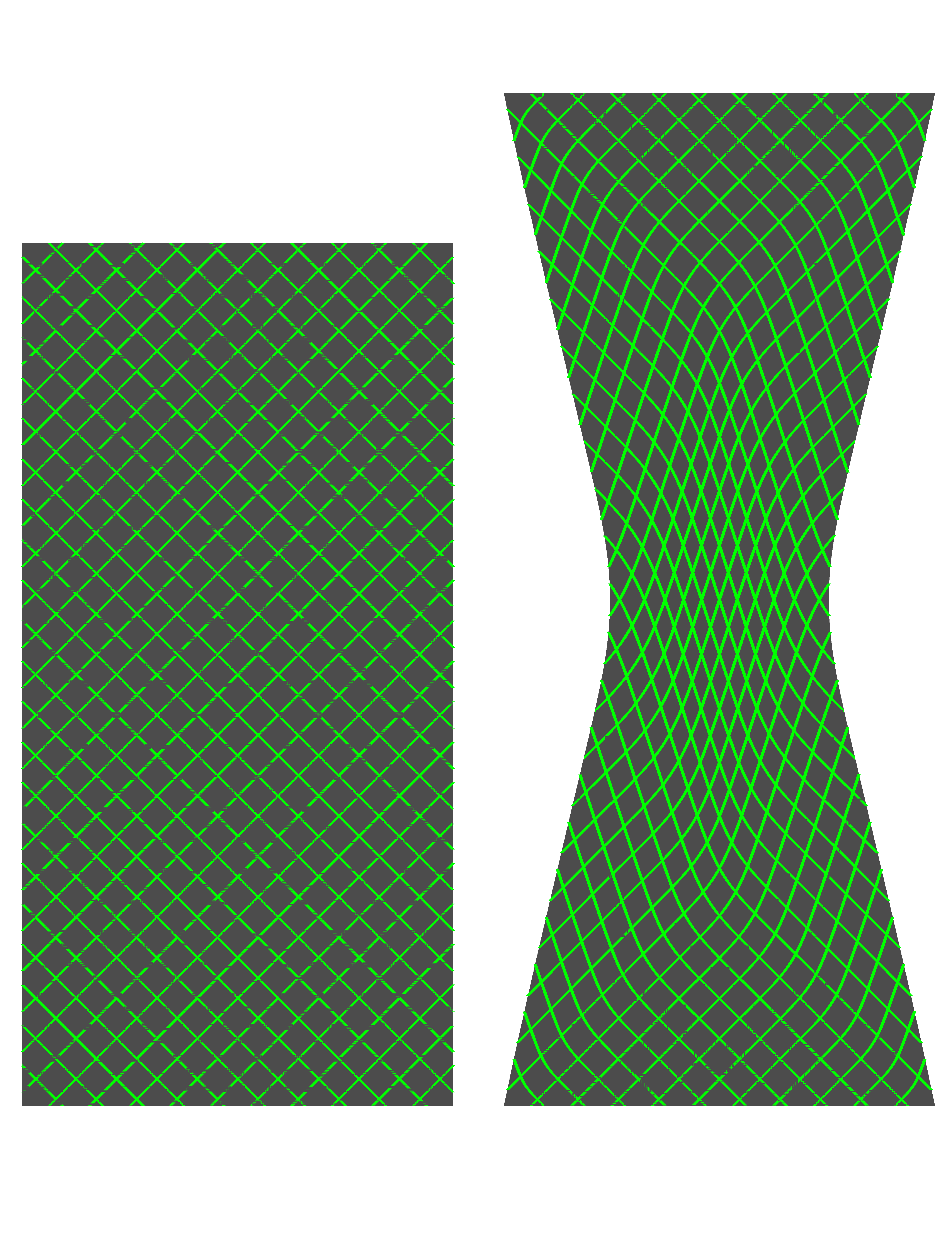}}
%
%
%
\put(-3.5,-0.2){\includegraphics[height=0.48\textwidth]{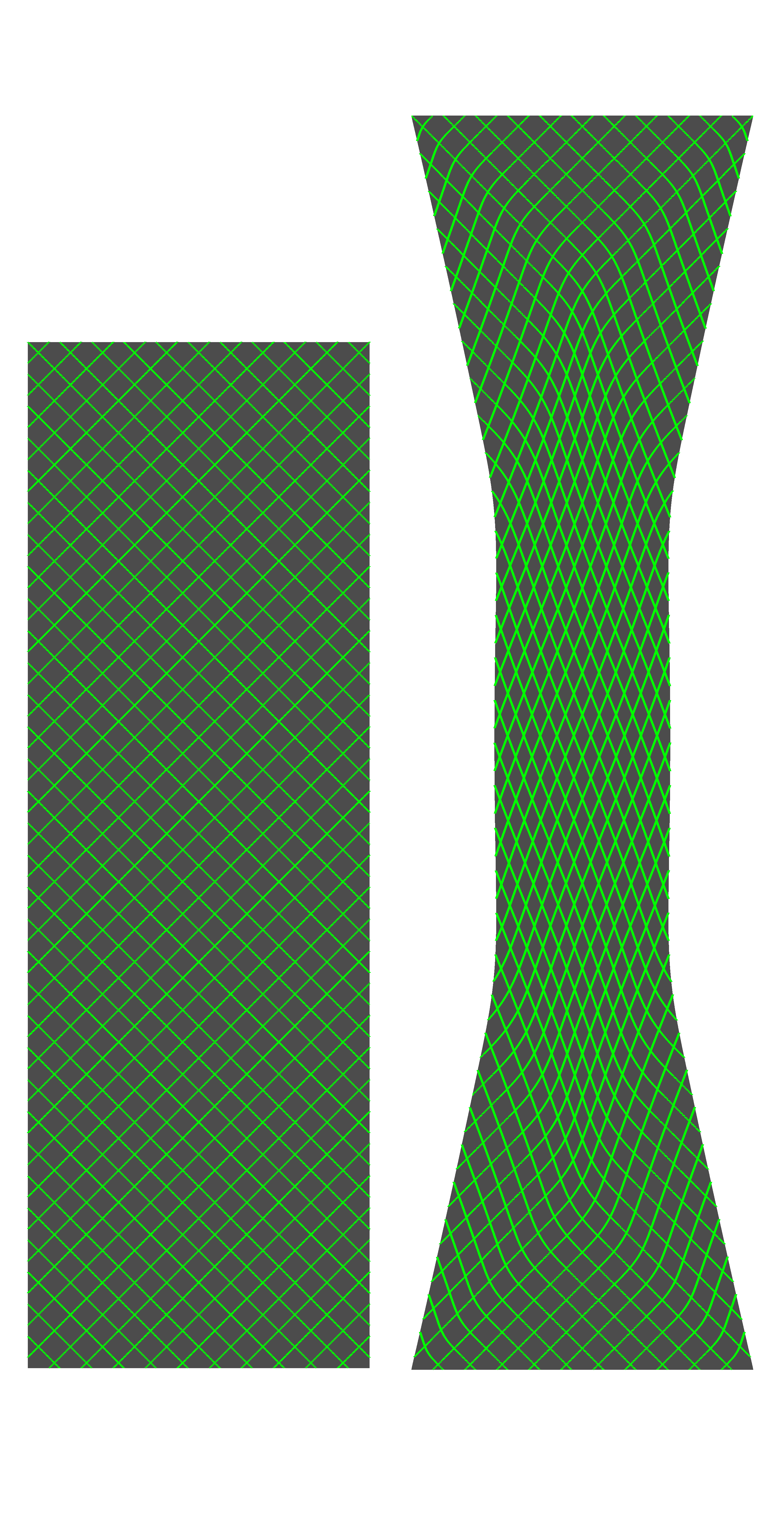}}
\put(0.6,0){\includegraphics[height=0.37\textwidth]{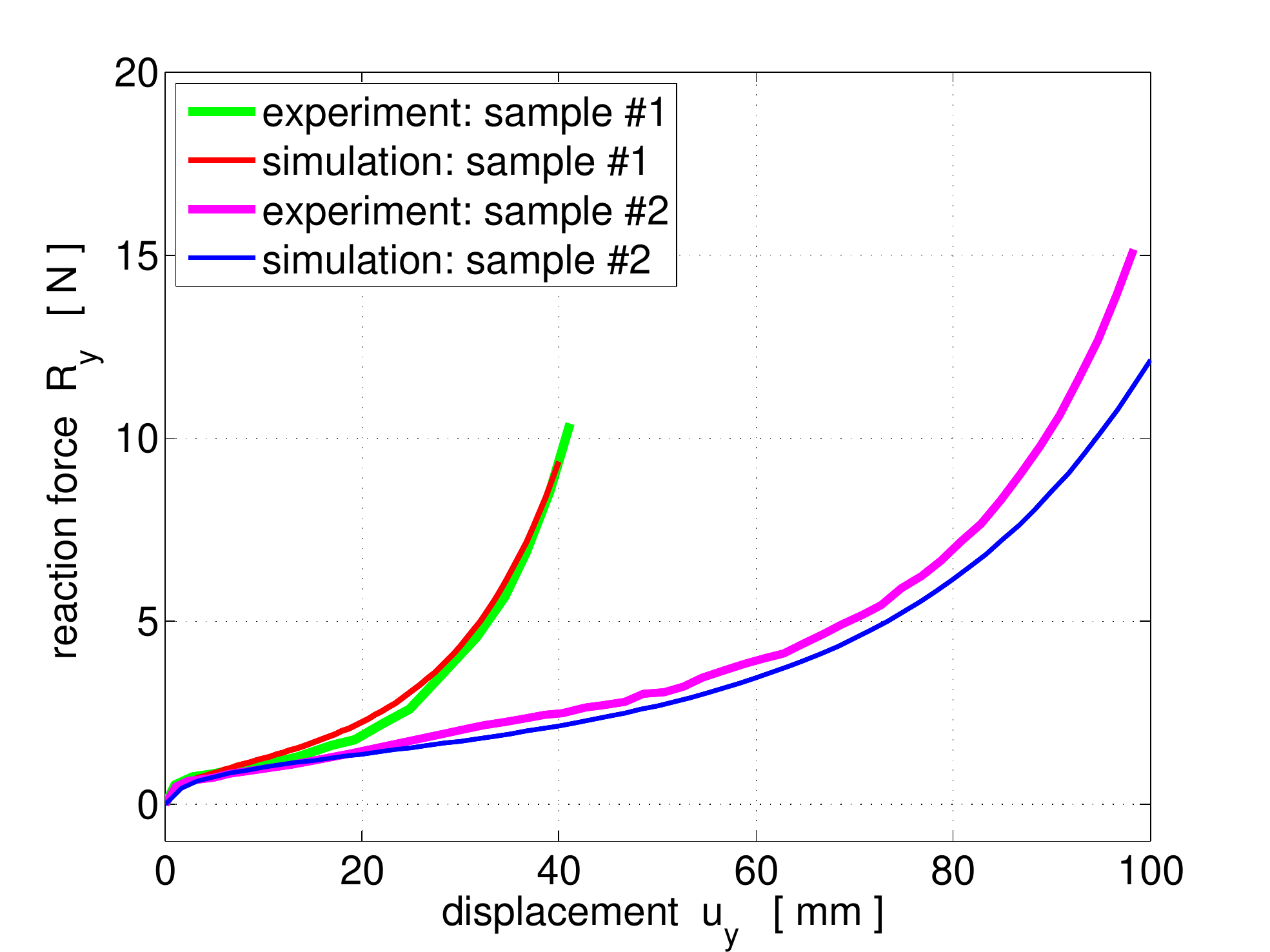}}
\put(-7.8,3.95){\tiny{$115\!\times\!230$mm$^2$}}
\put(-3.25,5.85){\tiny{$150\!\times\!450$mm$^2$}}

\put(-7.4,0.1){{\small{a.~Sample\,\#1}}}
\put(-2.7,0.1){{\small{b.~Sample\,\#2}}}
\put(1.0,0.1){{\small{c.}}}
\end{picture}
\caption[caption]{Bias extension test of plain weave fabrics: a-b.~{Initial and deformed} configurations of sample\,\#1 and \#2 at displacement $40$mm and $100$mm, respectively. The green lines show two fiber families. c.~Reaction-displacement curves  compared to the experimental data of Cao et al.~\cite{Cao2008}.}
\label{f:Bias_gamma_fitting}
\end{center}
\end{figure}

We first fit material model   \eqref{e:eg_WFs}--\eqref{e:eg_WFs2} to the experimental data of Cao et al.~\cite{Cao2008} for the bias extension test of balanced weave fabrics. In these fabrics, all fiber families are characterized by identical material properties.  Two initially rectangular samples, $\#1$ and $\#2$,  with dimension $115$mm$\times230$mm and $150$mm$\times 450$mm, respectively, are used for the test, {see Fig.~\ref{f:Bias_gamma_fitting}}. Two fiber families, initially  aligned by $\pm45^{\circ}$ w.r.t.~the edges, are considered. The two samples are discretized by  $16\times 32$ and $16\times48$ quadratic NURBS elements, respectively. The samples are stretched in the longer direction by applying Dirichlet boundary conditions on the shorter edges. 

Figs.~\ref{f:Bias_gamma_fitting}a-b show, that the two samples exhibit symmetric deformation in the bias extension test. Here, the material parameters of model \eqref{e:eg_WFs2} are obtained by fitting the load-displacement curve of sample\,$\#1$ to the  experimental data of Cao et al.~\cite{Cao2008}. The fitted curve is plotted in Fig.~\ref{f:Bias_gamma_fitting}c and the obtained parameters are listed in Tab.~\ref{t:WFconstant}. The model is then validated by comparing the corresponding experimental results to the simulation for sample $\#2$ as shown in Fig.~\ref{f:Bias_gamma_fitting}c. As seen, the proposed model demonstrates good prediction at small and medium deformations but deviates from the experimental data at larger strains. 


\begin{remark}
Note that the fit is based on a purely hyperelastic material model and hence does not capture plasticity. As plastic deformations usually play an important role in woven fabrics, in the form of fiber-fiber sliding, the presented model is only of limited use: It can help to understand the loading behavior, but it will not capture the unloading response correctly. {To this end,} a plasticity model should be included, which lies outside the  scope of the present work.
\end{remark}

\subsubsection{Bias extension of balanced weave fabrics: the role of in-plane bending}

Next, we investigate the influence of the in-plane bending stiffness $\beta_\mrg$ on the deformation, the load-displacement curve, and the finite element convergence behavior, using the bias extension test for sample\,\#1. 
We use the material parameters from Tab.~\ref{t:WFconstant}, but vary the in-plane bending stiffness $\beta_\mrg$. In order to quantify the shear bands in the specimen, we examine the sum of the geodesic curvatures for the two fiber families, as $|\kappa_\mrg^1|+|\kappa_\mrg^2|$.

\begin{figure}[H]
\begin{center} \unitlength1cm
\begin{picture}(0,5.9)
\put(-8.1,0){\includegraphics[width=0.50\textwidth]{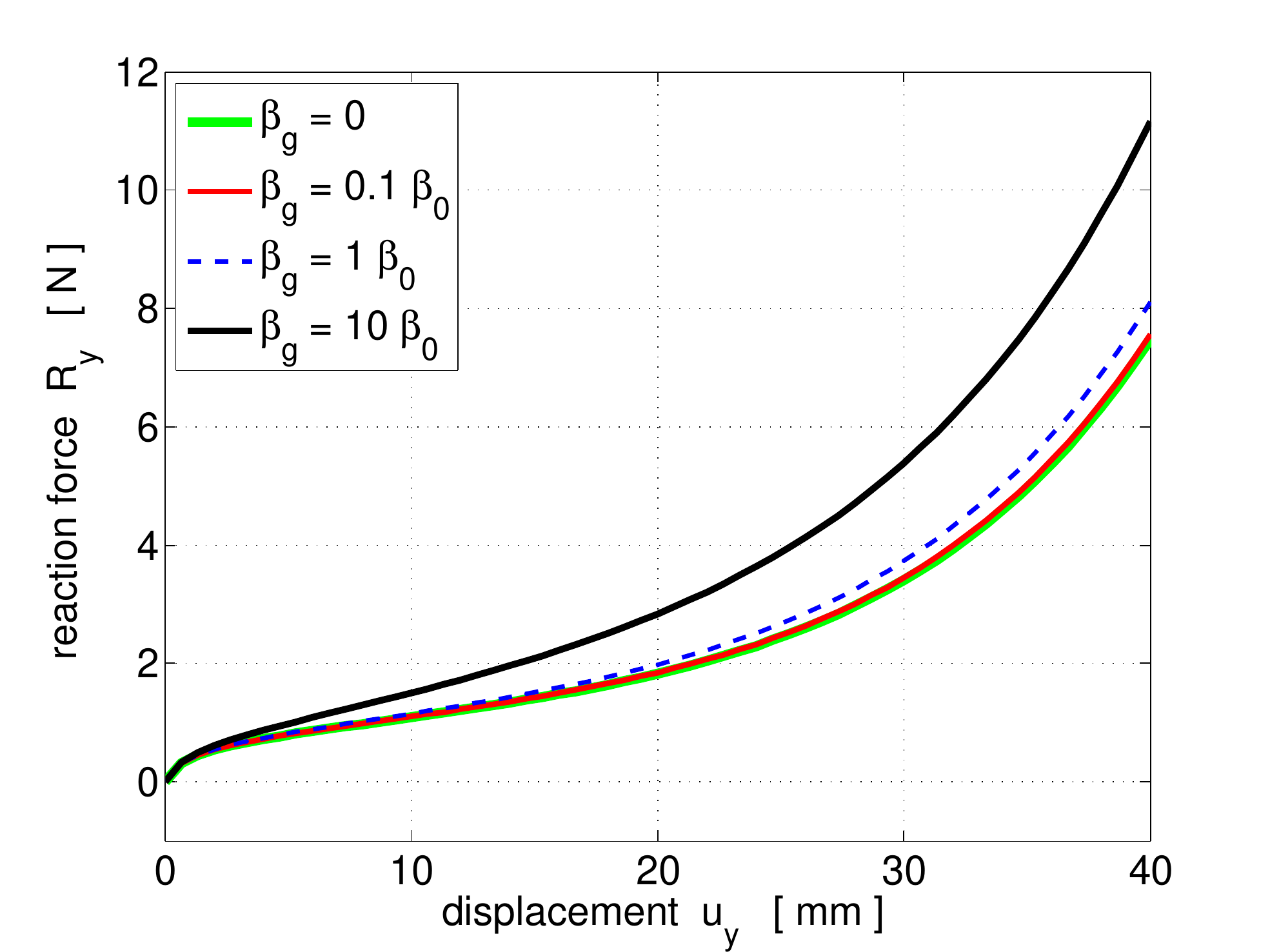}}
\put(0.2,0){\includegraphics[width=0.50\textwidth]{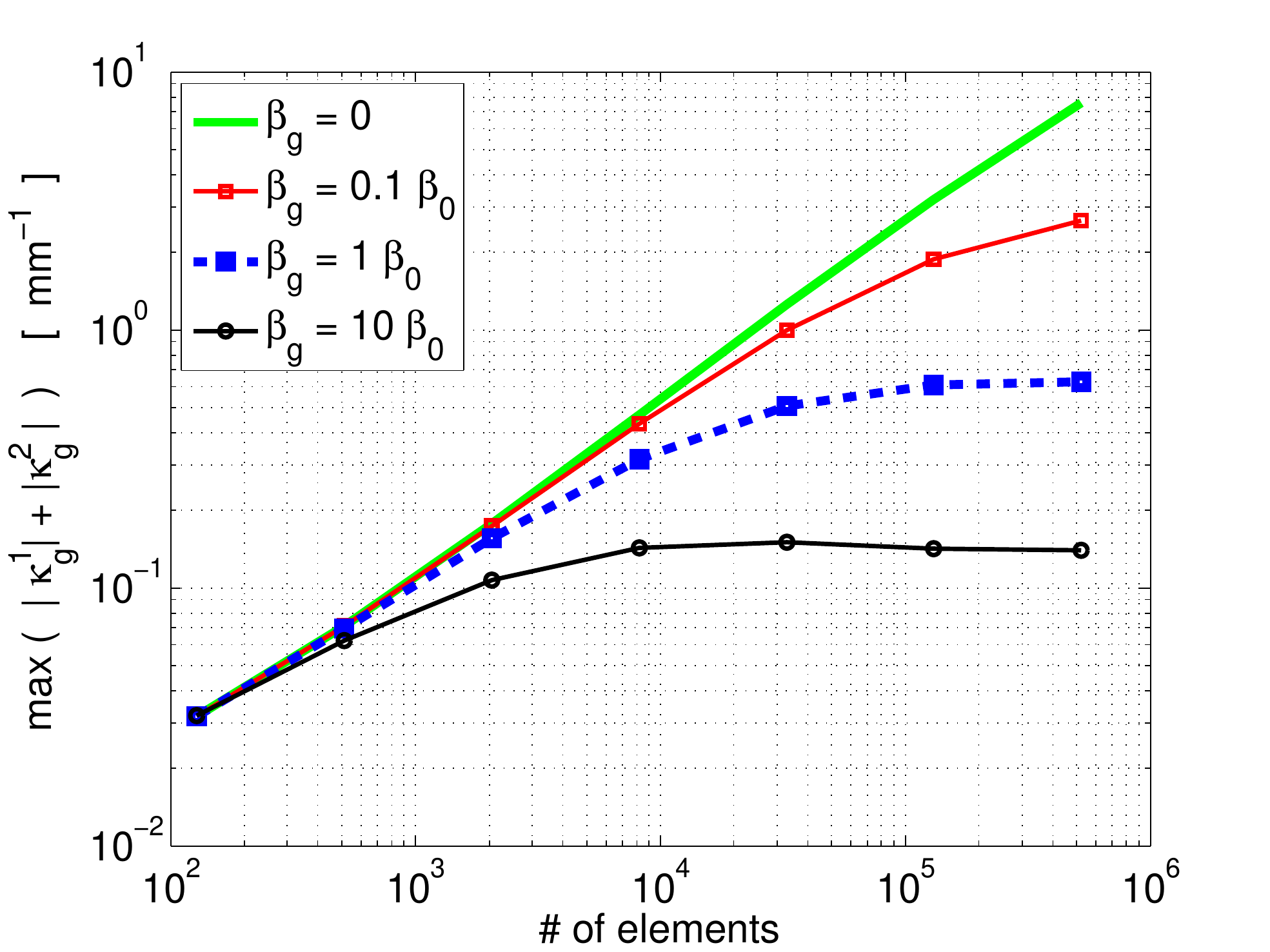}}

\put(-7.8,0){{\small{a.}}}
\put(0.4,0){{\small{b.}}}

\end{picture}
\caption[caption]{Bias extension of balanced weave fabric sample  \#1: a.~Load-displacement curves for various in-plane bending stiffnesses $\beta_\mrg$ using $32\times64$ quadratic NURBS elements.
b.~Convergence of the shear bands measured by max$(|\kappa_\mrg^1|+|\kappa_\mrg^2|)$ (over the sample domain) vs. mesh refinement.  Here,
$\beta_0 = 1.6$Nmm.   }
\label{f:Bias_conv_varyBeta}
\end{center}
\vspace{-0.7cm}
\end{figure}

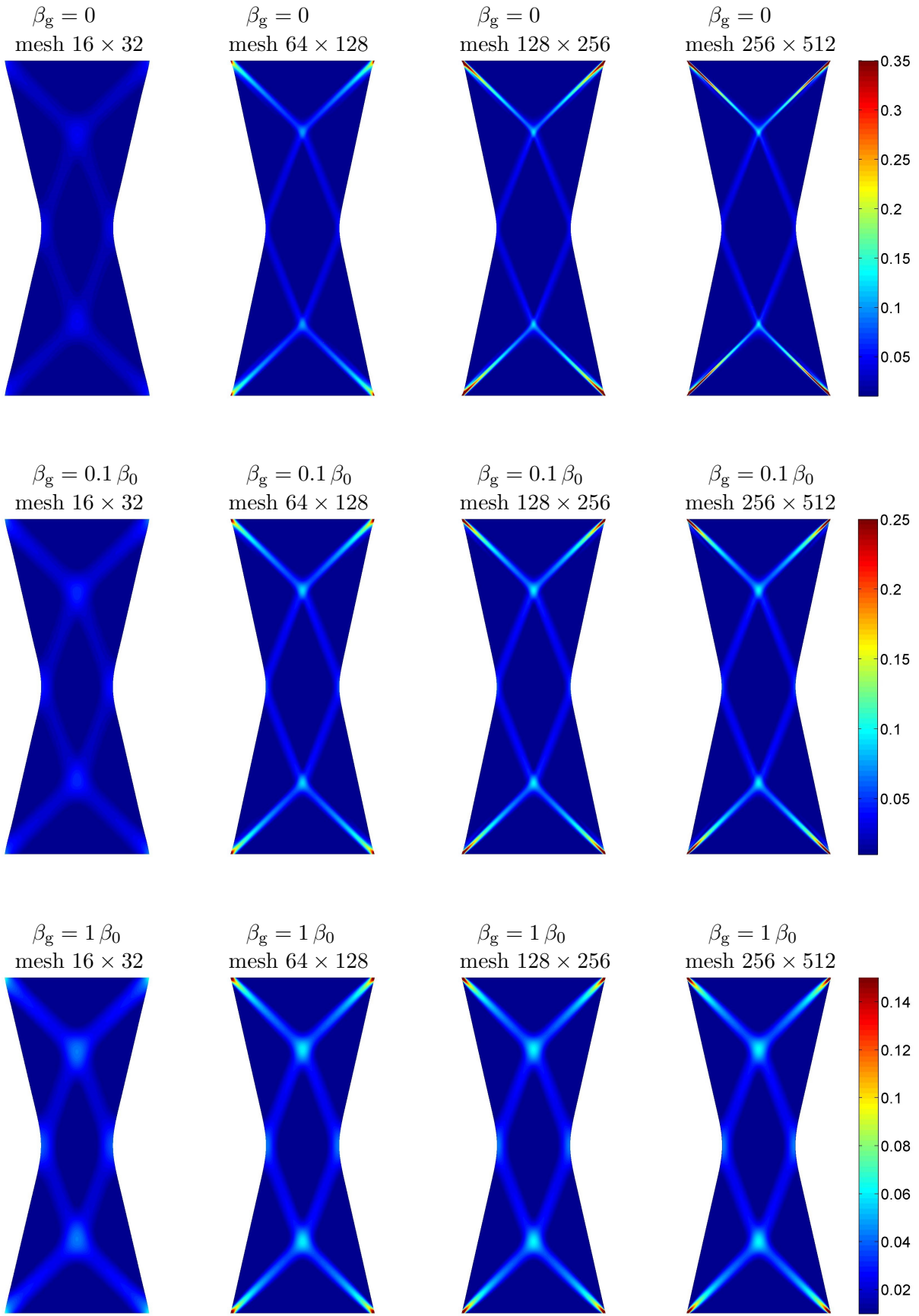
\begin{figure}[H]
\begin{center} \unitlength1cm
\begin{picture}(0,22.5)
\put(-8,15.5){\includegraphics[height=0.45\textwidth]{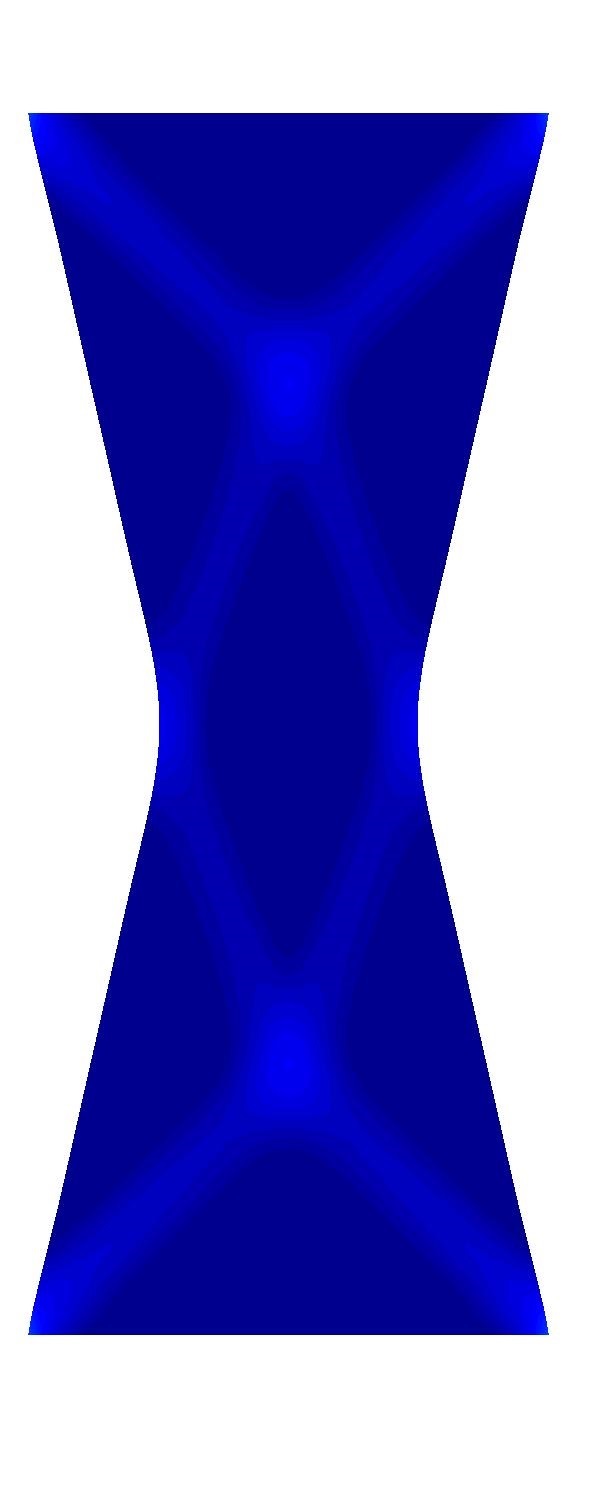}}
\put(-4.1,15.5){\includegraphics[height=0.45\textwidth]{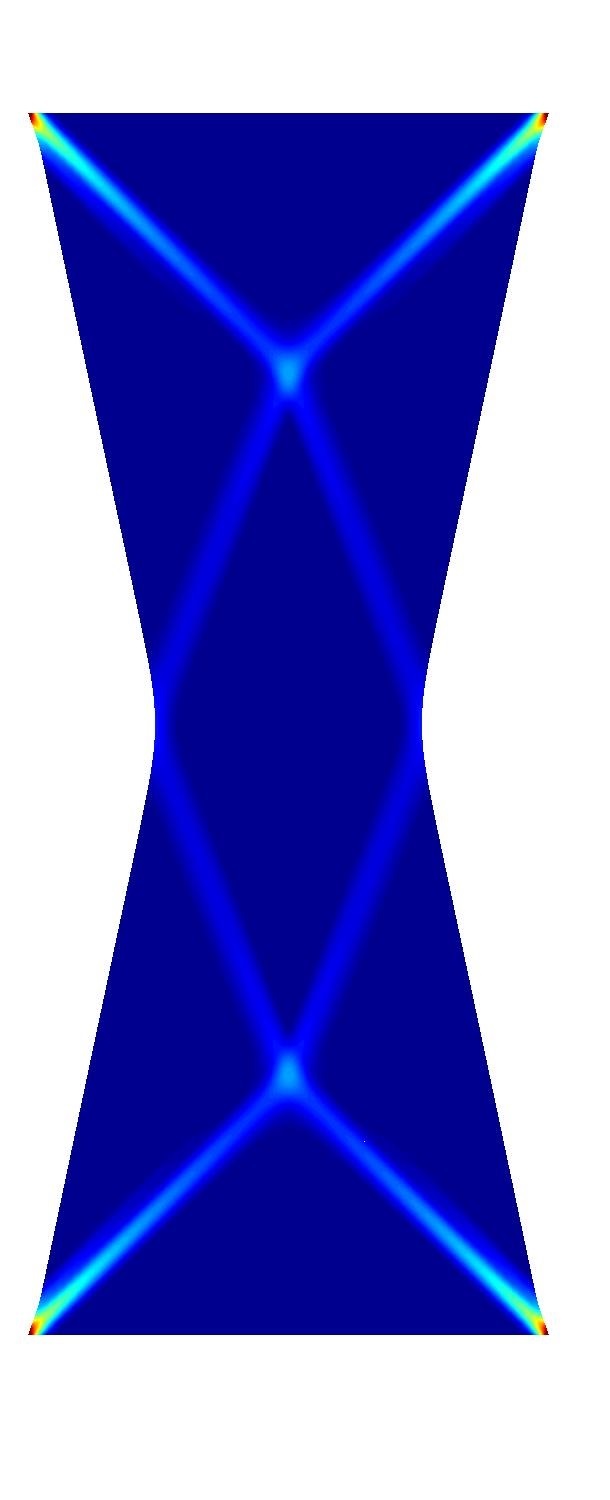}}
\put(-0.1,15.5){\includegraphics[height=0.45\textwidth]{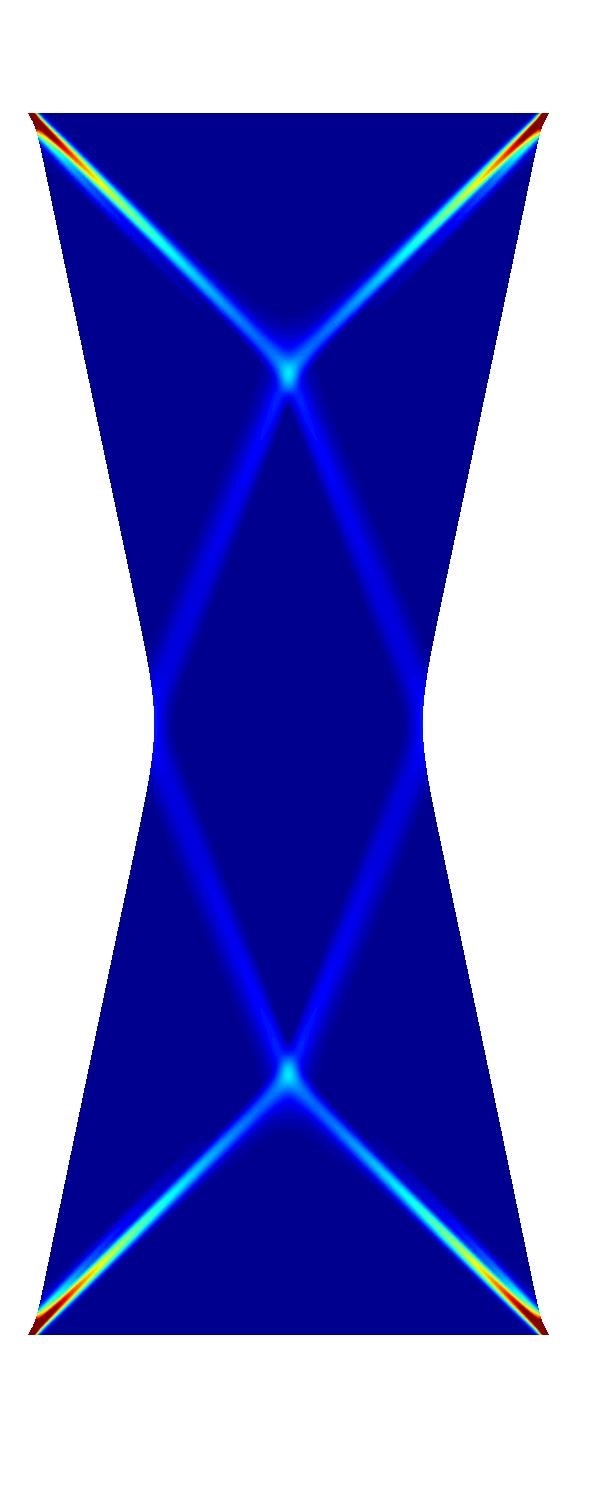}}
\put(3.8,15.5){\includegraphics[height=0.45\textwidth]{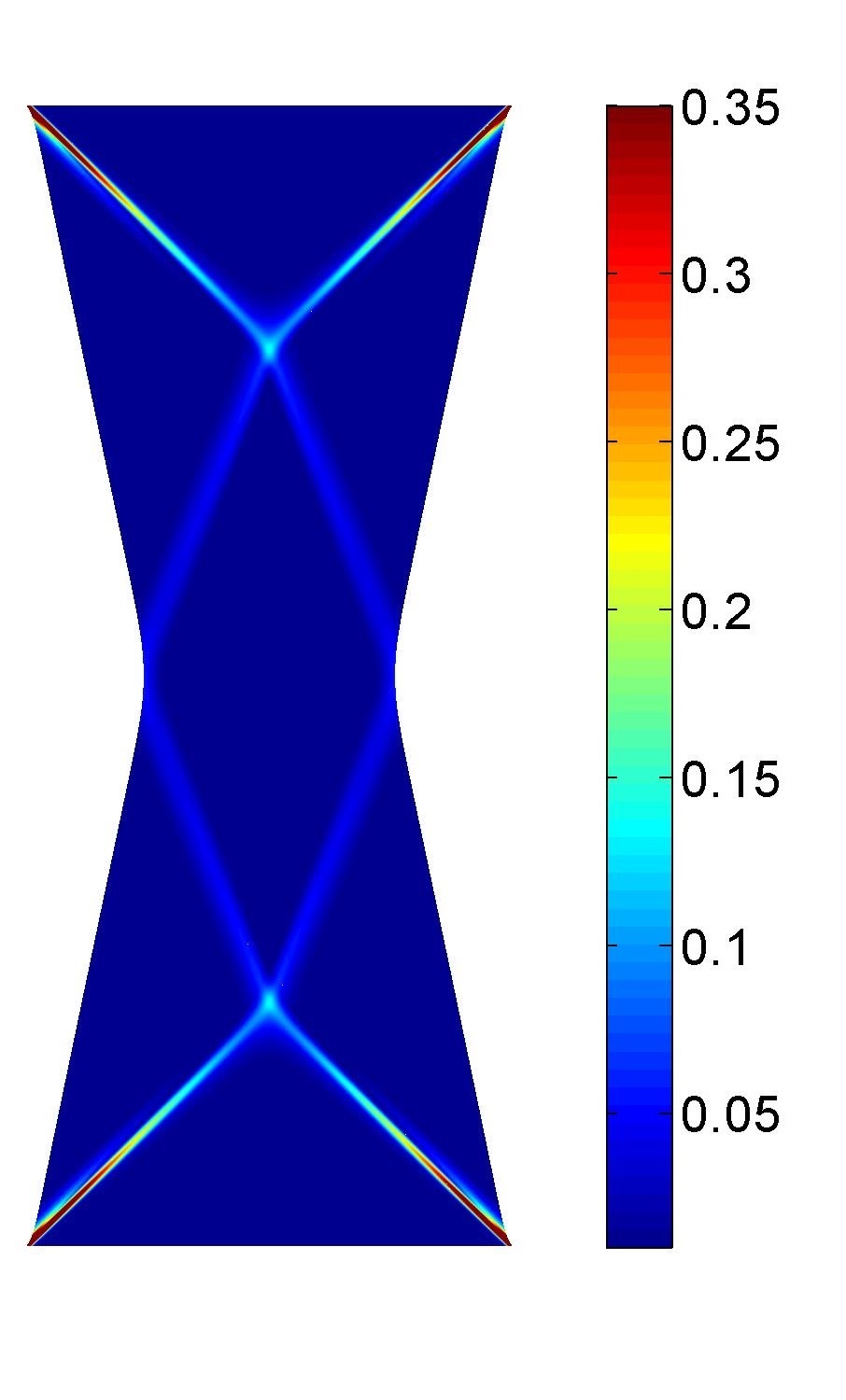}}

\put(-7.4, 22.8){$\beta_\mrg = 0$}
\put(-7.7, 22.3){mesh $16\times 32$}

\put(-3.7, 22.8){$\beta_\mrg = 0$}
\put(-4.0, 22.3){mesh $64\times 128$}

\put(0.3, 22.8){$\beta_\mrg = 0$}
\put(0.0, 22.3){mesh $128\times 256$}

\put(4.3, 22.8){$\beta_\mrg = 0$}
\put(3.9, 22.3){mesh $256\times 512$}

\put(-8,7.5){\includegraphics[height=0.45\textwidth]{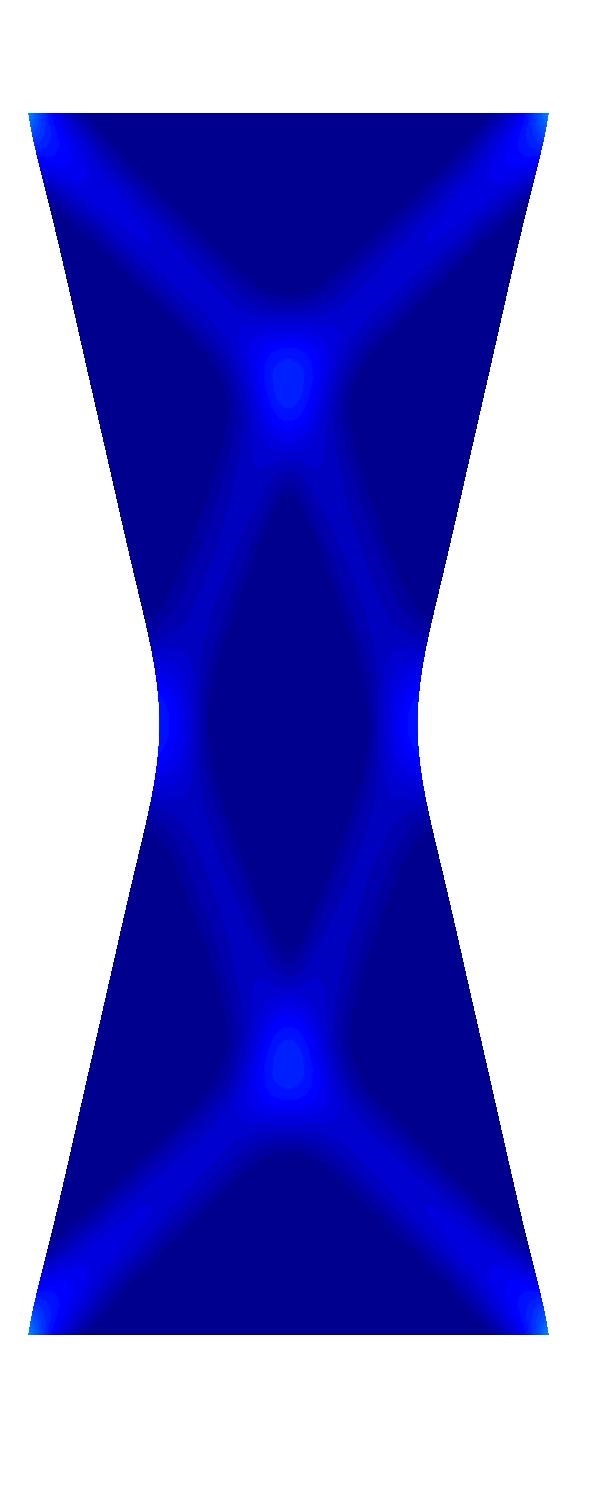}}
\put(-4.1,7.5){\includegraphics[height=0.45\textwidth]{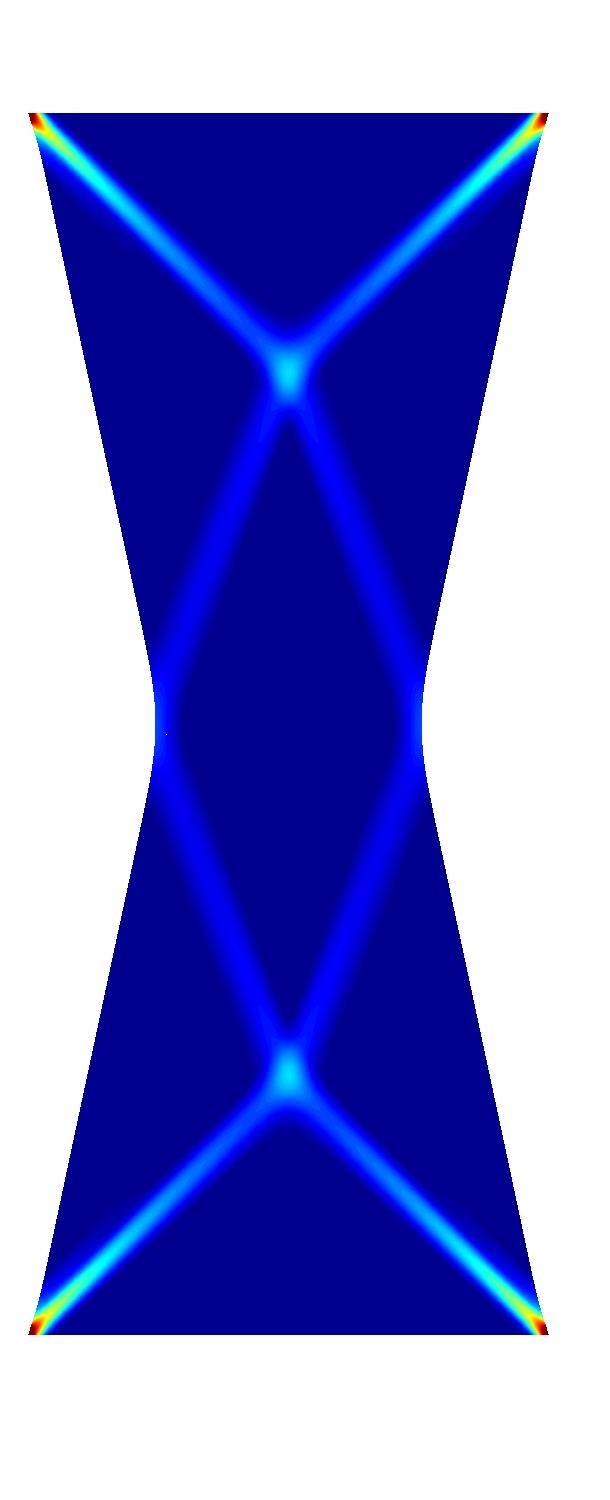}}
\put(-0.1,7.5){\includegraphics[height=0.45\textwidth]{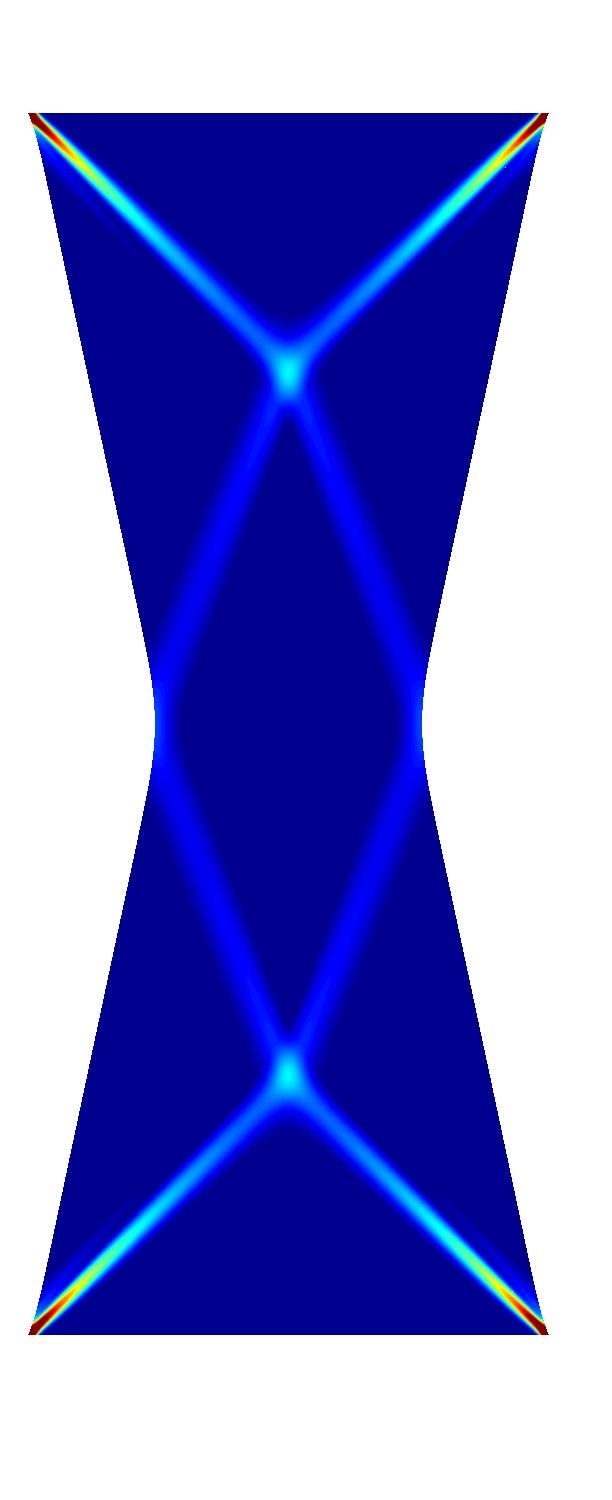}}
\put(3.8, 7.5){\includegraphics[height=0.45\textwidth]{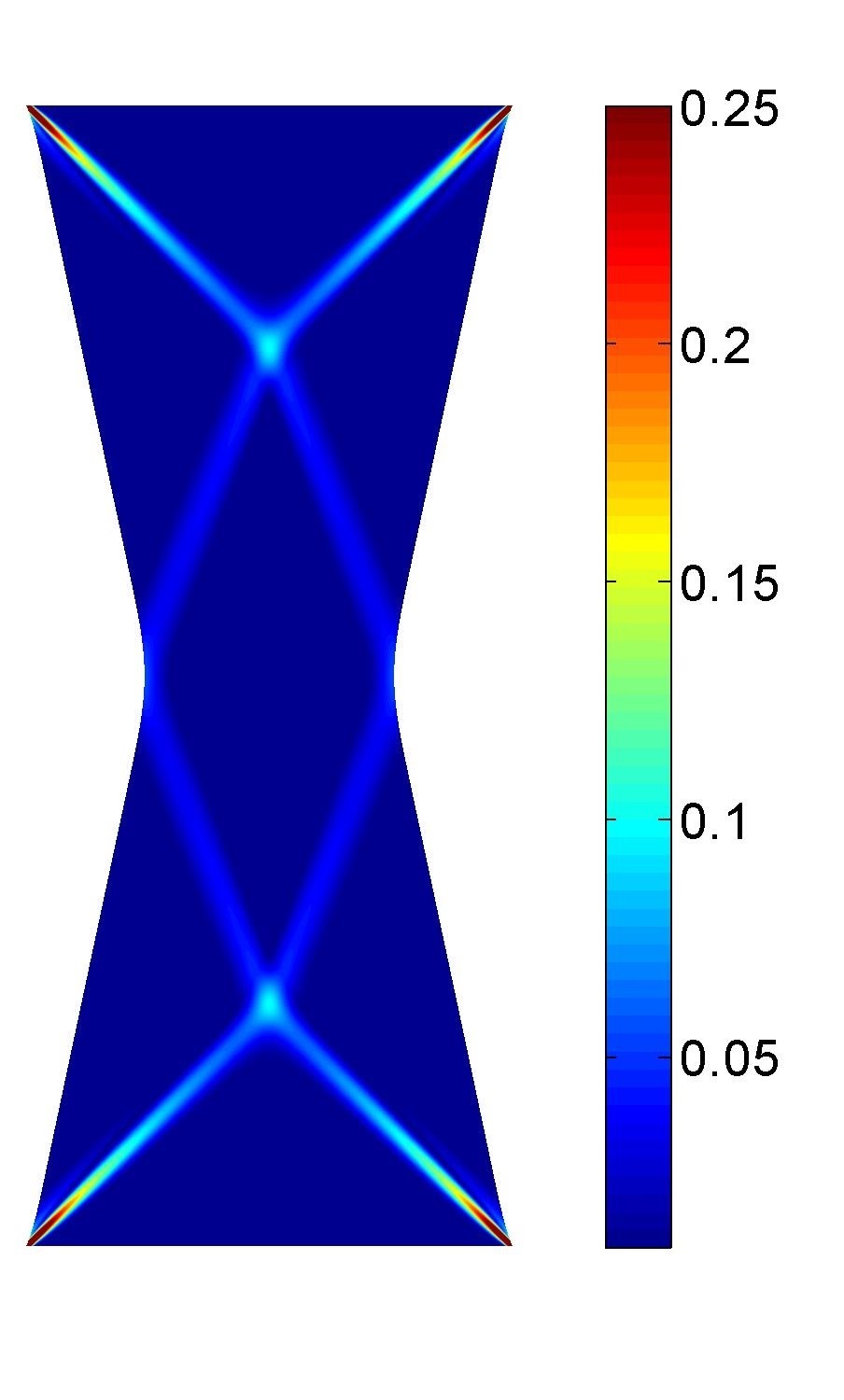}}

\put(-7.4, 14.8){$\beta_\mrg = 0.1\,\beta_0$}
\put(-7.7, 14.3){mesh $16\times 32$}

\put(-3.7, 14.8){$\beta_\mrg = 0.1\,\beta_0$}
\put(-4.0, 14.3){mesh $64\times 128$}

\put(0.3, 14.8){$\beta_\mrg = 0.1\,\beta_0$}
\put(0.0, 14.3){mesh $128\times 256$}

\put(4.3, 14.8){$\beta_\mrg = 0.1\,\beta_0$}
\put(3.9, 14.3){mesh $256\times 512$}

\put(-8,-0.5){\includegraphics[height=0.45\textwidth]{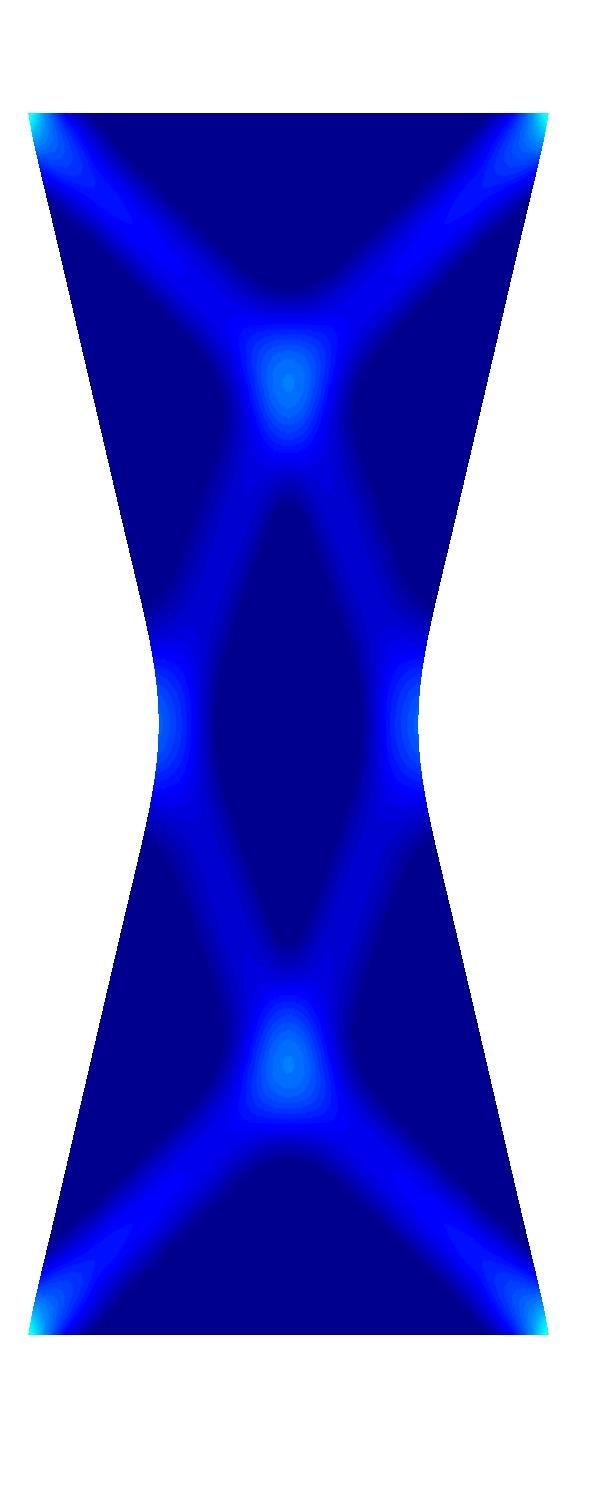}}
\put(-4.1,-0.5){\includegraphics[height=0.45\textwidth]{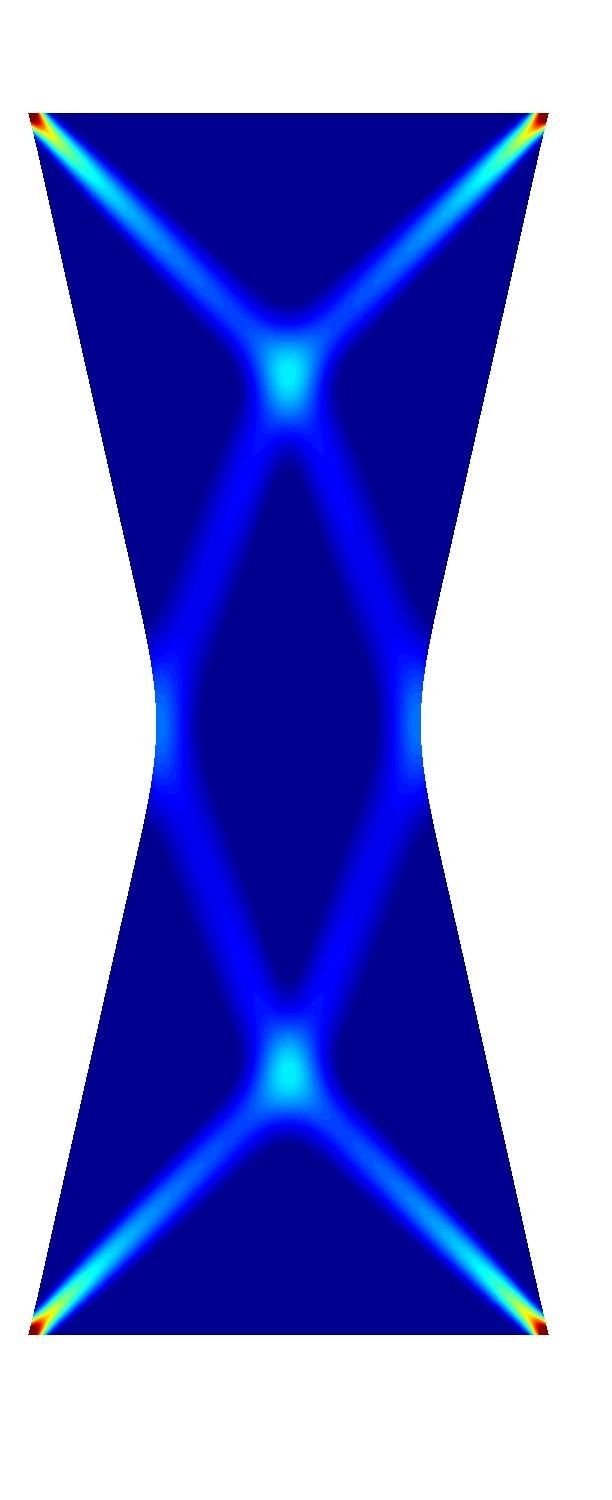}}
\put(-0.1,-0.5){\includegraphics[height=0.45\textwidth]{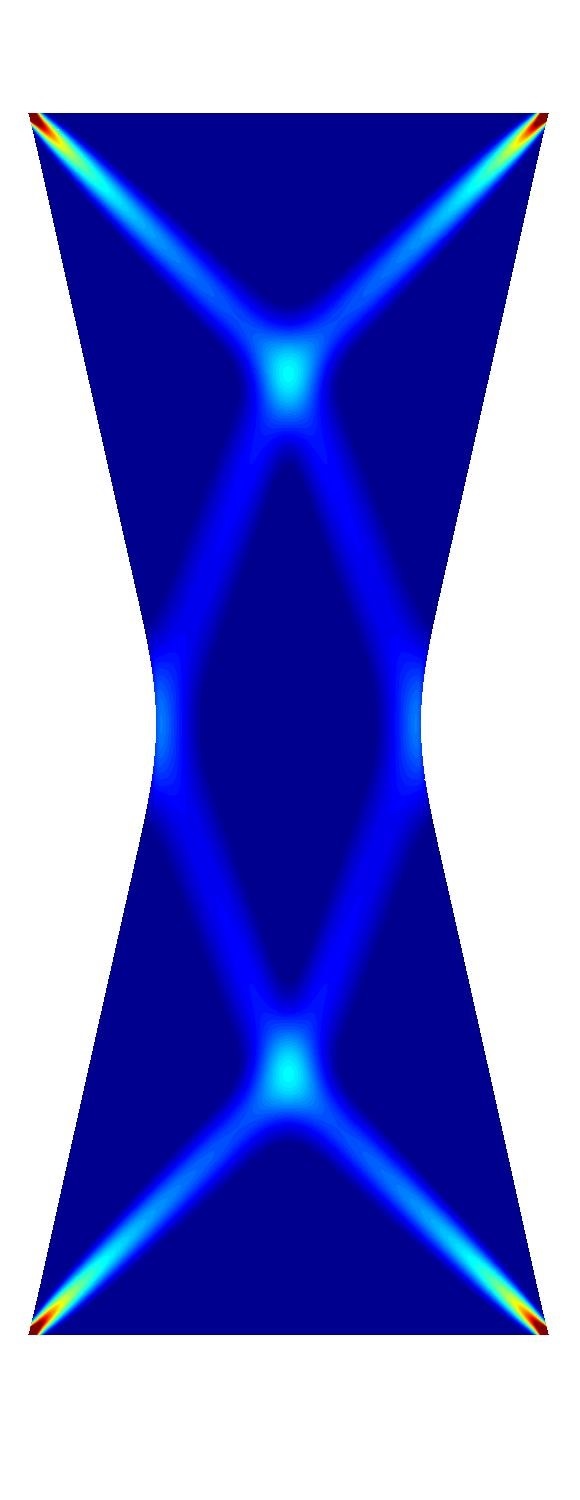}}
\put(3.8,-0.5){\includegraphics[height=0.45\textwidth]{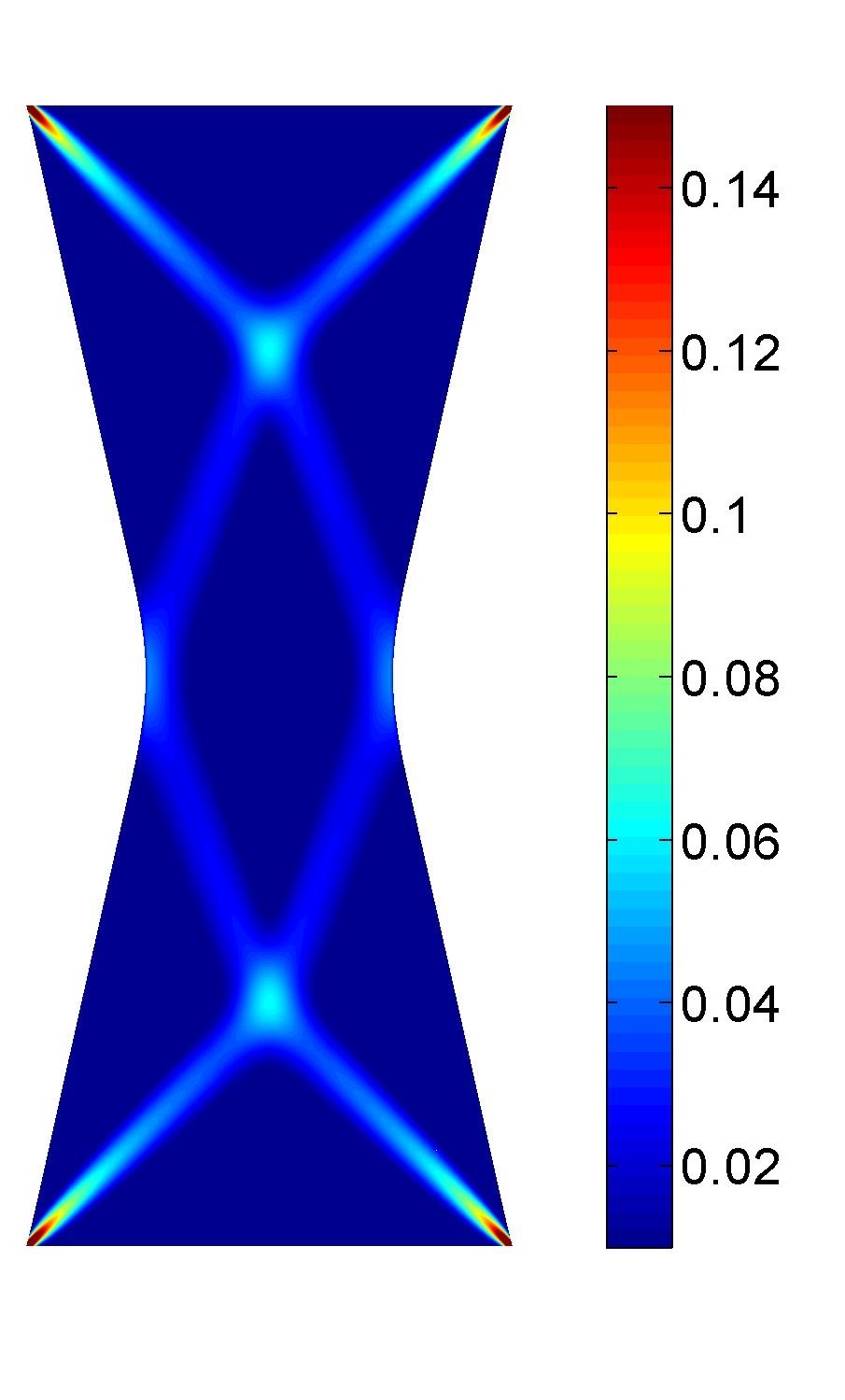}}
\put(-7.4, 6.8){$\beta_\mrg = 1\,\beta_0$}
\put(-7.7, 6.3){mesh $16\times 32$}

\put(-3.7, 6.8){$\beta_\mrg = 1\,\beta_0$}
\put(-4.0, 6.3){mesh $64\times 128$}

\put(0.3, 6.8){$\beta_\mrg = 1\,\beta_0$}
\put(0.0, 6.3){mesh $128\times 256$}

\put(4.3, 6.8){$\beta_\mrg = 1\,\beta_0$}
\put(3.9, 6.3){mesh $256\times 512$}

\end{picture}
\caption[caption]{Bias extension of  balanced weave fabric sample  \#1: $|\kappa_\mrg^1|+|\kappa_\mrg^2|$ (units  [mm$^{-1}$]), a measure of shear bands, for various in-plane bending stiffnesses $\beta_\mrg$ (from top to bottom), and for various FE meshes  (from left to right).~The shear bands only converge for non-zero $\beta_\mrg$.~Here, $\beta_0 = 1.6$Nmm. }
\label{f:Bias_conv_varyBeta_config}
\end{center}
\end{figure}

Fig.~\ref{f:Bias_conv_varyBeta}a shows the influence of $\beta_\mrg$ on the load-displacement curve. 
Accordingly, larger values of $\beta_\mrg$ lead to the significantly stiffer response.
Fig.~\ref{f:Bias_conv_varyBeta}b shows the FE mesh convergence behavior of the shear bands measured by max$(|\kappa_\mrg^1|+|\kappa_\mrg^2|)$. The quantity  $|\kappa_\mrg^1|+|\kappa_\mrg^2|$ is also shown in Fig.~\ref{f:Bias_conv_varyBeta_config} to visualize the shear bands for various in-plane bending stiffnesses. As expected, in case of zero in-plane bending stiffness ($\beta_\mrg=0$),  the shear bands do not converge to a finite width.\footnote{Without in-plane bending stifness,  the theoretical shear band width becomes zero, which is unphysical.} On the other hand, for $\beta_\mrg>0$,  the shear bands converge to a finite width as 
observed in experiments (see e.g.~\cite{Boisse17}). The shear band width depends on the magnitude of the bending stiffness.  Fig.~\ref{f:Bias_conv_varyBeta_config} also shows that $\beta_\mrg$ visibly affects the width of the shear bands: they increase with $\beta_\mrg$. 

\begin{figure}[H]
\begin{center} \unitlength1cm

%

\begin{picture}(0,21.2)
%
\put(-8,14.5){\includegraphics[height=0.45\textwidth]{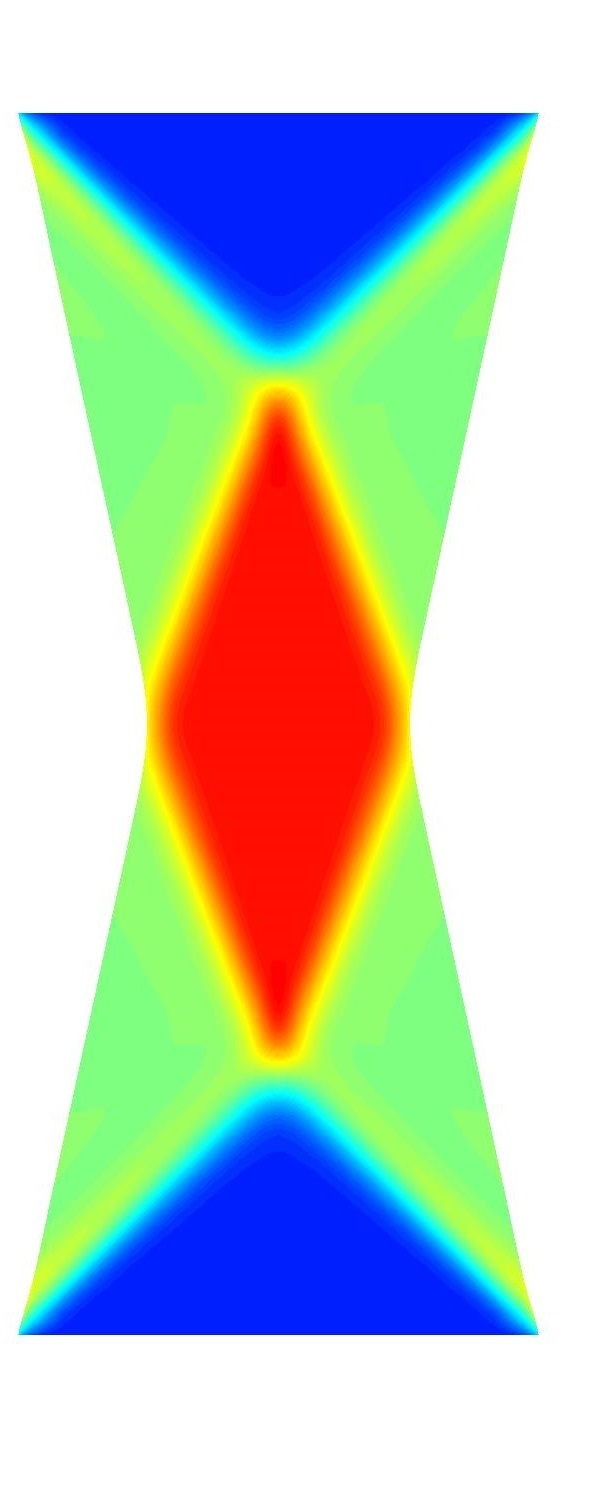}}
\put(-4.1,14.5){\includegraphics[height=0.45\textwidth]{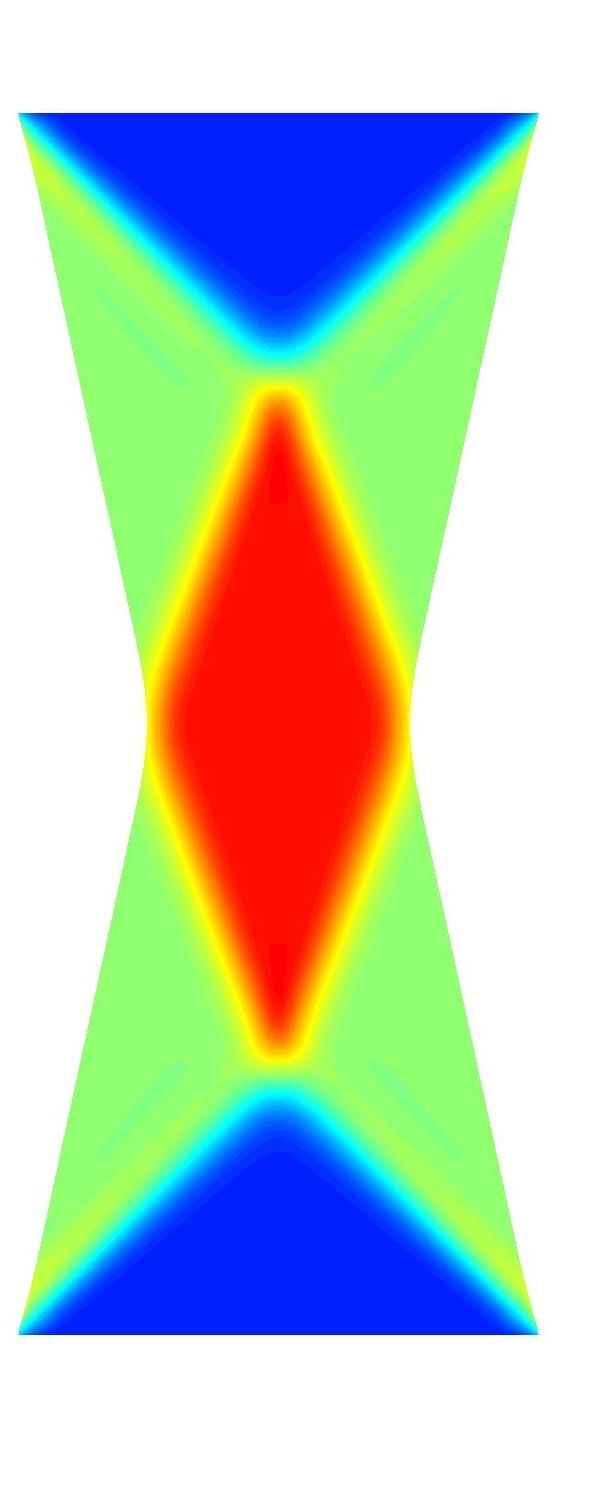}}
\put(-0.1,14.5){\includegraphics[height=0.45\textwidth]{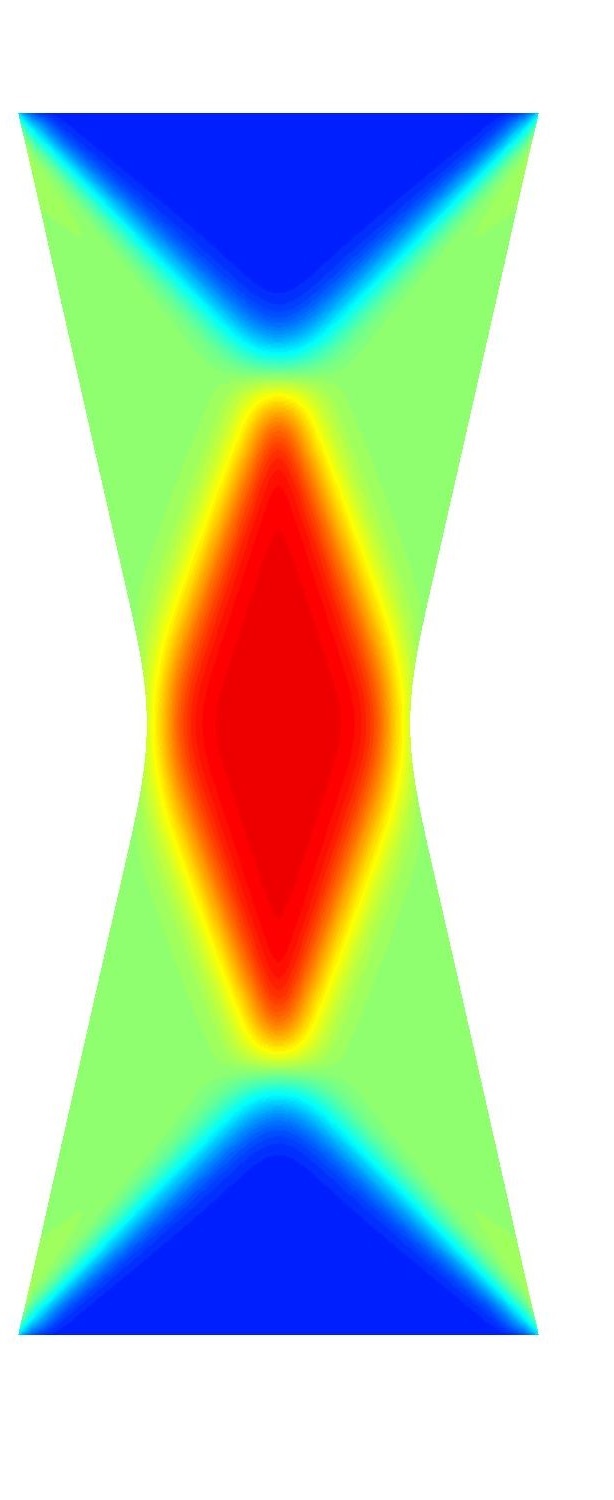}}
\put(3.8,14.5){\includegraphics[height=0.45\textwidth]{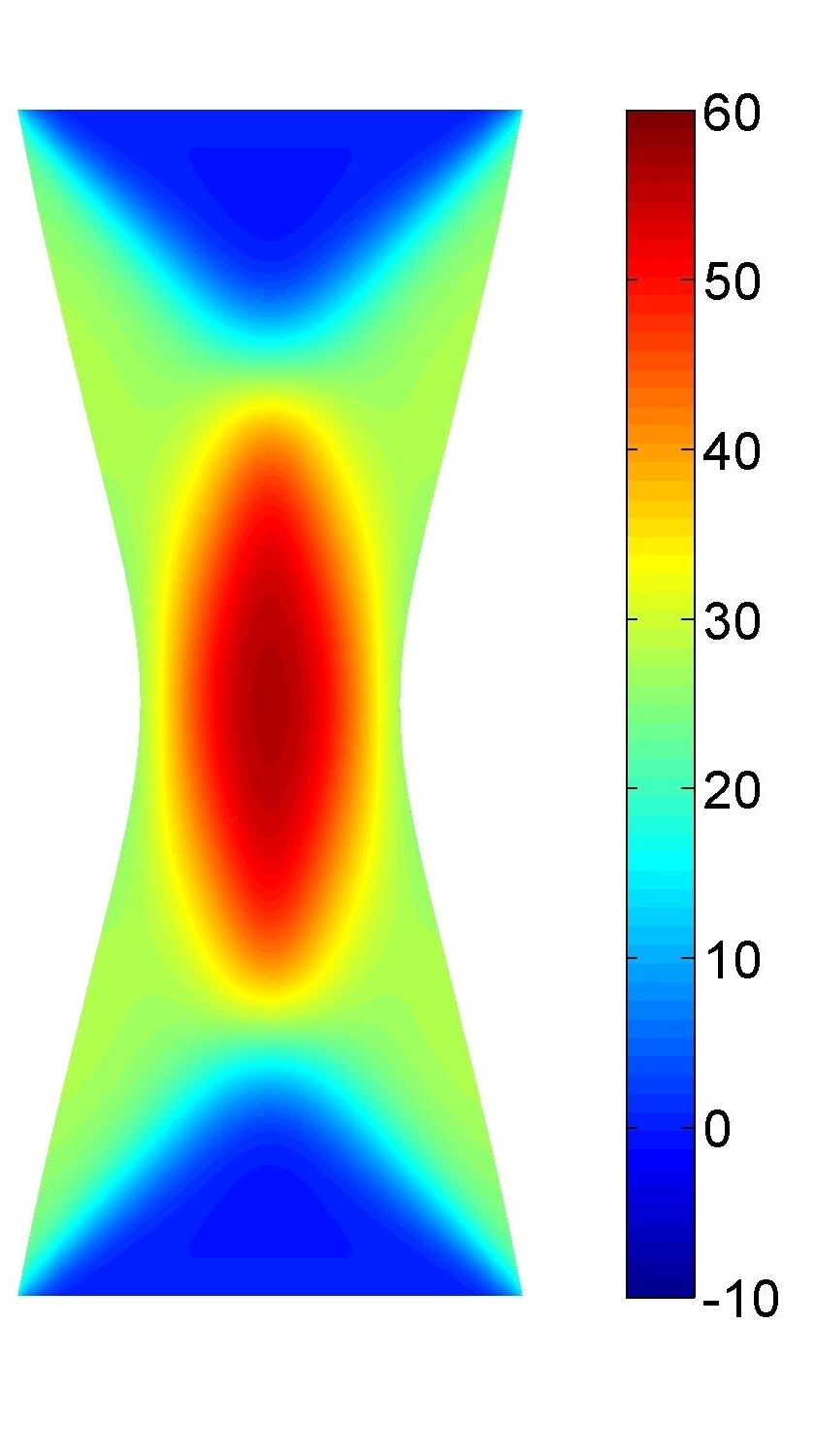}}

\put(-7.4, 21.3){$\beta_\mrg = 0$}

\put(-3.7, 21.3){$\beta_\mrg = 0.1\,\beta_0$}

\put(0.3, 21.3){$\beta_\mrg = 1\,\beta_0$}

\put(4.3, 21.3){$\beta_\mrg = 10\,\beta_0$}
\put(6.55, 21.4){ $\theta$ [deg.]}
\put(-8,7.0){\includegraphics[height=0.45\textwidth]{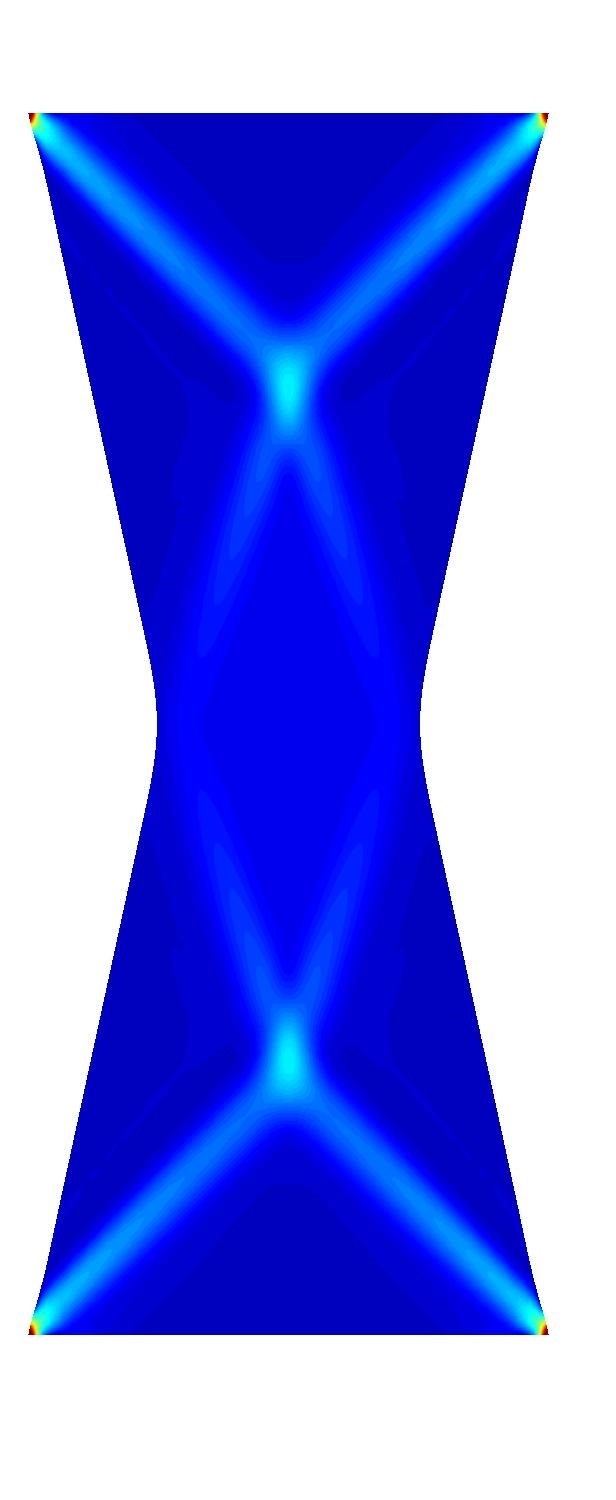}}
\put(-4.1,7.0){\includegraphics[height=0.45\textwidth]{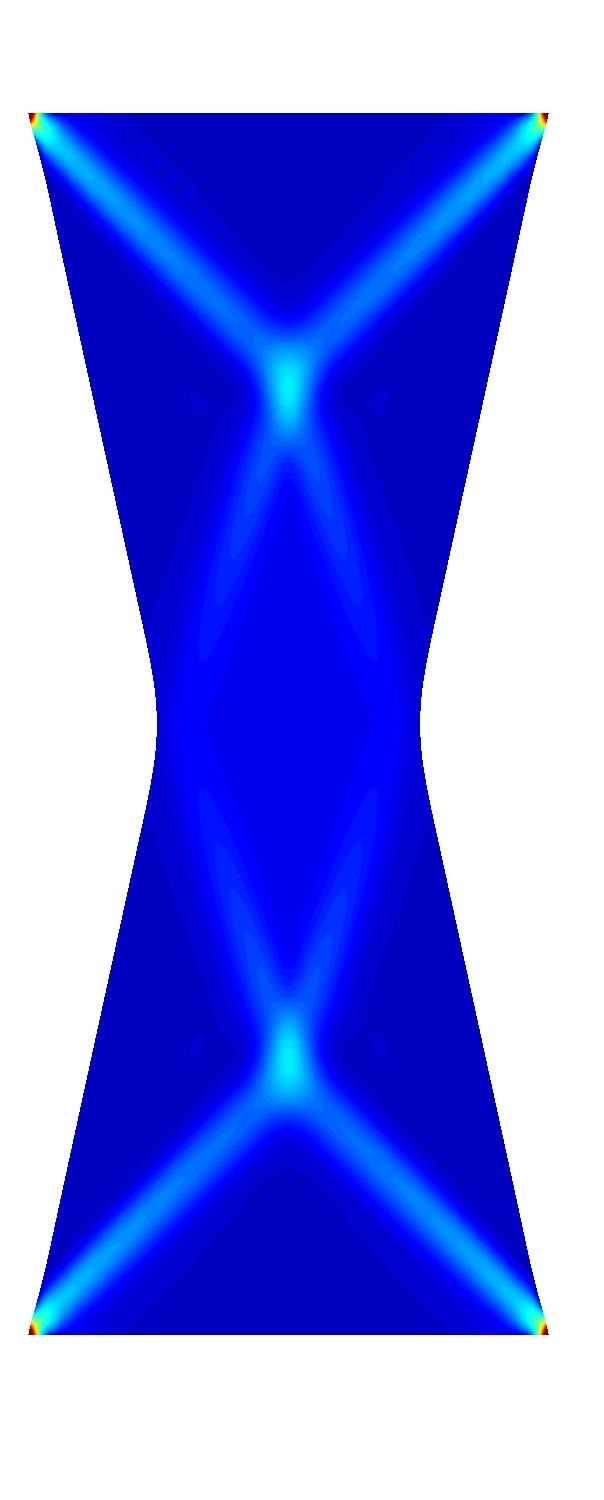}}
\put(-0.1,7.0){\includegraphics[height=0.45\textwidth]{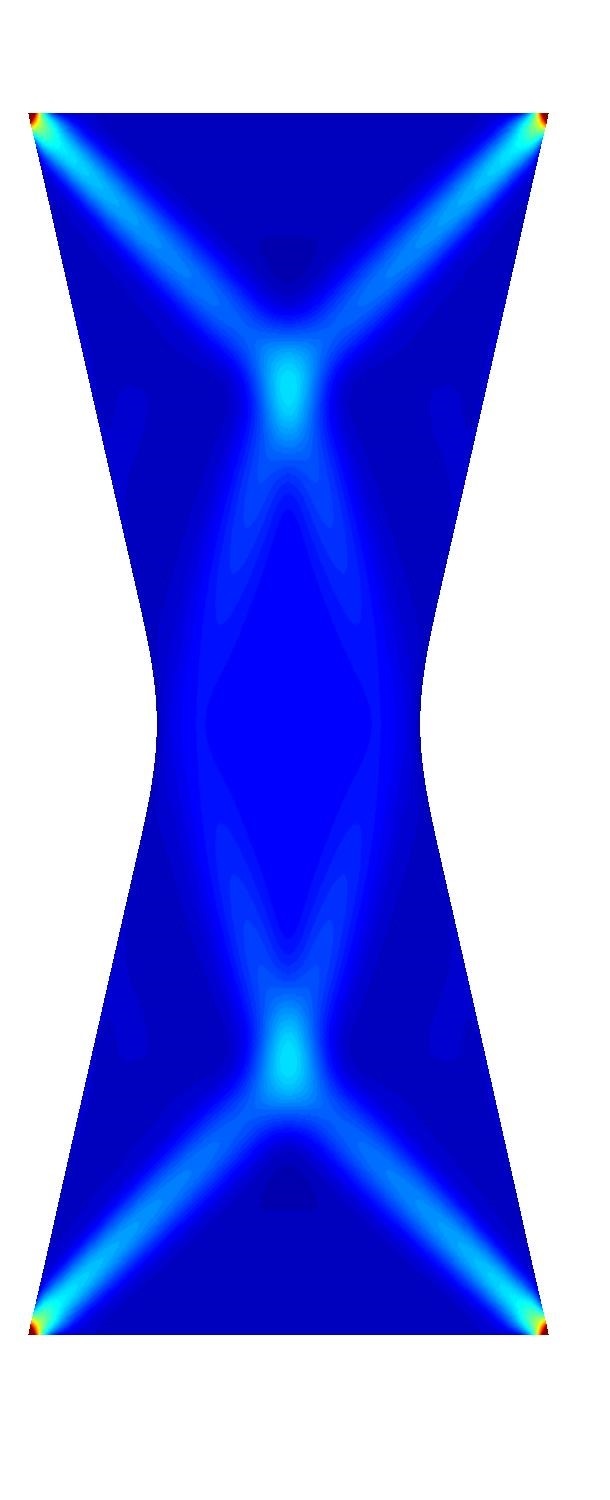}}
\put(3.8, 7.0){\includegraphics[height=0.45\textwidth]{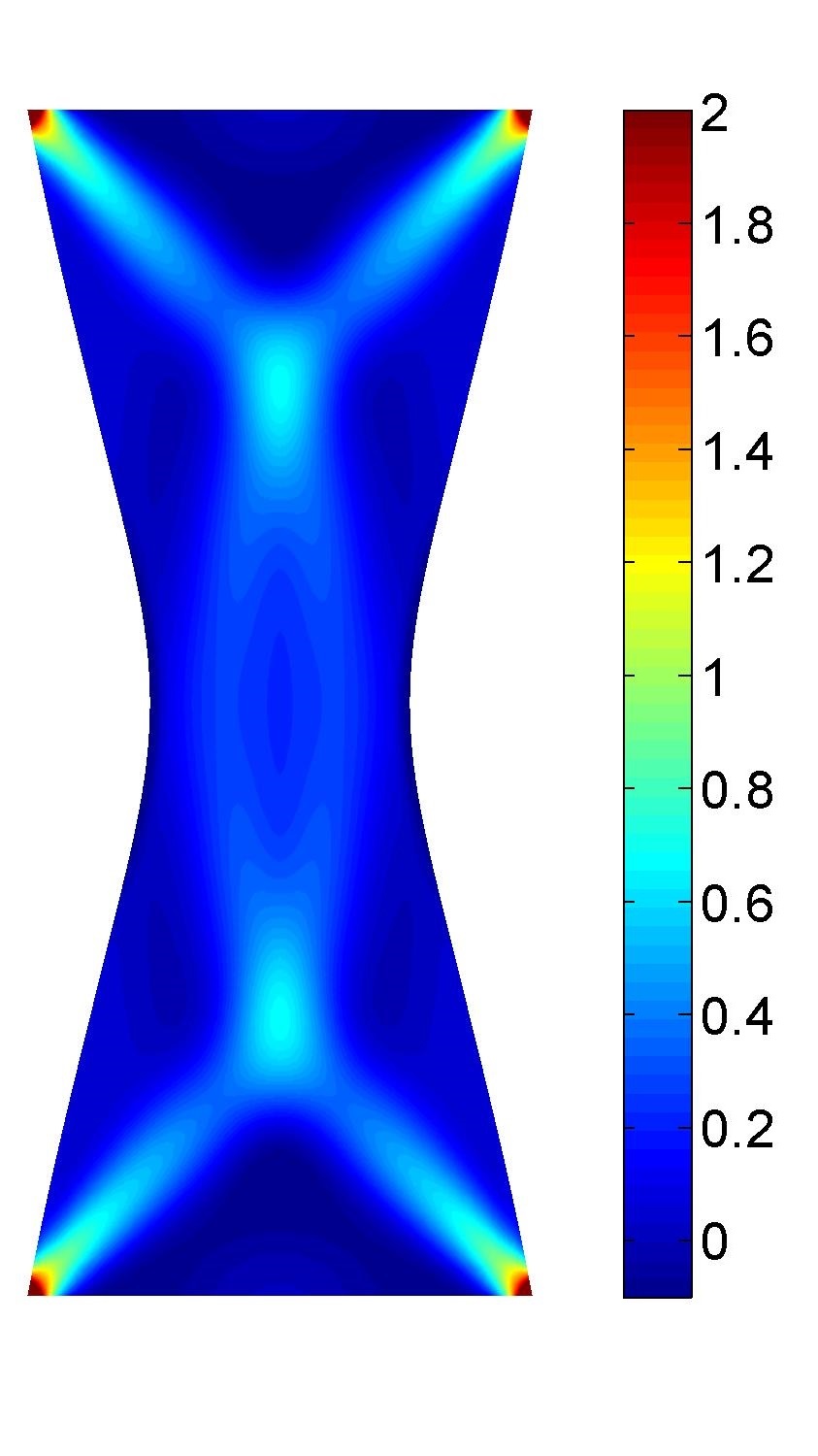}}

\put(-7.4, 13.8){$\beta_\mrg = 0$}

\put(-3.7, 13.8){$\beta_\mrg = 0.1\,\beta_0$}

\put(0.3, 13.8){$\beta_\mrg = 1\,\beta_0$}

\put(4.3, 13.8){$\beta_\mrg = 10\,\beta_0$}
\put(6.6, 13.8){ $\tr_{\!\mrs}\bsig$}

\put(-8,-0.5){\includegraphics[height=0.45\textwidth]{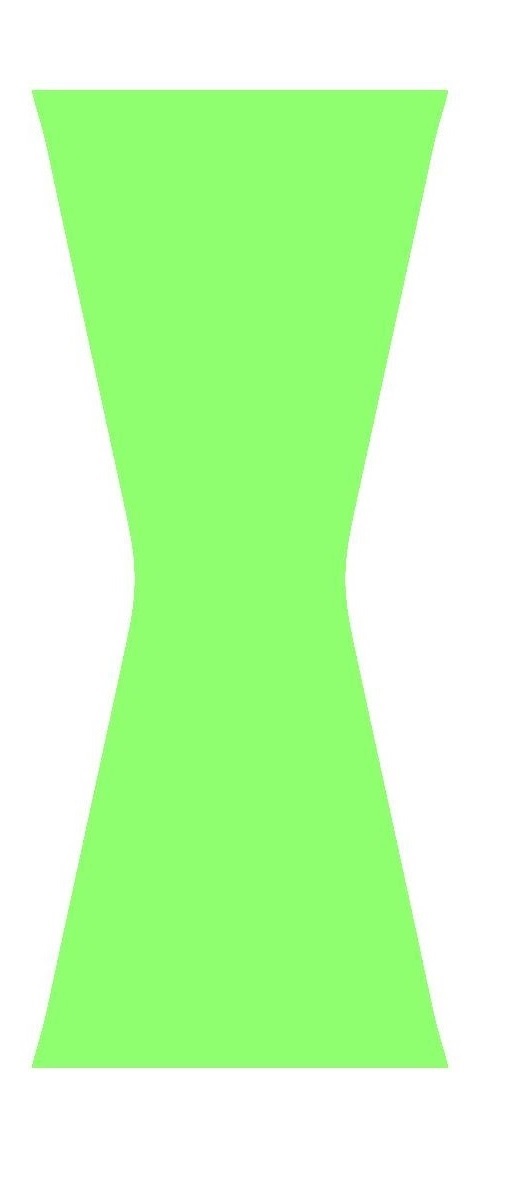}}
\put(-4.1,-0.5){\includegraphics[height=0.45\textwidth]{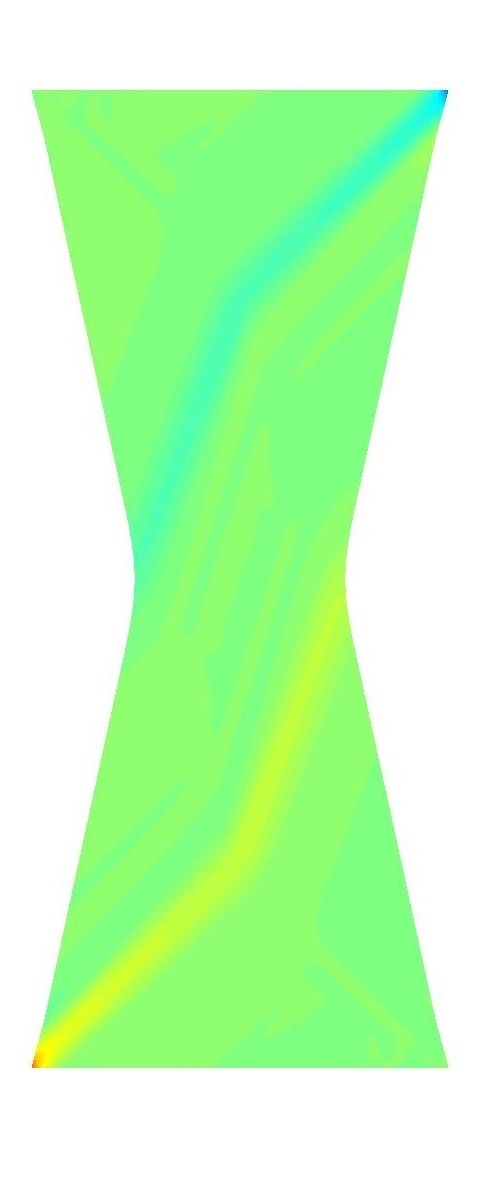}}
\put(-0.1,-0.5){\includegraphics[height=0.45\textwidth]{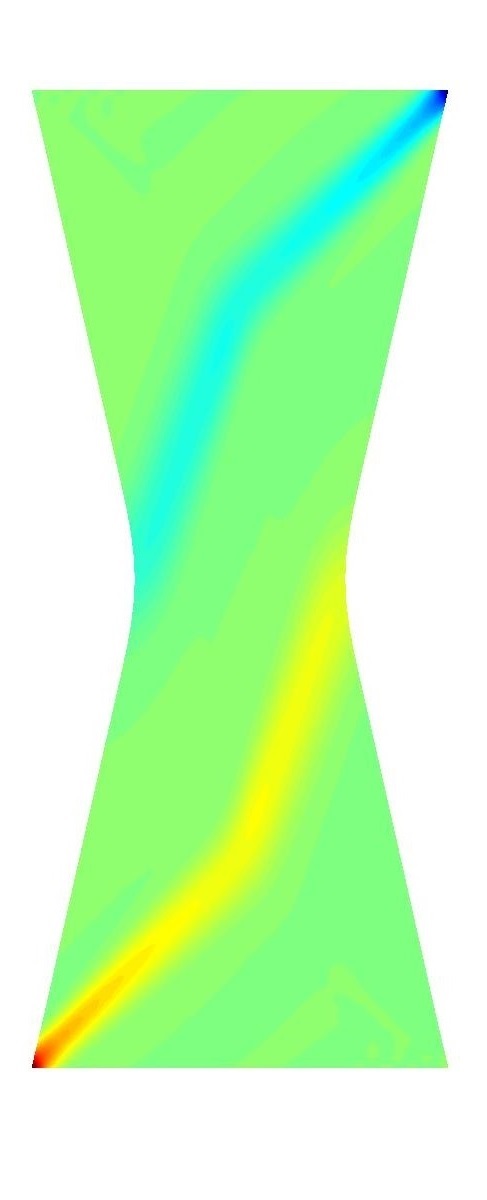}}
\put(3.8,-0.5){\includegraphics[height=0.45\textwidth]{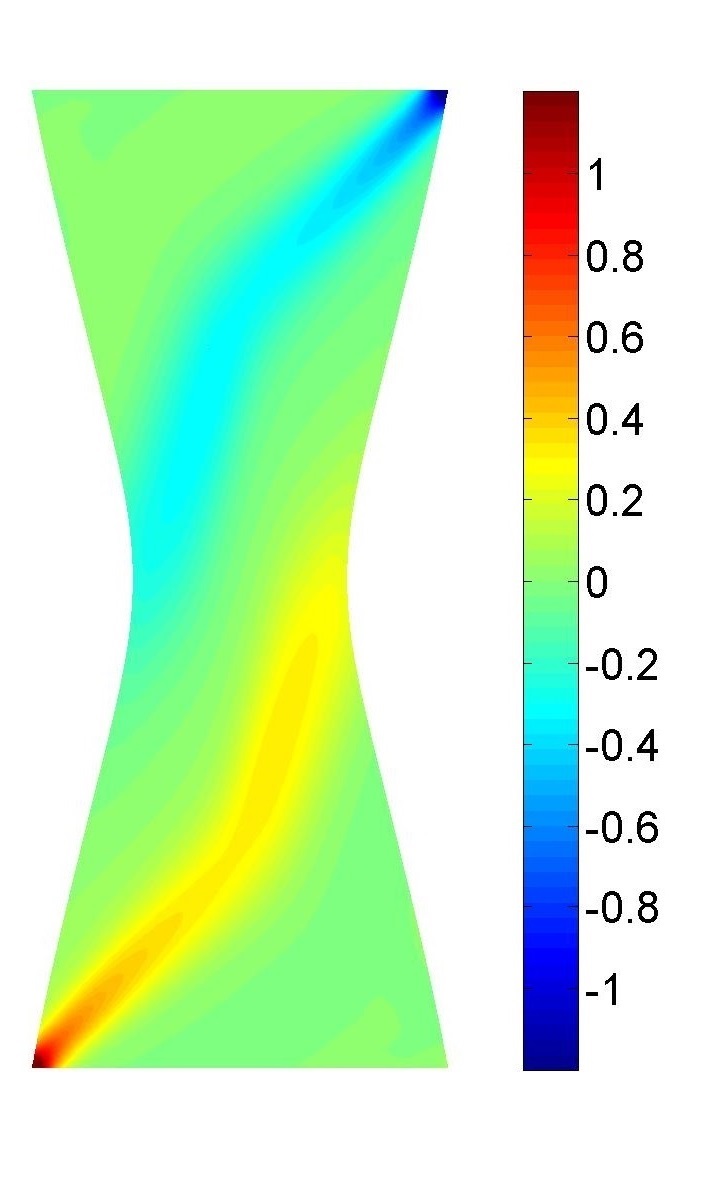}}
\put(-7.4, 6.3){$\beta_\mrg = 0$}

\put(-3.7, 6.3){$\beta_\mrg = 0.1\,\beta_0$}

\put(0.3, 6.3){$\beta_\mrg = 1\,\beta_0$}

\put(4.3, 6.3){$\beta_\mrg = 10\,\beta_0$}

\put(6.6, 6.3){ $\tr_{\!\mrs}\bmubar_1$}

\end{picture}
\caption[caption]{Bias extension of  balanced weave  fabric  sample \#1: 
 shear angle $\theta:=\mathrm{arccos}(\hat\gamma)-90^\circ$ in degrees (first row),
  the first stress invariant $I_1=\tr_{\!\mrs}\bsig$  (second row, units [N/mm]), and   the first moment invariant $\tr_{\!\mrs}\bmubar_1$ of fiber family $\#1$ (third row, units [N]), all for various in-plane bending stiffnesses $\beta_\mrg$, using  $\beta_0 = 1.6$Nmm and mesh $32\times64$.
 The value of  $\tr_{\!\mrs}\bmubar_1$ for $\beta_\mrg = 0.1\,\beta_0$ and  $\beta_\mrg = 1\,\beta_0$ has been scaled by $20$ and $5$, respectively,  w.r.t the actual value to increase visibility. (The distribution of $\tr_{\!\mrs}\bmubar_2$ of fiber family \#2  (not shown) is the mirror image of $\tr_{\!\mrs}\bmubar_1$). }
\label{f:Bias_gamma_varyBeta}
\end{center}
\end{figure}

 Fig.~\ref{f:Bias_gamma_varyBeta} shows the  shear angle (first row),
 stress invariant $\tr_{\!\mrs}\bsig$ (second row) and moment invariant $\tr_{\!\mrs}\bmubar_1$ (third row) for various values of $\beta_\mrg$.  Stress concentrations can be observed at the corners of the sample due to the high strains there. 


%

\subsubsection{Bias extension of unbalanced weave fabrics: the role of in-plane bending}
We further consider the influence of in-plane bending on the bias extension test for 
{unbalanced} weave fabrics -- i.e.~when the two fiber families have different material properties. This case can appear for example when the two families are made of different fiber materials, see e.g.~\cite{Madeo2016}. We assume here that  only the in-plane bending stiffness is  different, while all other parameters are equal. Sample\,\#2 and constitutive model \eqref{e:eg_WFs}--\eqref{e:eg_WFs2} with the parameters from Tab.~\ref{t:WFconstant} are  used again but now both $\beta^1_\mrg$ and $\beta^2_\mrg$ are varied. The difference in bending stiffness of the fiber families is characterized by the ratio 
{$r_\mrb:= \beta_\mrg^2/\beta_\mrg^1$.}

\begin{figure}[H]
\begin{center} \unitlength1cm
\begin{picture}(0,15.5)
\put(-7.9,7.5){\includegraphics[height=0.45\textwidth]{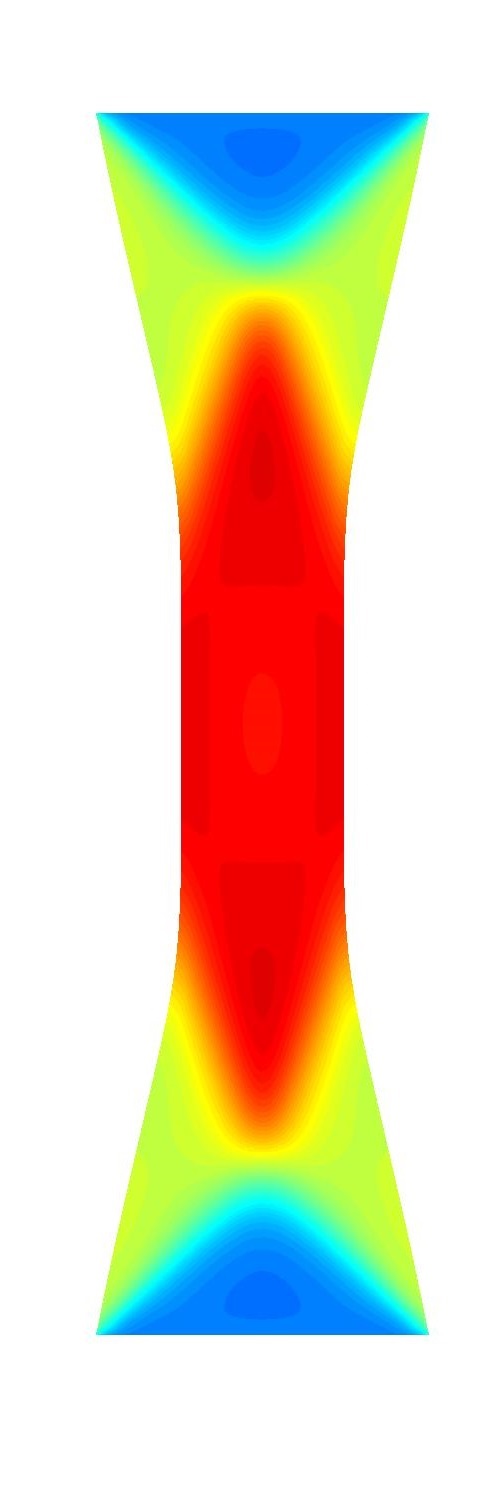}}
\put(-4.0,7.5){\includegraphics[height=0.45\textwidth]{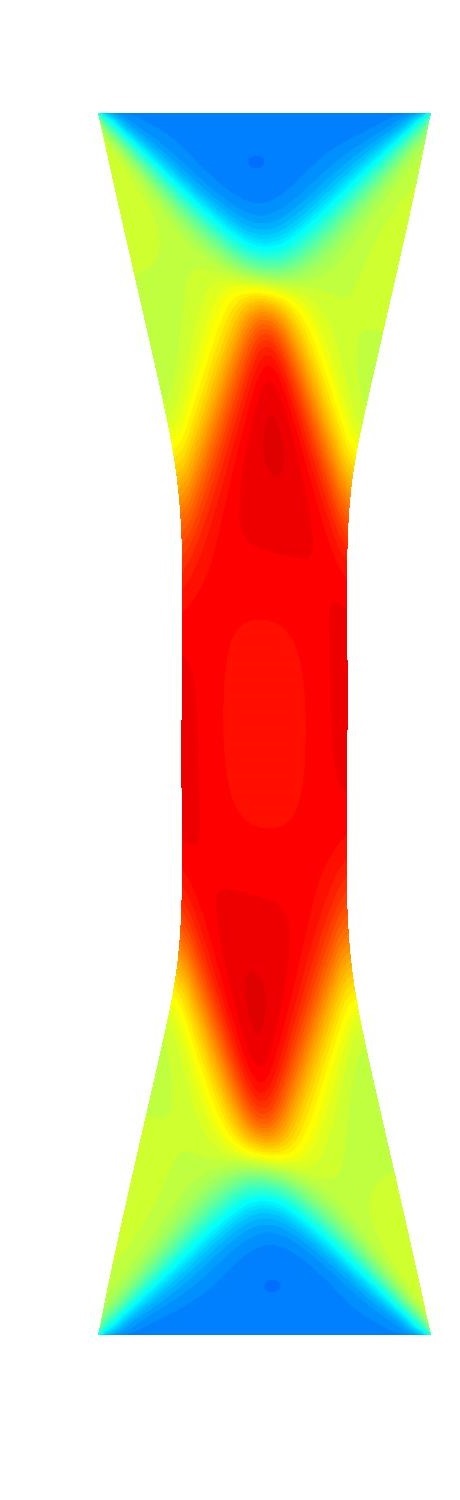}}
\put(-0.1,7.5){\includegraphics[height=0.45\textwidth]{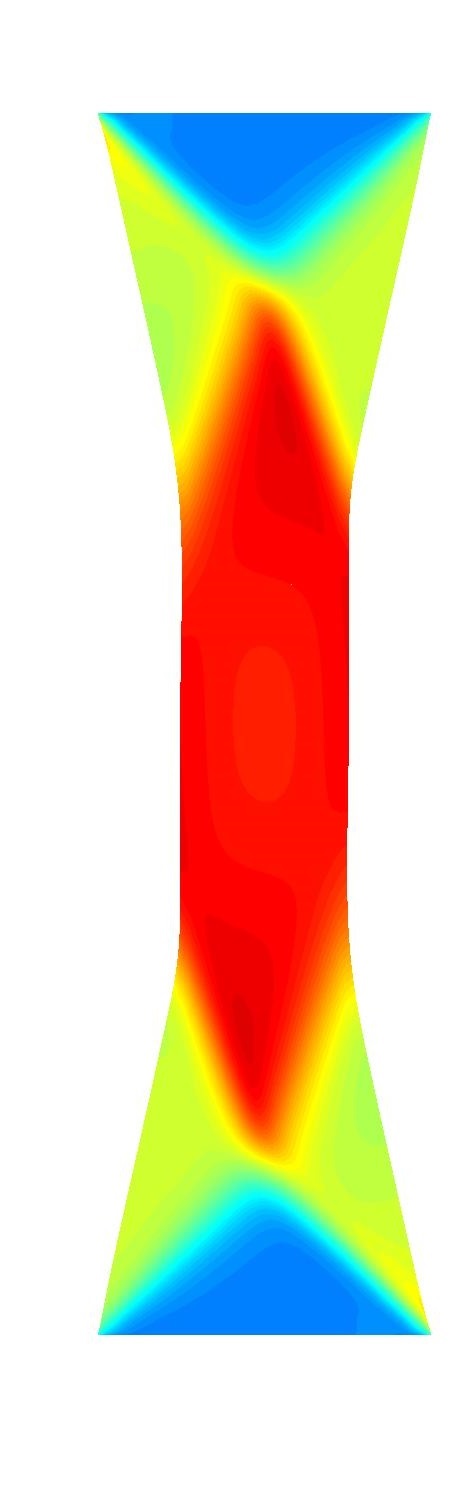}}
\put(3.8, 7.5){\includegraphics[height=0.45\textwidth]{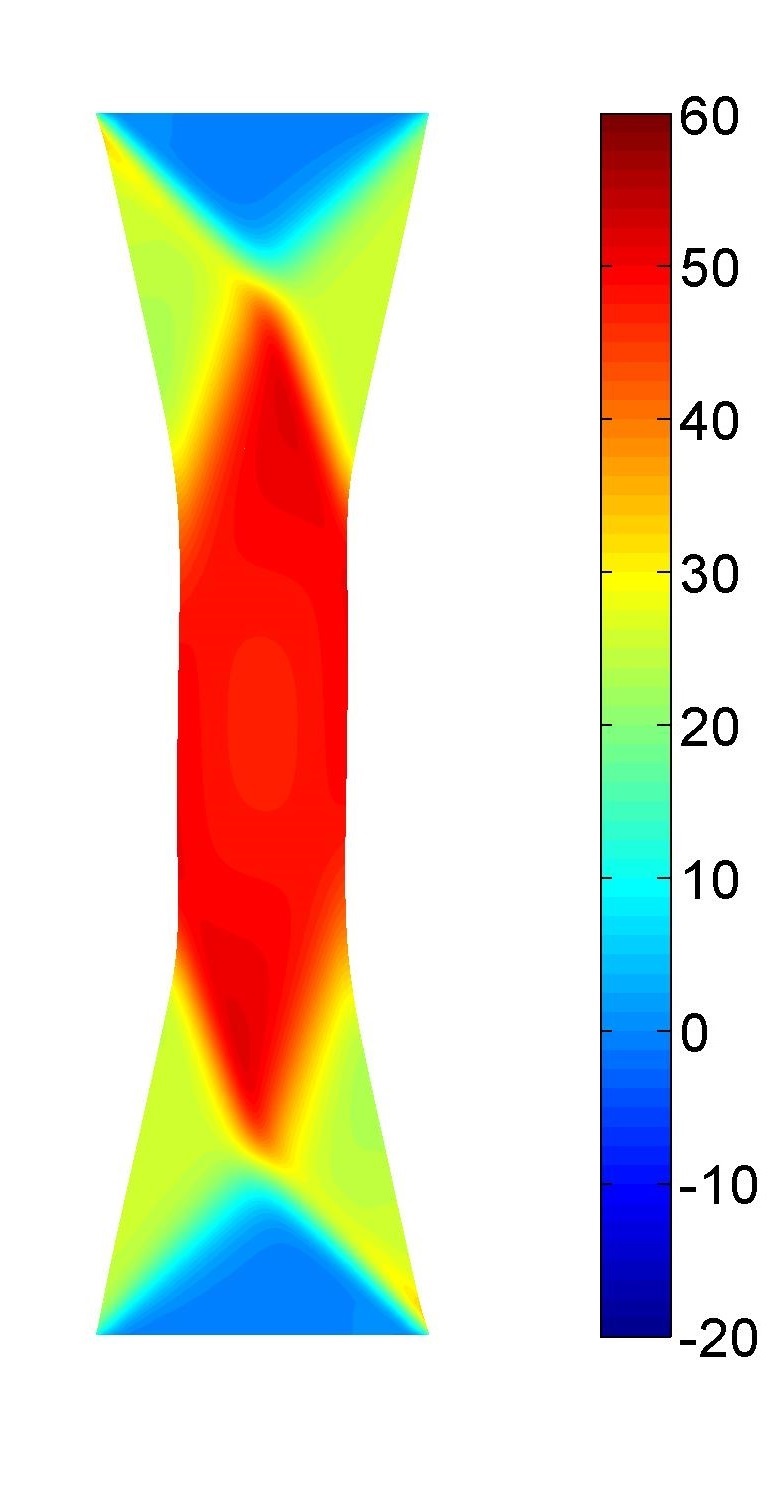}}
\put(-7.4, 14.8){$\beta^2_\mrg = 10\,\beta_0$}
\put(-7.4, 14.3){$ r_\mrb = 1 $}

\put(-3.5, 14.8){$\beta^2_\mrg = 10\,\beta_0$}
\put(-3.5, 14.3){$ r_\mrb = 2 $}

\put(0.4, 14.8){$\beta^2_\mrg = 10\,\beta_0$}
\put(0.4, 14.3){$ r_\mrb = 10 $}

\put(4.3, 14.8){$\beta^2_\mrg = 10\,\beta_0$}
\put(4.3, 14.3){$ r_\mrb = 20 $}
%
\put(-7.9,-0.5){\includegraphics[height=0.45\textwidth]{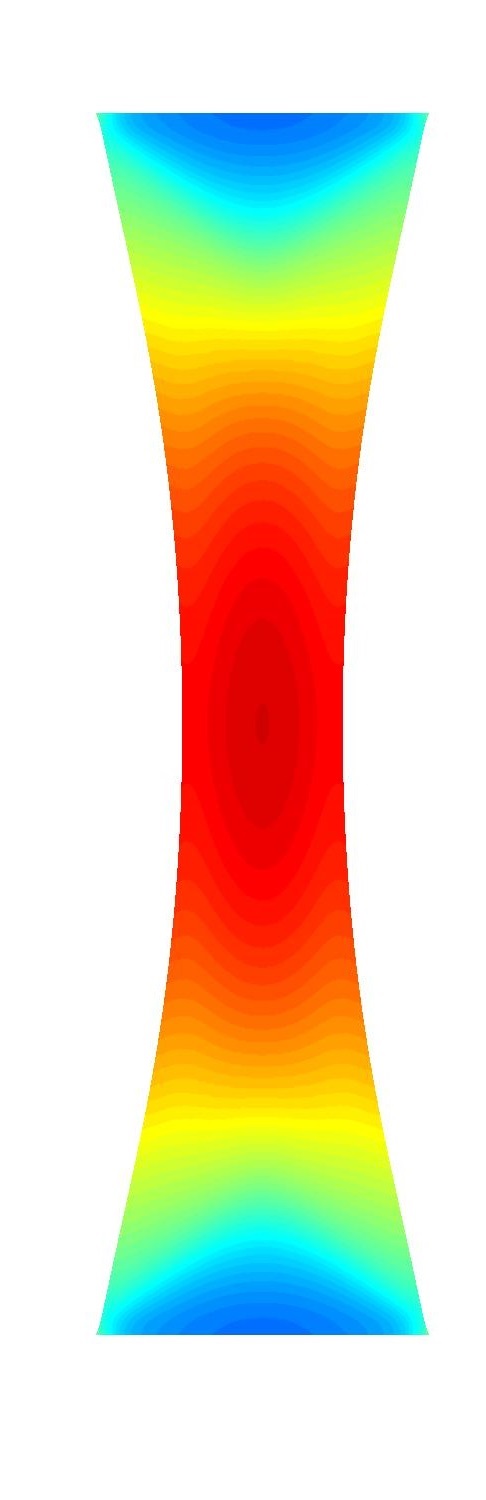}}
\put(-4.0,-0.5){\includegraphics[height=0.45\textwidth]{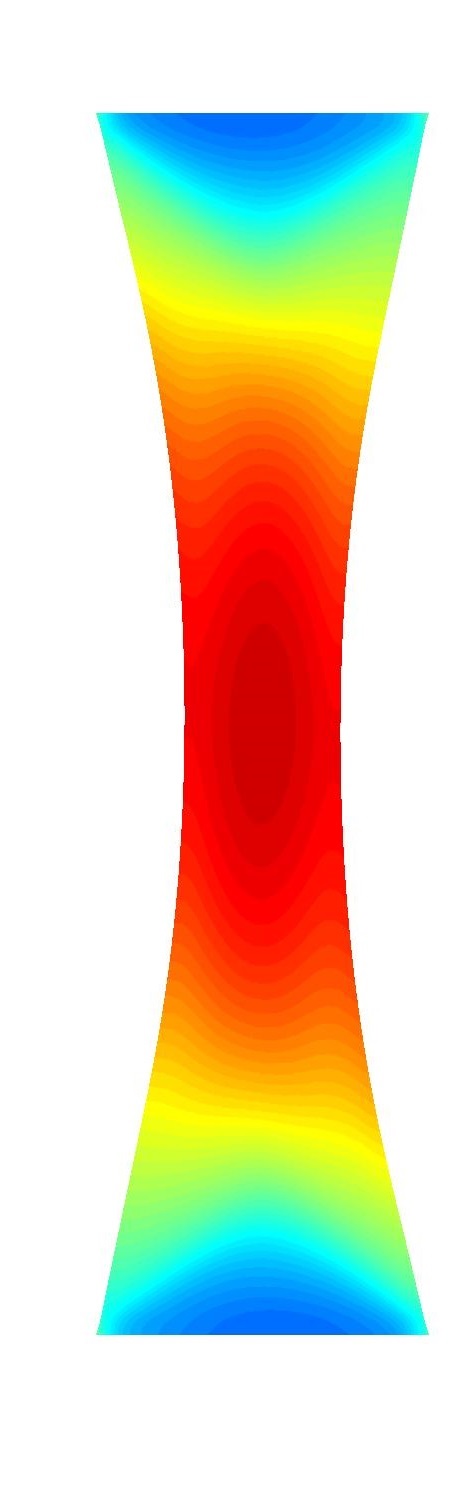}}
\put(-0.1,-0.5){\includegraphics[height=0.45\textwidth]{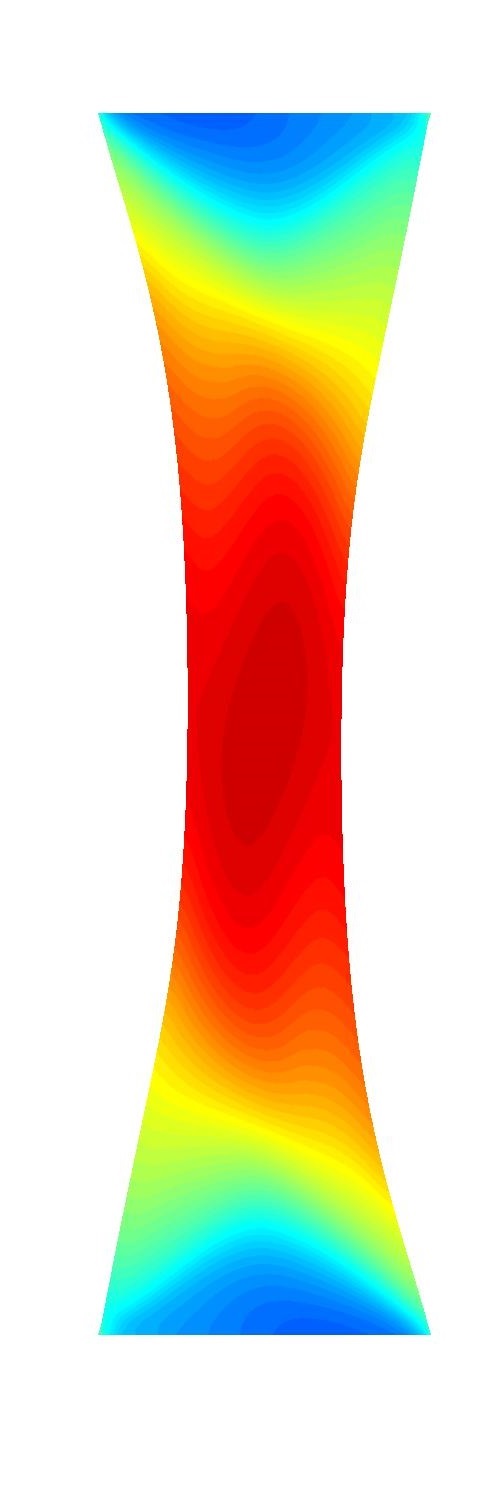}}
\put(3.8,-0.5){\includegraphics[height=0.45\textwidth]{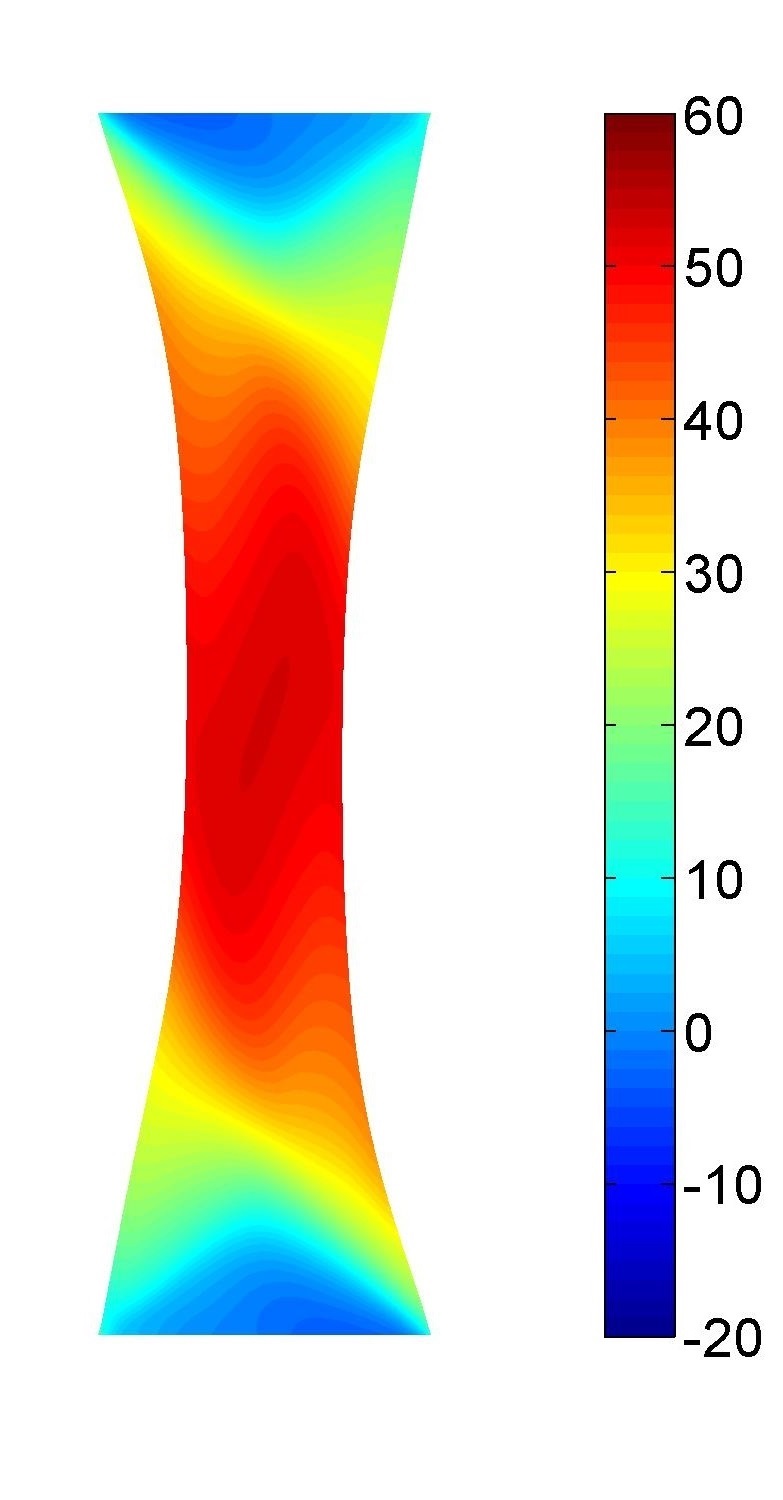}}
\put(-7.4, 6.8){$\beta^2_\mrg = 10^3\,\beta_0$}
\put(-7.4, 6.3){$ r_\mrb = 1$}

\put(-3.5, 6.8){$\beta^2_\mrg = 10^3\,\beta_0$}
\put(-3.5, 6.3){$ r_\mrb = 2 $}

\put(0.4, 6.8){$\beta^2_\mrg = 10^3\,\beta_0$}
\put(0.4, 6.3){$ r_\mrb = 10 $}

\put(4.3, 6.8){$\beta^2_\mrg = 10^3\,\beta_0$}
\put(4.3, 6.3){$ r_\mrb = 20 $}

\end{picture}
\caption[caption]{Bias extension of unbalanced weave fabric sample\,\#2: Deformed configuration
for various $r_\mrb=\beta^2_\mrg/\beta_\mrg^1$  (from left to right) and  various in-plane bending stiffnesses $\beta^2_\mrg$ (from top to bottom). The color shows the shear angle $\theta:=\mathrm{arccos}(\hat\gamma)-90^\circ$ in degrees. Here, $\beta_0 = 1.6$Nmm. }
\label{f:UnbBias_conv_varyBeta_config}
\end{center}
\end{figure}
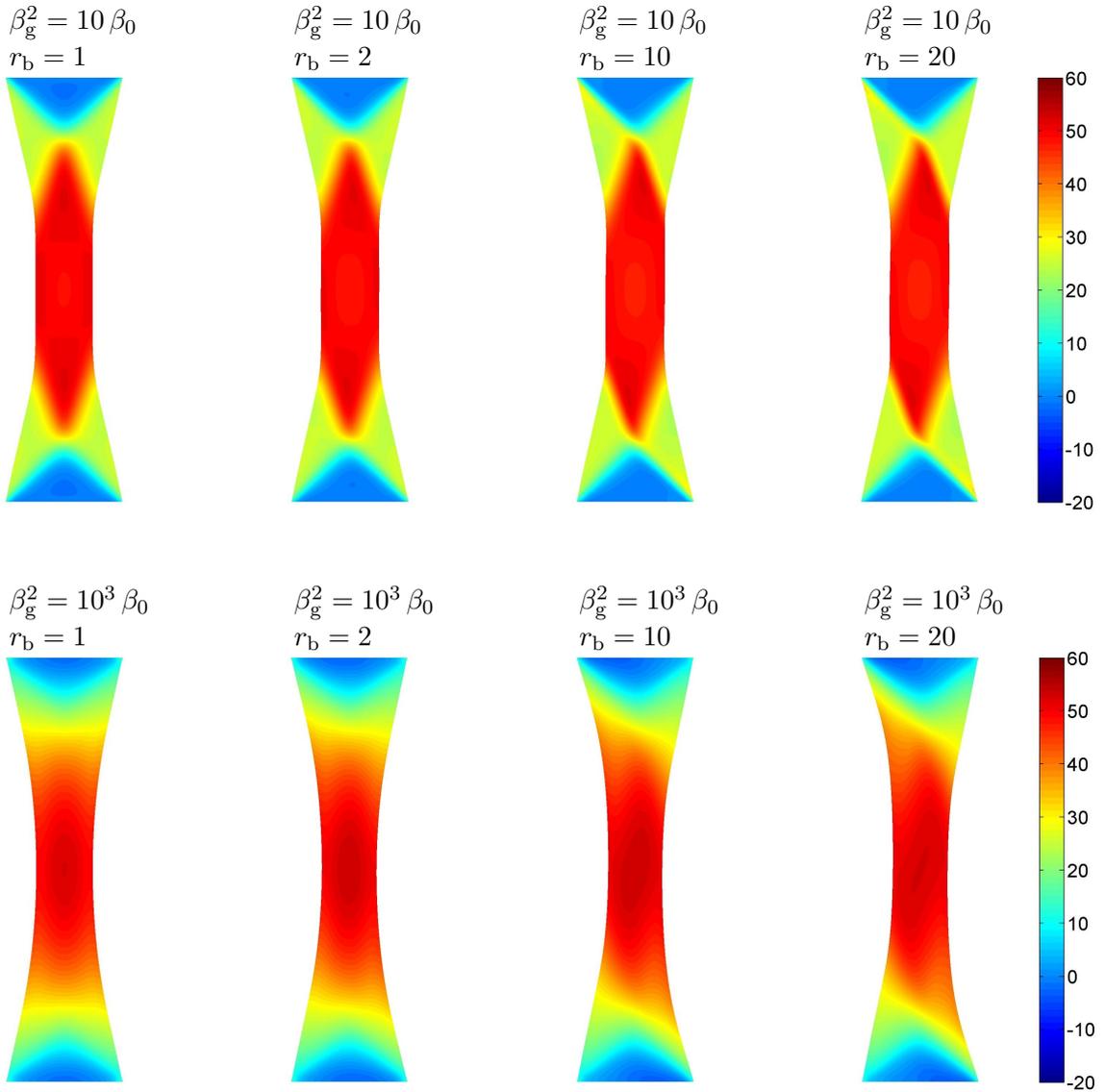

Fig.~\ref{f:UnbBias_conv_varyBeta_config} shows the deformed shapes of the sample for various bending stiffnesses $\beta^1_\mrg$ and $\beta^2_\mrg$. As expected,  unsymmetric sample shapes are obtained 
for  $r_\mrb>1$, especially when $\beta^2_\mrg$ is large.  This is in qualitative agreement with the experimental results given by Madeo et al.~\cite{Madeo2016}. 

%


\subsection{Torsion of dry fabrics} 
\begin{figure}[H]
\begin{center} \unitlength1cm
\begin{picture}(0,3.5)
\put(-7.8,0){\includegraphics[width=0.99\textwidth]{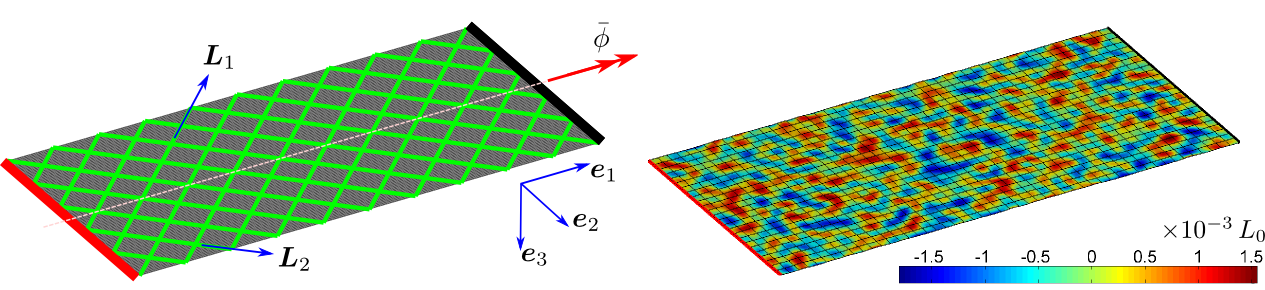}}

%

\put(-6.9,0){{\small{a.}}}
\put(1.0,0){{\small{b.}}}

\end{picture}
\caption[caption]{Torsion of dry fabrics : a.~Fabric specimen, boundary conditions, and fiber directions $\bL_1$ and $\bL_2$ of two fiber families. A rotation is applied around the center axis (dashed line) on the right edge. b.~Corresponding FE mesh with a small imperfection in the $X_3$-coordinate obtained by randomly displacing the control points  in the $\be_3$ direction following a normal distribution with standard deviation $1.1\times10^{-3}\,L_0$. }
\label{f:Torsion_setup}
\end{center}
\end{figure}

{\label{s:example_Tors}}
The second example  considers the torsion of a rectangular sheet with dimension $2L_0\times L_0$ as shown in Fig.~\ref{f:Torsion_setup}a. The left edge is fixed in all three directions, while the right edge is only fixed along $\be_1$.  The two longer edges  are free. A twisting angle $\bar\phi$ is applied around the center line on the right edge from $0^\circ$ to   $180^\circ$ with  $1^\circ$ per load step. The sheet contains dry fabrics with two fiber families, initially aligned by $\pm45^{\circ}$ w.r.t.~the $\be_1$ direction. Material model \eqref{e:eg_Wsimple}--\eqref{e:eg_W1} is used.  The sheet is discretized by $50\times 25$ quadratic NURBS elements.  In order to capture wrinkling of the sheet (if any), a random imperfection of $X_3$  is imposed  (following a standard distribution) as shown in Fig.~\ref{f:Torsion_setup}b. Further, to deal with possible out-of-plane instability due to wrinkling, a small viscosity ($\epsilon = \epsilon_0$)  is added for stabilization (see e.g.~\cite{duong-phd}).\footnote{{Apart from numerical damping, arc-length solvers can also be used for the treatment of  instabilities.}}

\vspace{-0.7cm}
\subsubsection{Nearly incompressible fibers}
We first consider the case with axially nearly inextensible and incompressible  fibers. Therefore, the  material parameters are taken as  $\epsilon_\mrL = 4\,\epsilon^\mre_{\mathrm{stab}} = 2000\,\epsilon_0$, $\epsilon_\mra = \epsilon_0$,  and  $\beta_\mrn = \beta_\mrg = \beta_\tau =  \epsilon_0\,L_0^2$, while $\mu$ and $\epsilon^\mrv_{\mathrm{stab}}$ are zero. 

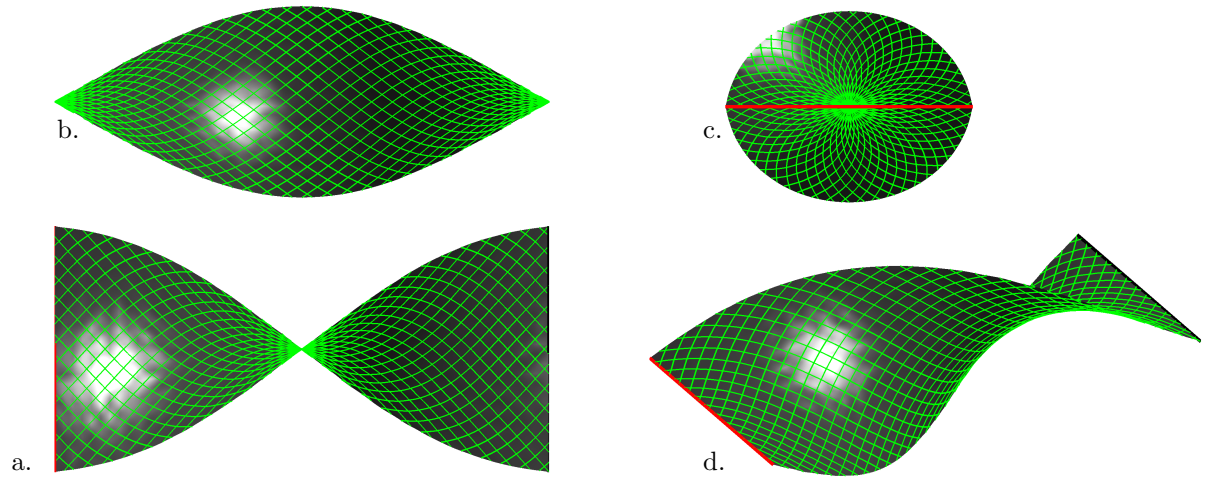
\begin{figure}[H]
\begin{center} \unitlength1cm
\begin{picture}(0,6.3)

\put(-8.2,-0.3){\includegraphics[width=0.5\textwidth]{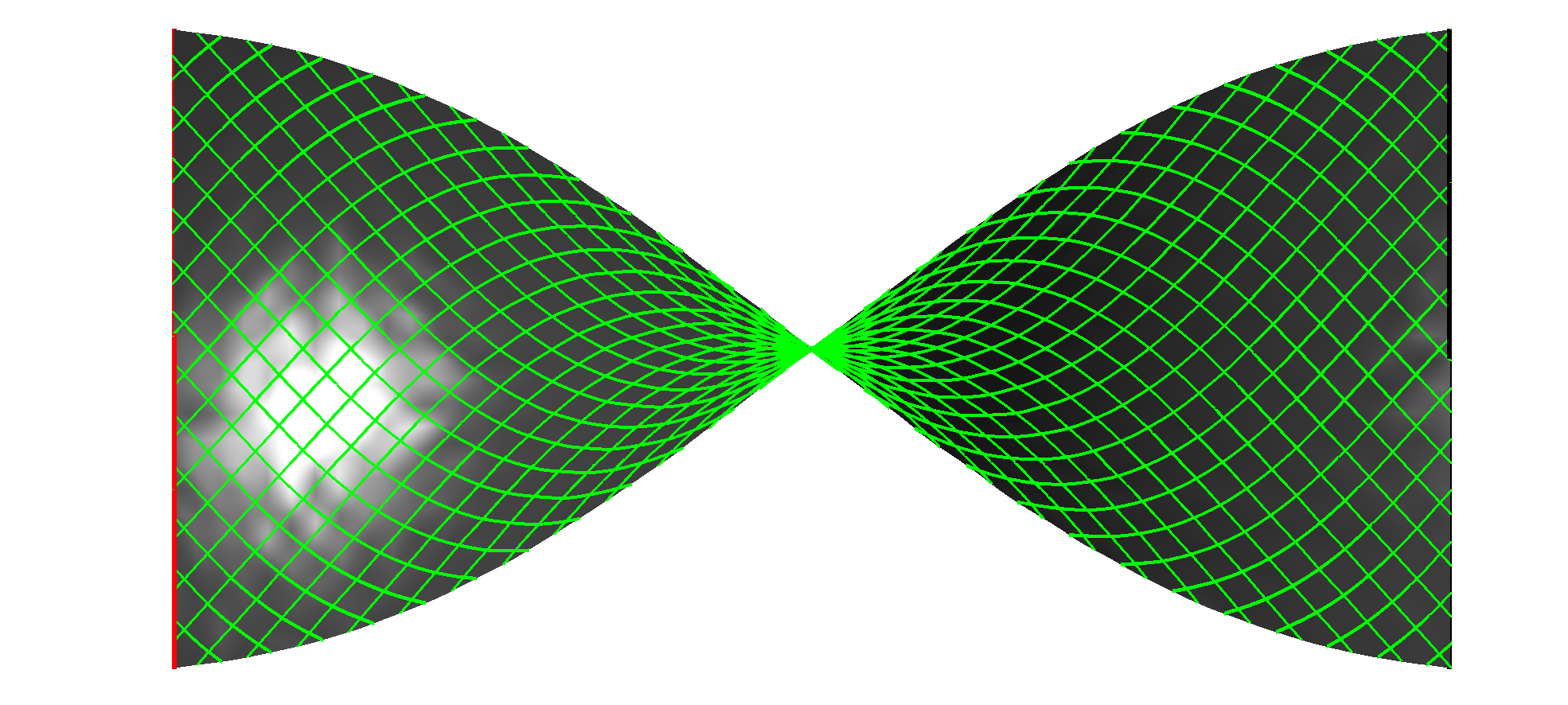}}

\put(0.0,-0.3){\includegraphics[width=0.5\textwidth]{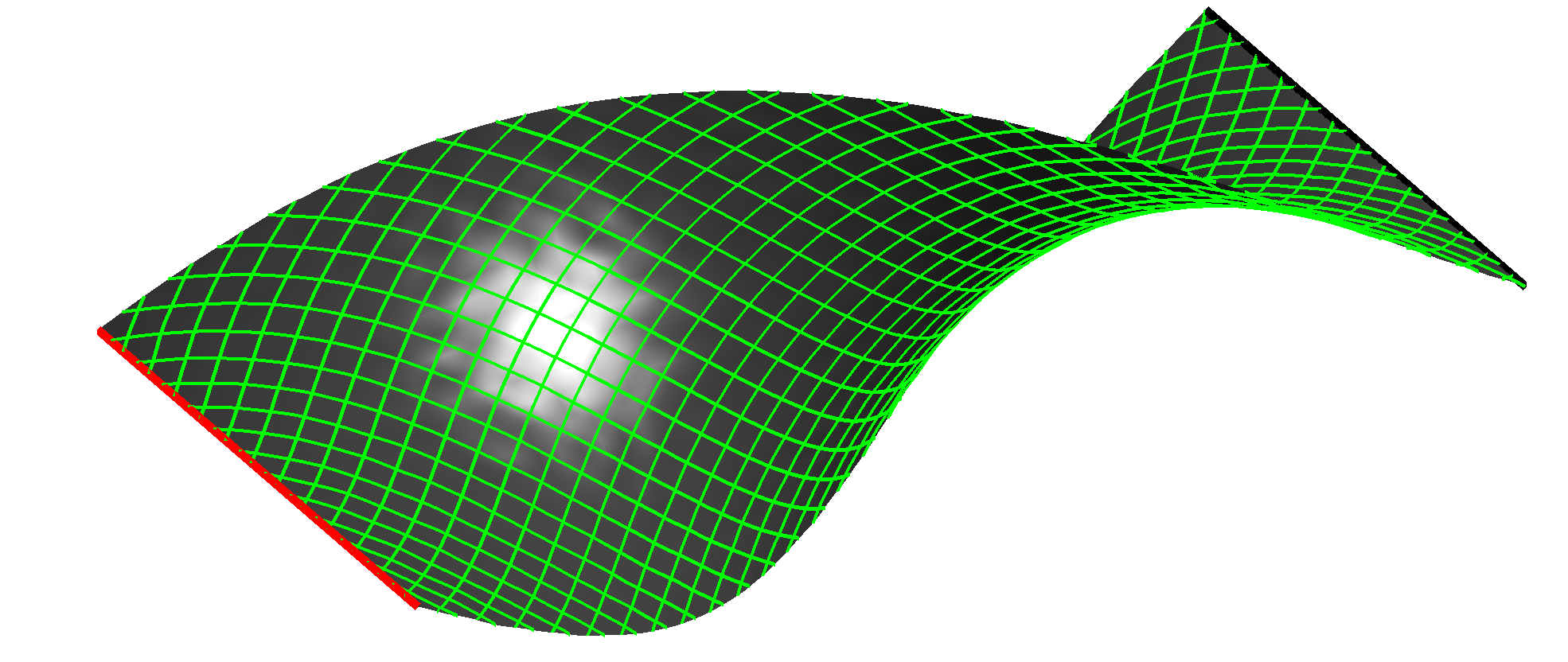}}


\put(-1.0,3.4){\includegraphics[width=0.5\textwidth]{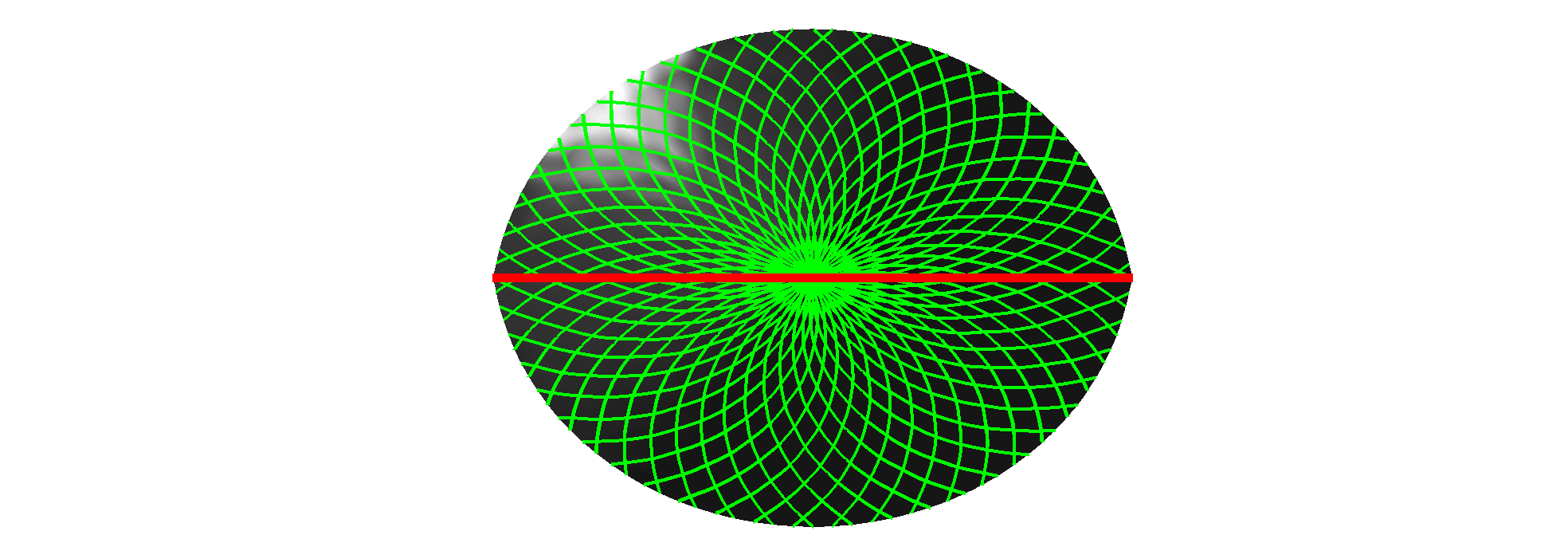}}

\put(-8.2,3.4){\includegraphics[width=0.5\textwidth]{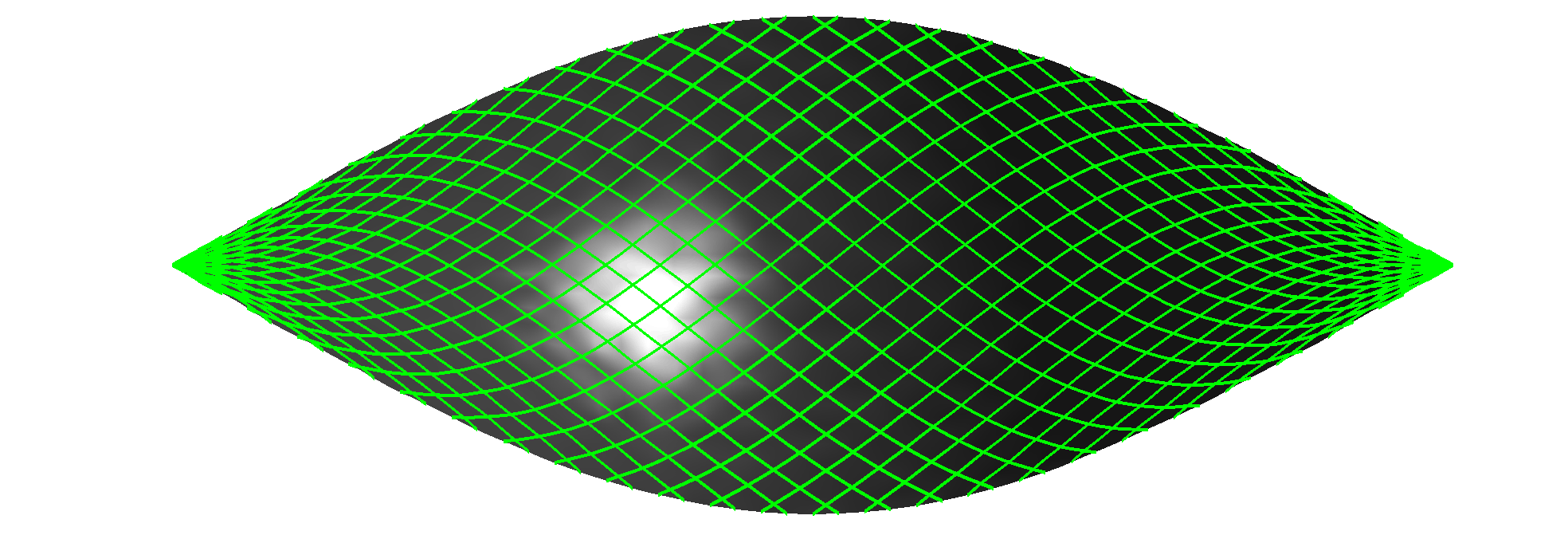}}

\put(-7.9,-0.1){{\small{a.}}}
\put(-7.3,4.3){{\small{b.}}}

\put(1.2,4.3 ){{\small{c.}}}
\put(1.2,-0.1 ){{\small{d.}}}

\end{picture}
\caption[caption]{Torsion of dry fabrics with nearly incompressible fibers: Deformed configuration at $\bar\phi = 180^{\circ}$ showing embedded fibers in a.~top view, b.~front view, c.~side view, and d.~3D view.
}
\label{f:Torsion_incompressible_x}
\end{center}

\end{figure}
Fig.~\ref{f:Torsion_incompressible_x}  shows the deformed configuration with the embedded fibers at $\bar\phi=180^\circ$. As seen, the deformed sheet behaves similar to an isotropic elastic shell, and no wrinkling occurs in spite of the geometrical imperfection in the out-of-plane direction.\footnote{A similar deformed shape is also obtained in the simulation result of Schulte et al.~\cite{Schulte2020}. 
}

\begin{figure}[H]
\begin{center} \unitlength1cm
\begin{picture}(0,6.3)
\put(-8.2,0){\includegraphics[width=0.52\textwidth]{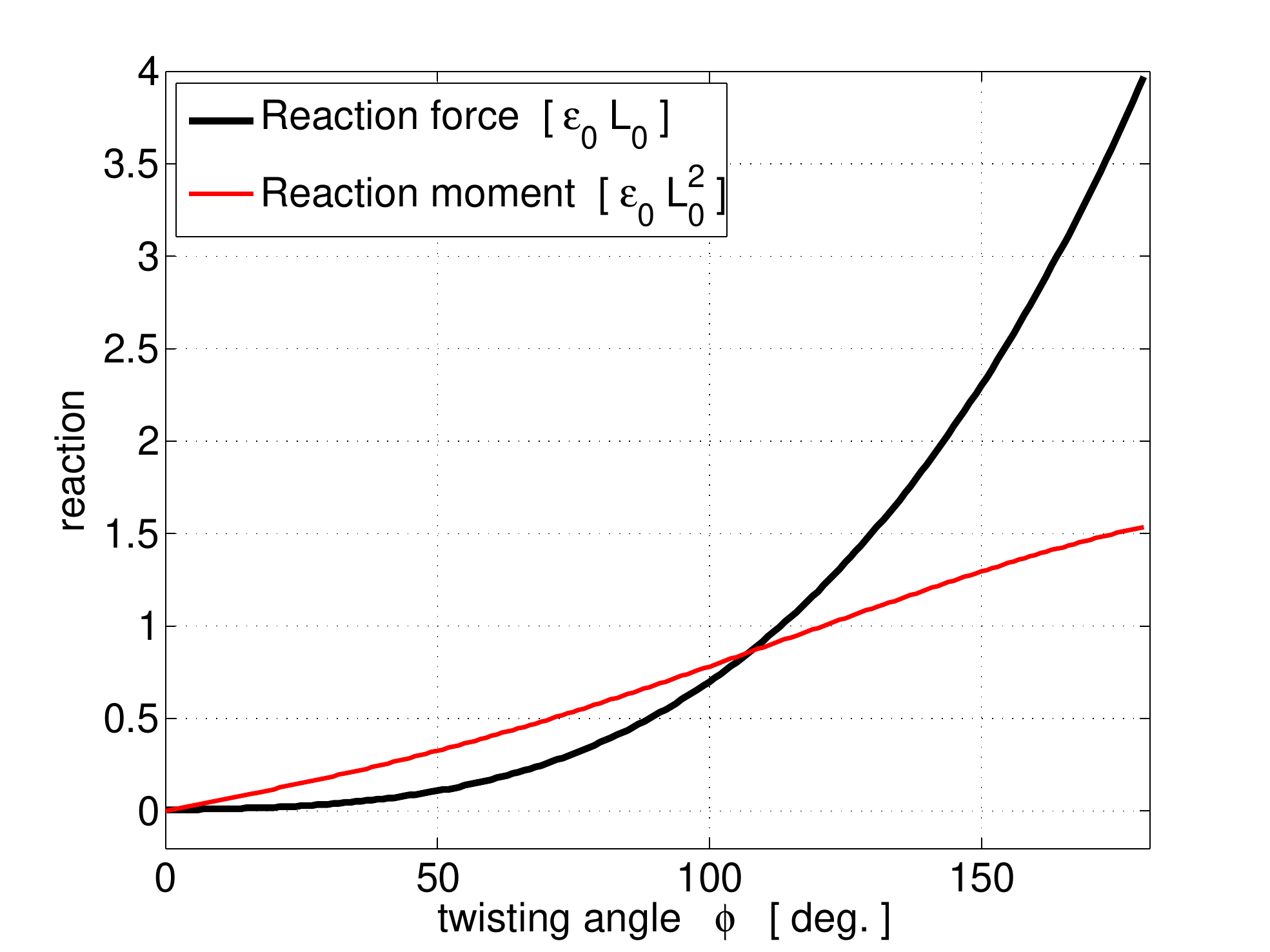}}
\put(-0.1,0){\includegraphics[width=0.52\textwidth]{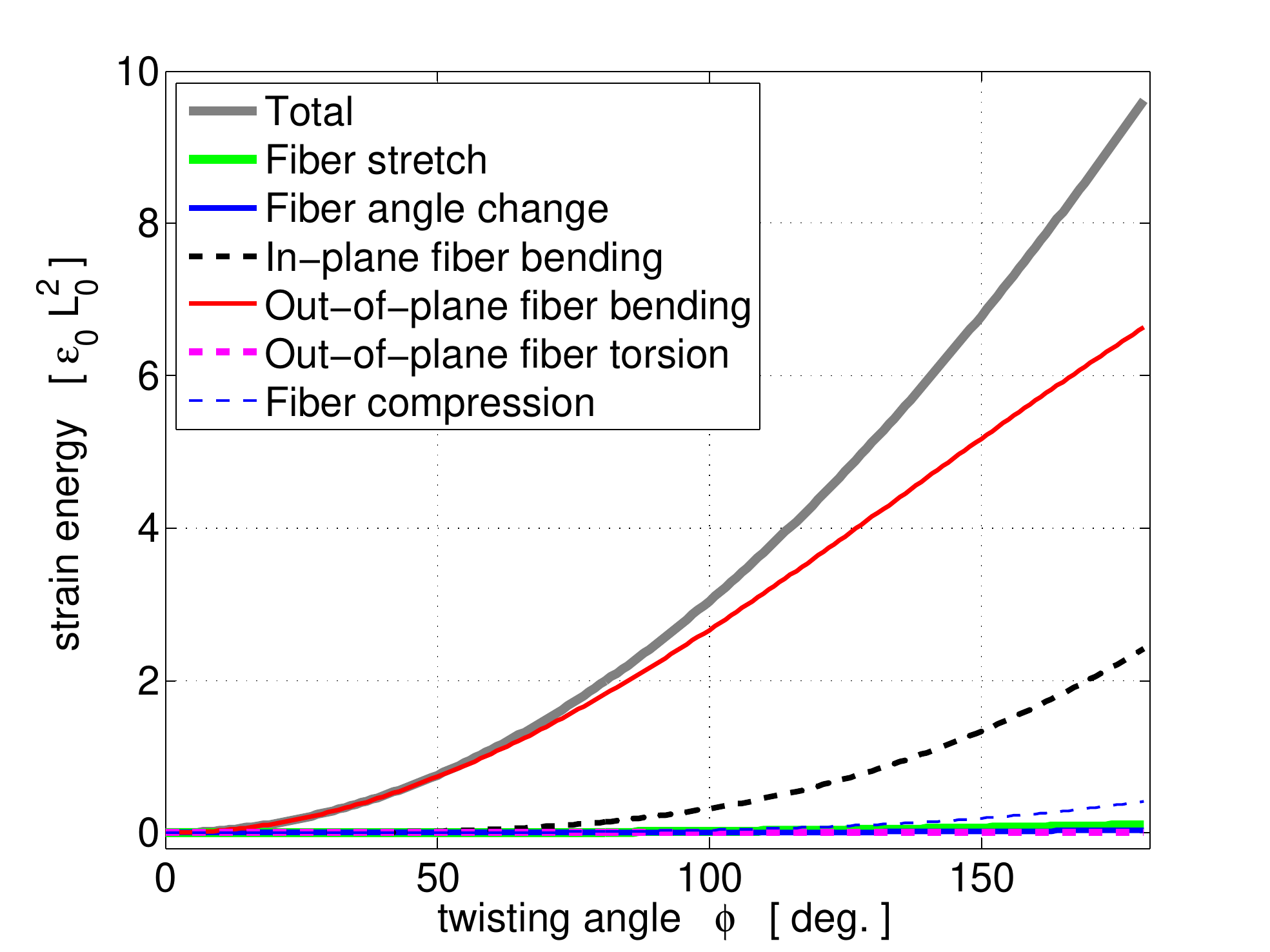}}


\put(-7.8,0){{\small{a.}}}
\put(0.2,0){{\small{b.}}}

\end{picture}
\caption[caption]{Torsion of dry fabrics with nearly incompressible fibers: a.~Suport reactions and b.~strain energies (integrated strain energy densities) of the sheet versus twisting angle $\bar\phi$. }
\label{f:Torsion_incompressible_disp}
\end{center}
\end{figure}

Fig.~\ref{f:Torsion_incompressible_disp}a plots the reaction-twisting curve, which shows that both the reaction force and the reaction moment are monotonically increasing. As seen in Fig.~\ref{f:Torsion_incompressible_disp}b, most of the strain energy initially goes into out-of-plane bending. However, as the twisting angle increases, the in-plane fiber bending energy increases significantly, while the other membrane energies remain relatively small.

\subsubsection{Effectively compressible fibers}
Next, we consider fabrics with effectively compressible fibers 
based on the phenomenological model of Sec.~\ref{s:two_material_model} that accounts for the microscale buckling of compressed fibers. To this end, we reset material parameter  $\epsilon^\mre_{\mathrm{stab}} = {5}\,\epsilon_0$ and $\epsilon^\mrv_{\mathrm{stab}}= 250\,\epsilon_0$, while the other parameters remain unchanged.

\begin{figure}[H]
\begin{center} \unitlength1cm
\begin{picture}(0,6.5)

\put(-1.0,5.35){\includegraphics[width=0.5\textwidth]{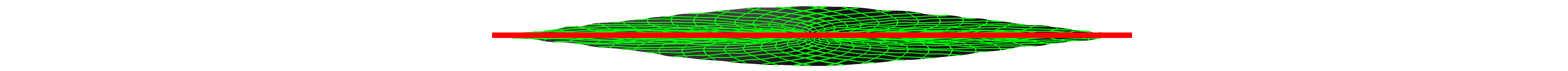}}

\put(-8.2,5.3){\includegraphics[width=0.5\textwidth]{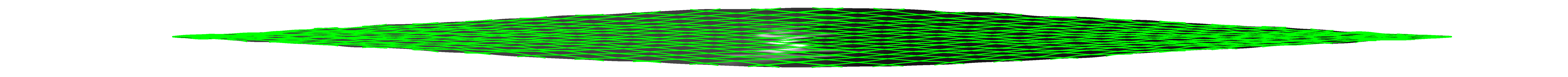}}

\put(-8.2,0.5){\includegraphics[width=0.5\textwidth]{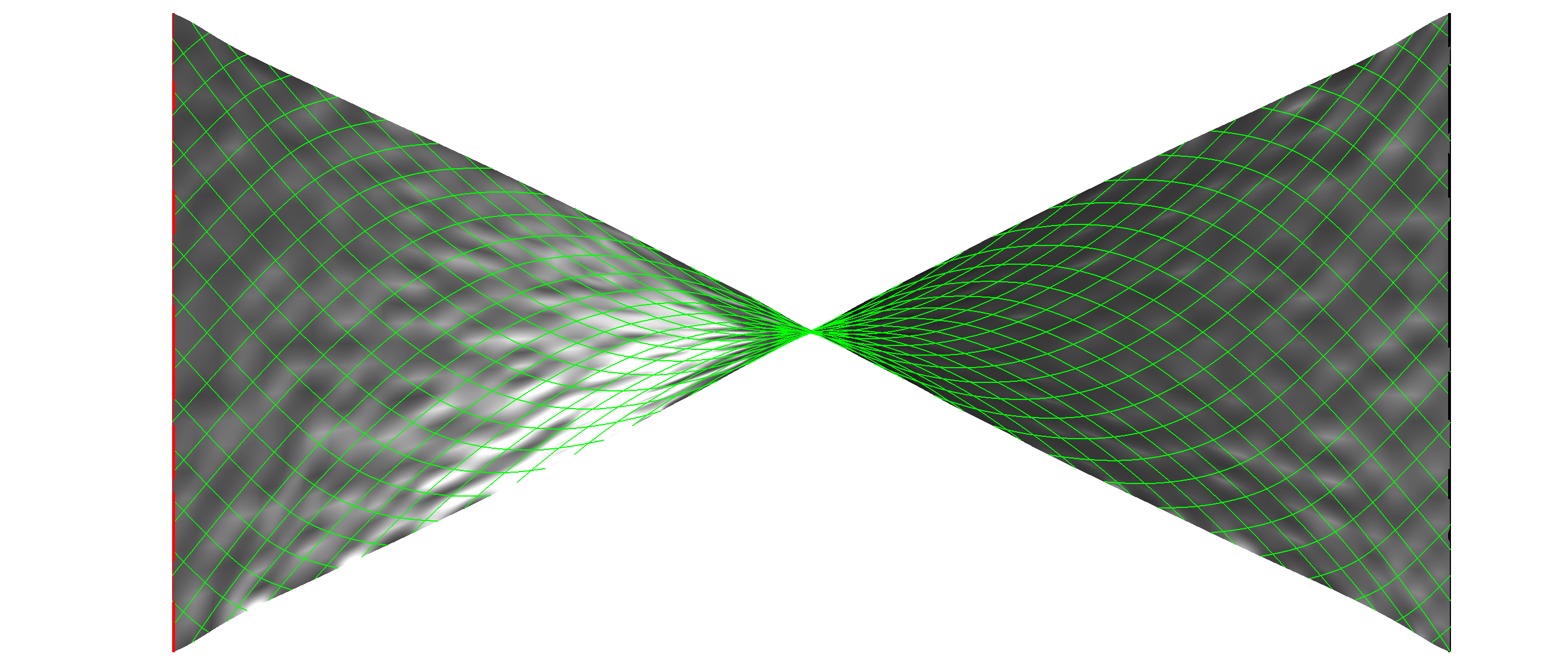}}

\put(1.1,0.){\includegraphics[width=0.45\textwidth]{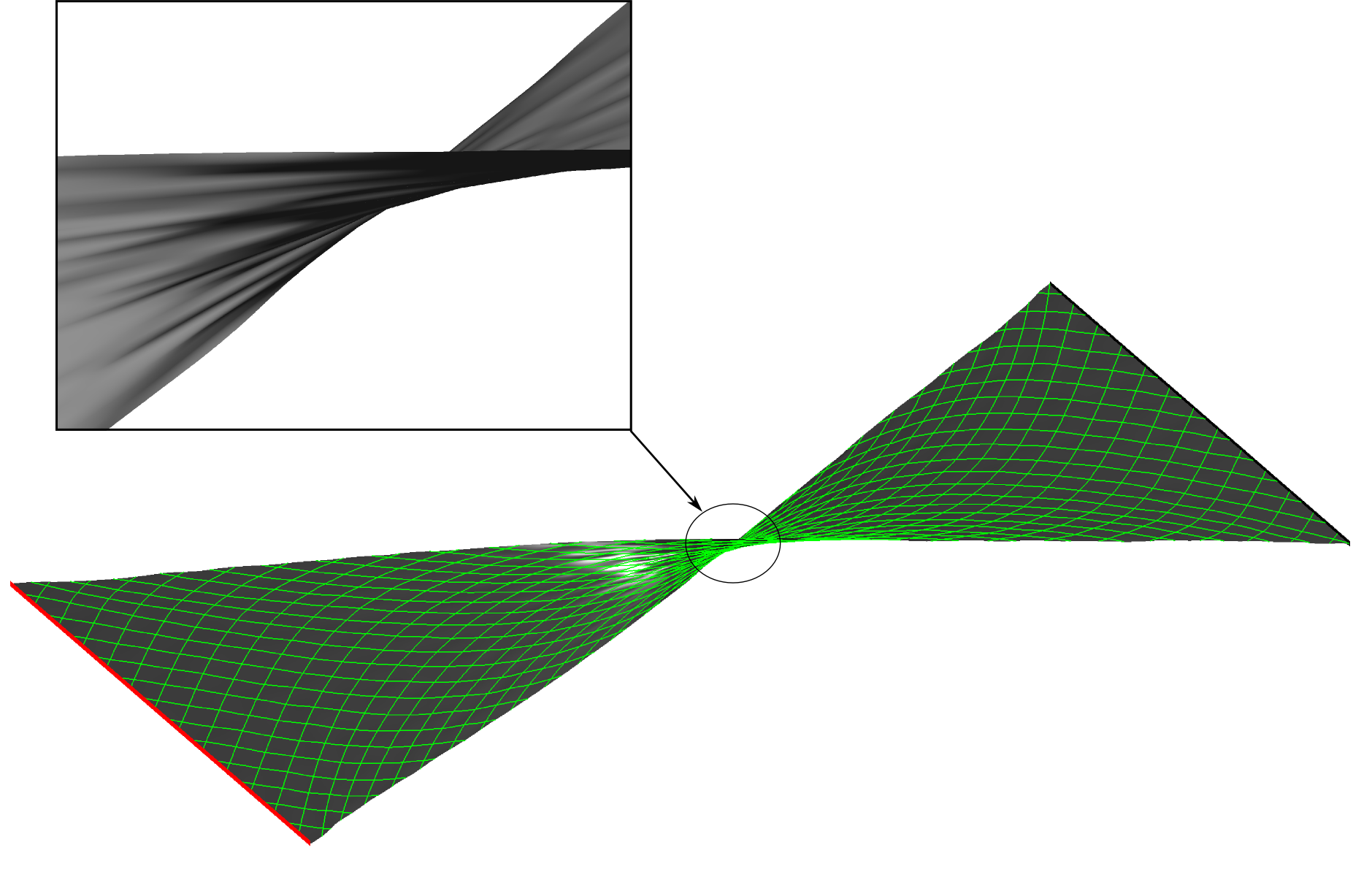}}

\put(-7.9,0.7){{\small{a.}}}
\put(-7.3,5.7){{\small{b.}}}

\put(1.2,5.7 ){{\small{c.}}}
\put(1.0,0.0 ){{\small{d.}}}

\put(1.5,4.0 ){e. }

\end{picture}
\caption[caption]{Torsion of dry fabrics with effectively compressible fibers: Deformed configuration at $\bar\phi = 180^{\circ}$ showing embedded fibers in a.~top view, b.~front view, c.~side view, d.~3D view, and e.~zoom into the sheet center.
}
\label{f:Torsion_compressible_x_fb}
\end{center}
\begin{center} \unitlength1cm
\begin{picture}(0,6.0)

\put(-7.8,0){\includegraphics[width=1\textwidth]{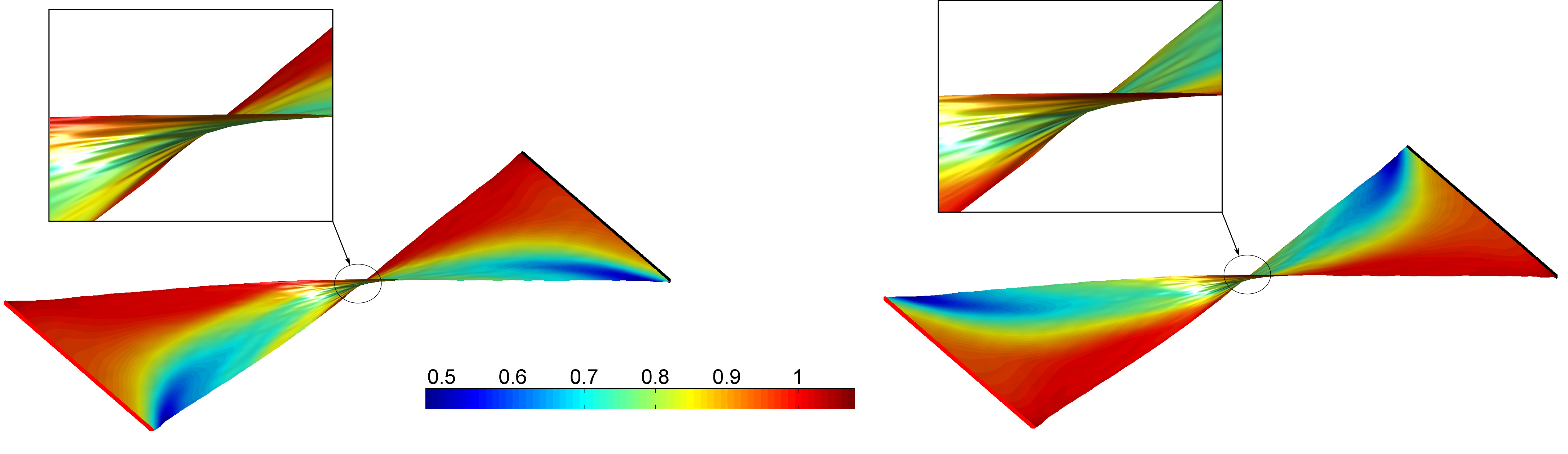}}

\put(-7.8,1){{\small{a.}}}
\put(1.2,1){{\small{b.}}}
\end{picture}
\caption[caption]{The torsion test for dry fabrics of compressible fibers: Deformed configuration at $\bar\phi = 180^{\circ}$ showing invariant  $\Lambda^1$ (a.) and $\Lambda^2$ (b.). 
%
%
}
\label{f:Torsion_compressible_x}
\end{center}

\begin{center} \unitlength1cm
\begin{picture}(0,6.5)
\put(-8.0,0){\includegraphics[width=0.52\textwidth]{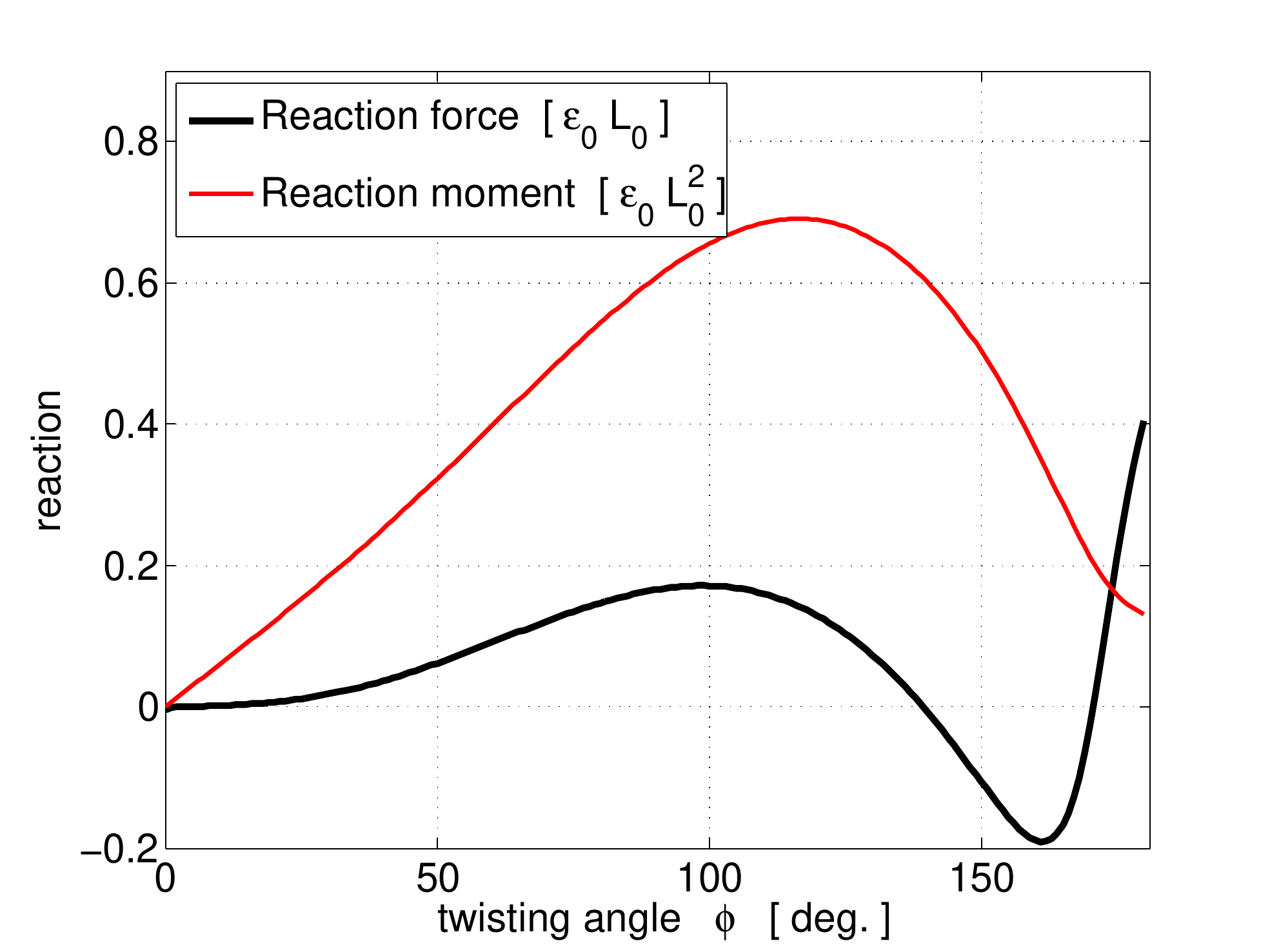}}

\put(0.1,0){\includegraphics[width=0.52\textwidth]{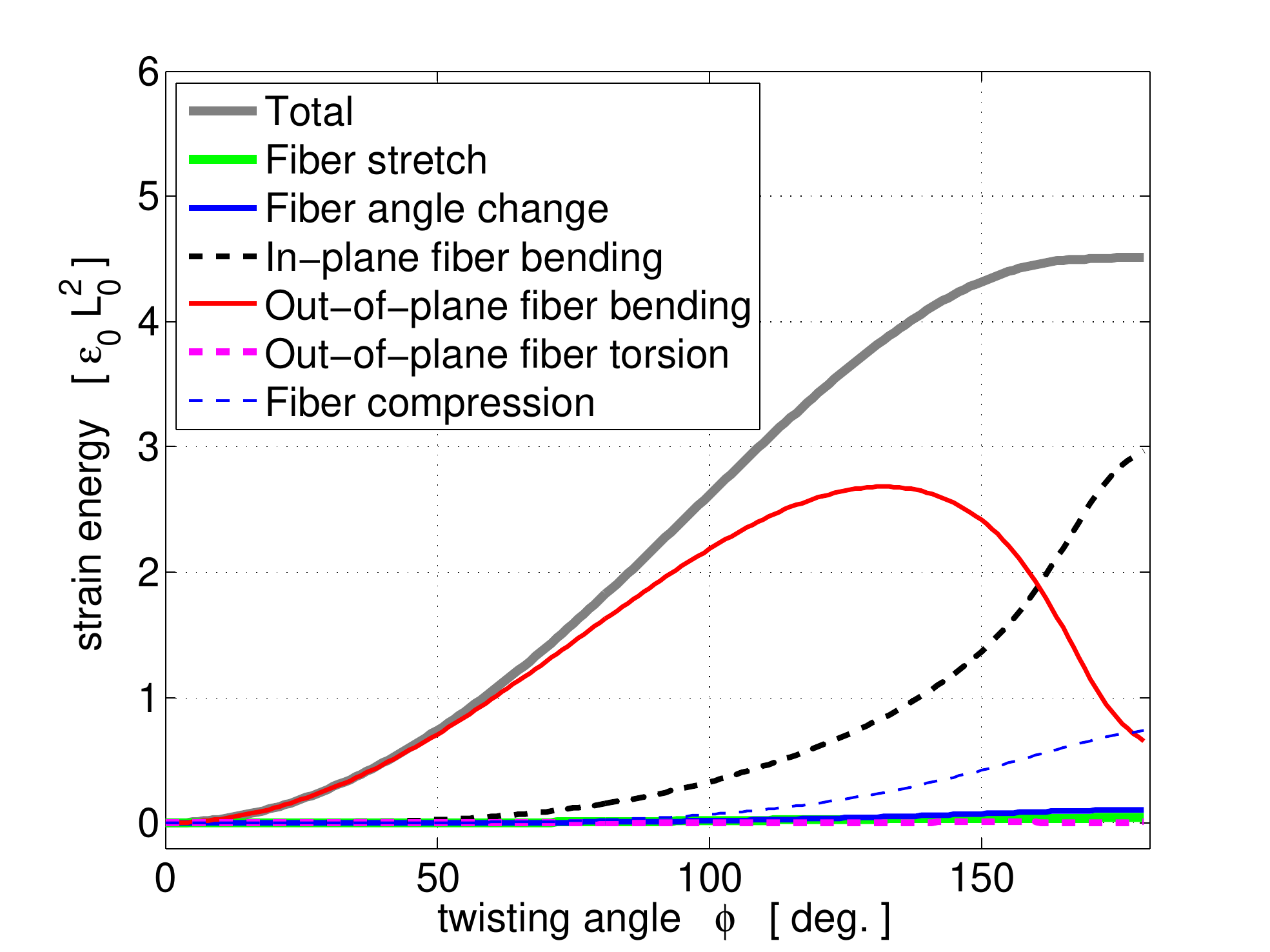}}

\put(-7.7,0){{\small{a.}}}
\put(0.5,0){{\small{b.}}}


\end{picture}
\caption[caption]{The torsion test for dry fabrics of effectively compressible fibers: a.~Reactions and b.~strain energies (integrated strain energy densities) of the sheet  versus twisting angle $\bar\phi$. }
\label{f:Torsion_compressible_disp}
\end{center}
\end{figure}

Fig.~\ref{f:Torsion_compressible_x_fb} shows the deformed configuration with embedded fibers at $\bar\phi=180^\circ$. As seen, the center of the sheet is compressed significantly in the lateral direction, and consequently macroscopic 
{wrinkling} can be {observed} there as shown in Fig.~\ref{f:Torsion_compressible_x_fb}e. As Fig.~\ref{f:Torsion_compressible_x} shows,  fibers (of either family) are   compressed  not only  at the sheet center, but also along {its diagonals}. This implies that microscopic fiber buckling can also occur along the diagonals. 

Figs.~\ref{f:Torsion_compressible_disp}a-b plot the reaction-twisting curves and the strain energies of the sheet, respectively. Similar to  the incompressible case, most of the strain energy initially goes into out-of-plane bending, but the out-of-plane energy loses convexity and is exceeded by the in-plane bending energy as the twisting angle increases. This implies that the fibers become unstable in  out-of-plane bending. The resulting deformation and the reaction-twisting curve  are shown in Fig.~\ref{f:Torsion_compressible_x} and  Fig.~\ref{f:Torsion_compressible_disp}a, respectively.


\section{Conclusion}{\label{s:conclusion}}

We have presented a nonlinear rotation-free isogeometric shell formulation that can capture the in-plane bending behavior of embedded fibers. The formulation is based on the generalized Kirchhoff-Love shell theory of Duong et al.~\cite{shelltextile}. Its finite element implementation can  be directly obtained from the isogeometric FE formulation of  Duong et al.~\cite{solidshell} by complementing it with the additional in-plane bending term. The construction of material models for the proposed shell formulation can follow that of classical shells, since inducing invariants for the relative in-plane curvature tensor $\bar\bK$ is very similar to that of the relative out-of-plane curvature tensor $\bK$. We have demonstrated this point by proposing two material models for fabrics in Sec.~\ref{s:matmodels}. The woven fabric model of Sec.~\ref{s:wovenfabricmodel}  shows good agreement with existing experiment data for the bias extension test. With this material model, the influence of the in-plane bending stiffness on the bias extension test has been investigated for both balanced and unbalanced weave fabrics. 
The proposed shell formulation can also admit a wide range of other material models  
including those expressed directly in surface energy form and those obtained from thickness integration of 3D material models. 
In order to suppress possible material instabilities due to fiber compression, 
we have added a stabilization term within the proposed shell formulation.
Finally, the accuracy and robustness of the proposed formulation is verified by several numerical examples, characterized by both homogenous  and  inhomogenous deformation  in Sec.~\ref{s:num_examples1} and \ref{s:num_examples2}, respectively. Our formulation can be extended to capture inter-ply and intra-ply sliding of yarns, which will be a subject of future work.
\appendix
\section{Tangent matrices of the external forces}
This appendix summarizes the linearization and discretization of the external virtual work term \eqref{e:Giie22} appearing in the weak form.

\subsection{Linearization of the external virtual work}{\label{s:linGextO}}

For the first term of $G_{\mathrm{ext}}$ in Eq.~\eqref{e:Giie22}, no linearization is required as  $\bff_{\!0}\,\dif A $ is constant for dead loading. The linearization of the remaining terms in Eq.~\eqref{e:Giie22} can be 
{found as} (see also \cite{membrane,shelltheo,solidshell})
\eqb{llllll}
\Delta G_\mathrm{ext} \is \ds\int_{\sS} \delta\bx\,p\,\big(\bn\otimes\ba^\alpha - \ba^\alpha\otimes\bn\big)\,\Delta\ba_\alpha\,\dif a   \\[5mm]
\plus \ds\int_{\partial_t\sS} \delta\bx\,\,\bt\otimes\ba_\xi\,\frac{1}{\norm{\ba_\xi}^2}\,\Delta\ba_\xi\,\dif s\, -\,  \delta\bx\,m_\nu\,\big(\ba^\alpha\otimes\bn\big)\,\Delta\ba_\alpha~\\[5mm]
\plus   \ds\int_{\partial_m\sS} \delta\ba_\alpha\,\big(\nu^\alpha\,\ba^\beta\otimes\bn + \nu^\beta\,\bn\otimes\ba^\alpha\big)\,\Delta\ba_\beta \, m_\tau\, \dif s \\[5mm]
 \mi \ds\int_{\partial_m\sS} \delta\ba_\alpha\, \nu^\alpha\,\bn\otimes\ba_\xi\,\frac{1}{\norm{\ba_\xi}^2}\, \Delta\ba_\xi\, m_\tau\,\dif s \, +\, \ds\int_{\partial_m\sS} \delta\ba_\alpha\, \tau^\alpha\,\bn\otimes\bnu\, \frac{1}{\norm{\ba_\xi}}\,\Delta\ba_\xi\,m_\tau\,\dif s~\\[5mm]
\plus \ds \int_{\partial_{\bar{m}}\sS} \delta\ba_\alpha\, \big[  \ell^\alpha\,c^\beta\, \bn\otimes\bn - \ellab\,(\bell\otimes\bc + \bc\otimes\bell) \big]\,\Delta\ba_\beta \,\bar{m}\,\dif s ~\\[5mm]
\plus \ds\int_{\partial_{\bar{m}}\sS} \delta\ba_\alpha\,\ell^{\alpha}\,\bc\otimes\ba_\xi \,\Delta\ba_\xi\,\frac{\bar m}{\norm{\ba_\xi}^2}\,\dif s ~,
\label{e:linext}
\eqe
where 
{$p$ is assumed to be constant,} and
$\xi$ is the convective coordinate along the boundary, so that $\btau = \ba_\xi /\norm{\ba_\xi}$ and $\Delta\dif s = (\ba_\xi / \norm{\ba_\xi}^2)\cdot\Delta\ba_\xi\,\dif s $. 
 For the last term in Eq.~\eqref{e:linext}, we have used  Eq.~\eqref{e:lin_bn_bc_bell} and $\Delta \ell^\alpha = -\ellab\,\bell\cdot\Delta\ba_\beta$ (see \cite{shelltextile}).

\subsection{{Discretization of the external virtual work}{\label{s:disGextO}}}
The tangent matrices in Eq.~\eqref{e:kexte} follow from applying discretization to Eq.~\eqref{e:linext}, which gives
\eqb{llllll}
\mk^e_{\mathrm{ext}p} \is \ds\int_{\sS} \mN^\mrT\,p\,\big(\bn\otimes\ba^\alpha - \ba^\alpha\otimes\bn\big)\,\mN_{,\alpha}\,\dif a ~, \\[5mm] 
\mk^e_{\mathrm{ext}t} \is  \ds\int_{\partial_t\sS}\mN^\mrT\,\bt\otimes\ba_\xi\,\frac{1}{\norm{\ba_\xi}^2}\, \mN_{,\xi}\,\dif s~,\\[7mm]
\mk^n_{\mathrm{ext}\nu} \is - m_\nu\,\big(\ba_A^\alpha\otimes\bn_A\big)\,N_{A,\alpha}~,\\[5mm]
\mk^e_{\mathrm{ext}m} \is  \ds\int_{\partial_m\sS} \mN^\mrT_{,\alpha}\,\big(\nu^\alpha\,\ba^\beta\otimes\bn + \nu^\beta\,\bn\otimes\ba^\alpha\big)\,\mN_{,\beta} \, m_\tau\, \dif s ~\\[3mm]
\mi \ds\int_{\partial_m\sS} \mN^\mrT_{,\alpha}\, \nu^\alpha\,\bn\otimes\ba_\xi\, \mN_{,\xi}\,\frac{ m_\tau}{\norm{\ba_\xi}^2} \,\dif s \,
+ \left[\ds\int_{\partial_m\sS} \mN^\mrT_{,\alpha}\, \tau^\alpha\,\bn\otimes\bnu\, \mN_{,\xi}\, \frac{m_\tau}{\norm{\ba_\xi}} \,\dif s\right]~, \\[6mm]
\mk^e_{\mathrm{ext}\bar{m}} \is \ds \int_{\partial_{\bar{m}}\sS} \mN^\mrT_{,\alpha}\, \big[  \ell^\alpha\,c^\beta\, \bn\otimes\bn - \ellab\,(\bell\otimes\bc + \bc\otimes\bell) \big]\,\mN_{,\beta} \,\bar{m}\,\dif s~ \\[3mm]
\plus \ds\int_{\partial_{\bar{m}}\sS} \mN^\mrT_{,\alpha}\,\ell^{\alpha}\,\bc\otimes\ba_\xi \,\mN_\xi\,\frac{\bar m}{\norm{\ba_\xi}^2}\,\dif s ~.
\label{e:disGext}
\eqe

\begin{remark}
 The tangent matrices in Eq.~\eqref{e:disGext} account for live loading. In case  $\bt\,\dif s$, $m_\tau\,\dif s$, and $\bar m \,\dif s$ are constant, the last term in $\mk^e_{\mathrm{ext}t}$, the last two terms in $\mk^e_{\mathrm{ext}{m}}$, and the last term in $\mk^e_{\mathrm{ext}\bar{m}}$ vanish, respectively.
\end{remark}

\begin{remark}
Note, that the last term in $\mk^e_{\mathrm{ext}{m}}$ (in square bracket) accounts for the variation of $\bnu$, which is missing  in \cite{solidshell} (cf.~Appendix A, Eq.~(128)).
\end{remark}

\section{Efficient FE implementation}{\label{s:efficient}}

This appendix presents an efficient implementation of Eq.~\eqref{e:DiscretizedFEforce} for the Newton-Raphson method. The implementation can be viewed as an extension of Duong et al.~\cite{solidshell} by the in-plane bending term.


Since Eq.~\eqref{e:DiscretizedFEforce} holds for all nodal variations $\delta\mx$, it leads, after the application of the essential boundary conditions, to the system of nonlinear  equations 
\eqb{lll}
\mf (\muu) := \mf_{\mathrm{int}} - \mf_{\mathrm{ext}} =  \boldsymbol{0}~, 
\label{e:fe_global}
\eqe
where $\muu$ are the nodal unknowns, and  $\mf_{\mathrm{int}}$ and $\mf_{\mathrm{ext}}$ are the global FE forces obtained from assembling the elemental FE forces given by Eqs.~\eqref{e:FEforces} and \eqref{e:fext}, respectively. 

During Newton-Raphson, Eq.~\eqref{e:fe_global} is solved iteratively for the increment $\Delta \muu$ from
\eqb{lll}
 \mK (\muu^{i-1})\,\Delta \muu^{i} \is -\mf(\muu^{i-1})~,\\[3mm]
\muu^i \is \muu^{i-1} + \Delta \muu^i~,
\eqe
where $\mK$ is the (reduced) global stiffness matrix. It is obtained by assembling the elemental tangent matrices, i.e.~
\eqb{lll}
\mK = \ds \sum_{e=1}^{n_{\mathrm{el}}}\, \big( \mk^e_{\mathrm{mat}}  + \mk^e_{\mathrm{geo}} -  \mk^e_{\mathrm{ext}})~,
\eqe
and then eliminating the constrained degrees-of-freedom. Here, $\mk^e_{\mathrm{mat}}$, $\mk^e_{\mathrm{geo}}$, and $\mk^e_{\mathrm{ext}}$ are defined by Eqs.~\eqref{e:kmate}, \eqref{e:geoKij}, and \eqref{e:kexte}, respectively.  

\subsection{Implementation of FE force vectors}
Due to the symmetry of the stress and moment tensors, they can be represented compactly in Voigt notation as
 \eqb{lll}
 \hat{\btau} := [\tau^{11}~,~ \tau^{22}~,~  \,\tau^{12}]^\mrT ~, \quad
 \hat{\bM}_0 := [M_0^{11}~,~ M_0^{22}~,~  \,M_0^{12}]^\mrT ~, \quad
 \hat{\widerbar{\bM}}_0  : = [\bar{M}_0^{11}~,~ \bar{M}_0^{22}~,~  \,\bar{M}_0^{12}]^\mrT ~.
\eqe
Defining the  $3n_{\mathrm{ne}}\times 1$ arrays
\eqb{lll}
\mL_{\alpha\beta}^a \dis \mN^\mrT_{,\alpha}\,\ba_\beta~, \\[2mm]
\mL_{\alpha}^n \dis \mN^\mrT_{,\alpha}\,\bn~, \\[2mm]
\mG_{\alpha\beta}^n \dis \mN_{;\alpha\beta}^\mrT\,\bn \\[2mm]
\mG_{\alpha\beta}^a \dis  - \mN^\mrT_{,\alpha}\,\bar\bc_{,\beta} - {\mC}_{,\beta}^\mrT\,\ba_\alpha~,
  \label{e:appenx_auxN}
\eqe
where $n_{\mathrm{ne}}$ is the number of control points per element, and organizing them into the arrays
 \eqb{lll}
 \hat{\mL}_a \dis [\mL^a_{11}~,~ \mL^a_{22}~,~  \mL^a_{12}+\mL^a_{21}]~, \\[2mm]
 \hat{\mG}_n \dis [\mG^n_{11}~,~ \mG^n_{22}~,~  \mG^n_{12}+\mG^n_{21}]~,\\[2mm]
  \hat{\mG}_a \dis [\mG^a_{11}~,~ \mG^a_{22}~,~  \mG^a_{12}+\mG^a_{21}]~,
  \label{e:appenx_voigt}
\eqe
the FE forces \eqref{e:FEforces} can be implemented as
\eqb{lll}
\mf^e_\mathrm{int\tau} 
=  \ds\int_{\Omega^e_0} \hat{\mL}_a\, \hat{\btau}  \,\dif A~, \quad \quad
\mf^e_{\mathrm{int}M}  
=  \ds\int_{\Omega^e_0} \hat{\mG}_n\, \hat{\bM}_0  \, \dif A~,\quad \quad
\mf^e_{\mathrm{int}\bar{M}}  
=  \ds\int_{\Omega^e_0} \hat{\mG}_a\, \hat{\widerbar\bM}_0  \, \dif A~.
\eqe
Note, that for classical shell formulations without fiber bending, the last term and its associated tangent matrices are simply dropped.

\subsection{Implementation of material stiffness matrices}
To implement the material stiffness matrices, the nine material tangents $c^{\alpha\beta\gamma\delta}$,  $d^{\alpha\beta\gamma\delta}$, $e^{\alpha\beta\gamma\delta}$, $f^{\alpha\beta\gamma\delta}$,
$\bar{d}^{\alpha\beta\gamma\delta}$, 
$\bar{e}^{\alpha\beta\gamma\delta}$,  $\bar{f}^{\alpha\beta\gamma\delta}$,
$\bar{g}^{\alpha\beta\gamma\delta}$,  and $\bar{h}^{\alpha\beta\gamma\delta}$ are arranged  into the $3\times 3$ matrices $\mC$,  $\mD$, $\mE$, $\mF$,  $\bar\mD$,  $\bar\mE$, $\bar\mF$, $\bar\mG$, $\bar\mH$, respectively, as, for instance,
\eqb{l}
\mC := \left[\begin{matrix}
c^{1111} & c^{1122} & c^{1112} \\ 
c^{2211} & c^{2222} & c^{2212}  \\ 
c^{1211} & c^{1222} & c^{1212}\\ 
\end{matrix} \right] .
\eqe
Note, that for hyperelastic material models, we further have $\mC=\mC^\mrT$, $\mE = \mD^\mrT$,  $\mF=\mF^\mrT$, $\bar\mE = \bar\mD^\mrT$,  $\bar\mF=\bar\mF^\mrT$, $\bar\mG = \bar\mH^\mrT$.
With these and Eq.~\eqref{e:appenx_voigt}, the material stiffness matrices in Eq.~\eqref{e:matKij} can be implemented as
\eqb{llll}
\begin{aligned}
\mk^e_\mathrm{\tau\tau} 
\quad\is \ds\int_{\Omega^e_0} \hat{\mL}_a\,\mC\,\hat{\mL}_a^{\mrT}\,\dif A~, \\[2mm]
\mk^e_{\tau M}  
\quad\is  \ds\int_{\Omega^e_0}  \hat{\mL}_a \,\mD\, \hat{\mG}_n^\mrT\, \dif A~, \\[2mm] 
\mk^e_{M\tau}
\quad \is  \ds\int_{\Omega^e_0}  \hat{\mG}_n \,\mE\, \hat{\mL}_a^\mrT\, \dif A~, \\[2mm] 
\mk^e_{MM} \quad \is  \ds\int_{\Omega^e_0}  \hat{\mG}_n\,\mF\,\hat{\mG}_n^{\mrT} \, \dif A~,
\end{aligned}
\quad\quad\quad
\begin{aligned}
\mk^e_{\tau\bar{M}} \quad\is \ds\int_{\Omega_0^e} \hat{\mL}_a\,\bar\mD\,\hat{\mG}_a^{\mrT}  \,\dif A~,\\[2mm]
\mk^e_{\bar{M}\tau} \quad\is \ds\int_{\Omega_0^e} \hat{\mG}_a\, \bar\mE\, \hat{\mL}_a^\mrT \,\dif A~,\\[2mm]
\mk^e_{\bar{M}\bar{M}}\quad \is \ds\int_{\Omega_0^e} \hat{\mG}_a\,\bar\mF\,\hat{\mG}_a^{\mrT} \,\dif A~,\\[2mm]
\mk^e_{M\bar{M}}\quad \is \ds\int_{\Omega_0^e}  \hat{\mG}_n\,\bar\mG\,\hat{\mG}_a^{\mrT}     \,\dif A~,\\[2mm]
\mk^e_{\bar{M}M}\quad \is \ds\int_{\Omega_0^e} \hat{\mG}_a\,\bar\mH\,\hat{\mG}_n^{\mrT} \,\dif A~.
\end{aligned}
\label{e:mat_implement}
\eqe
Compared to classical Kirchhoff-Love shell theory, the five terms on the right hand side are additional terms due to in-plane bending.

\subsection{Implementation of geometrical matrices}
Using \eqref{e:appenx_auxN} and \eqref{e:appenx_voigt}, the geometric stiffness matrices in Eq.~\eqref{e:matGeosim} can be implemented as  
\eqb{lllll}
\mk^e_\mathrm{\tau} 
 = \plus \ds\int_{\Omega^e_0} \Big(\tau^{11}\, \mN^\mrT_{,1}\,\mN_{,1} +   \tau^{22}\, \mN^\mrT_{,2}\,\mN_{,2} +  \tau^{12}\, \mN^\mrT_{,1}\,\mN_{,2} + \tau^{21}\, \mN^\mrT_{,2}\,\mN_{,1} \Big)\,\dif A~,\\[8mm]
\mk^e_\mathrm{M} 
= \mi \ds\int_{\Omega^e_0} b_M\, \Big[a^{11}\, \mL^n_1\,\mL^{n\mrT}_1 +   a^{22}\, \mL^n_2\,\mL^{n\mrT}_{2} +  a^{12}\, \big( \mL^n_1\,\mL^{n\mrT}_{2} +  \mL^n_2\,\mL^{n\mrT}_{1} \big) \Big]\,\dif A~\\[4mm]
\mi \ds\int_{\Omega^e_0}  \Big(\mL^n_1\, (\ba^{1})^\mrT + \mL^n_2\, (\ba^2)^{\mrT} \Big)\,\Big(M_0^{11}\, \mN_{;11} + M_0^{22}\, \mN_{;22} + 2\,M_0^{12}\, \mN_{;12}\Big)\,\dif A~\\[4mm]
\mi \ds\int_{\Omega^e_0}  \Big(M_0^{11}\, \mN^\mrT_{;11} + M_0^{22}\, \mN^\mrT_{;22} + 2\,M_0^{12}\, \mN^\mrT_{;12}\Big)\,\Big( \ba^1\,\mL^{n\mrT}_1 + \ba^2\,\mL^{\mrT}n_2 \Big)\,\dif A~,\\[8mm]
\mk^e_\mathrm{\bar{M}} = \mi \ds\int_{\Omega^e_0}  \Big(\bar{M}_0^{11}\, \mN^\mrT_{,1}\,\mC_{,1} +   \bar{M}_0^{22}\, \mN^\mrT_{,2}\,\mC_{,2} +  \bar{M}_0^{12}\, \mN^\mrT_{,1}\,\mC_{,2} + \bar{M}_0^{21}\, \mN^\mrT_{,2}\,\mC_{,1} \Big)\,\dif A~\\[4mm]
 \mi \ds\int_{\Omega^e_0}  \Big(\bar{M}_0^{11}\, \mC^\mrT_{,1}\,\mN_{,1} +   \bar{M}_0^{22}\, \mC^\mrT_{,2} \,\mN_{,2}+  \bar{M}_0^{12}\, \mC^\mrT_{,2}\,\mN_{,1} + \bar{M}_0^{21}\, \mC^\mrT_{,1}\,\mN_{,2} \Big)\,\dif A~\\[4mm]
 \mi \ds\int_{\Omega^e_0}  \Big( \mN^\mrT_{,1}\, \bP^{11}\, \mN_{,1} +   \mN^\mrT_{,2}\,\bP^{22}\,\mN_{,2} +   \mN^\mrT_{,1}\,\bP^{12}\,\mN_{,2} +  \mN^\mrT_{,2}\,\bP^{21}\,\mN_{,1} \Big)\,\dif A~\\[4mm]
 \mi \ds\int_{\Omega^e_0}  \Big[ \mN^\mrT_{,1}\, \mQ^{1} +   \mN^\mrT_{,2}\,\mQ^{2} + (\mQ^{1})^\mrT \,\mN_{,1}+   (\mQ^{2})^\mrT\,\mN_{,2} \Big]\,\dif A~,
 \label{e:kgeo_imp}
\eqe
where we have defined  
$b_M:= b_{\alpha\beta}\,M_0^{\alpha\beta}$, and $\mQ^{\alpha}:= \bQ^{\alpha\beta\gamma}\,\mN_{,\beta\gamma}$~. The last term $\mk^e_\mathrm{\bar{M}}$ is associated with in-plane bending.

\begin{remark}
We note the term $ a^{12}\, \big( \mL^n_1\,\mL^{n\mrT}_{2} +  \mL^n_2\,\mL^{n\mrT}_{1} \big) $  in $\mk^e_\mathrm{M}$ is given incorrectly by $2\,a^{12}\, \mL^n_1\,\mL^{n\mrT}_{2}$ in Duong et al.~\cite{solidshell} (cf.~Appendix B.1, Eq.~(135.1)).
\end{remark}

\vspace{1cm}


{\Large{\bf Acknowledgements}}

The authors are grateful to the German Research Foundation (DFG)
for supporting this research under grants IT67/18-1 and SA1822/11-1. They thank Vu Ngoc Khi{\^e}m for his fruitful comments on the material modeling. 

\bigskip

\bibliographystyle{unsrt}

\bibliography{sauerduong,bibliography}

\end{document}